\documentclass[trackchanges]{aastex701}

\usepackage{makecell}
\usepackage{graphicx}
\usepackage{subcaption}
\usepackage{caption} 
\usepackage{booktabs}
\usepackage{amssymb}
\usepackage{amsmath}
\usepackage{float}
\usepackage{xtab}
\usepackage{orcidlink}
\usepackage{adjustbox}

\begin{document}

\title{Anisotropy of Satellite Galaxies-I: Contrasting Correlations \\ with Central Galaxy, Host Halo, and Large-Scale Filament Structures}

\correspondingauthor{Zhuoming Zhang, Weiguang Cui, Yun Chen}
\email{zhangzm@bao.ac.cn, weiguang.cui@uam.es, chenyun@bao.ac.cn}

\author[orcid=0000-0002-4518-6035]{Zhuoming Zhang}
\affiliation{National Astronomical Observatories, Chinese Academy of Sciences, Beijing 100101, China}
\affiliation{College of Astronomy and Space Sciences, University of Chinese Academy of Sciences, Beijing, 100049, China}
\email{} 

\author[orcid=0000-0002-2113-4863]{Weiguang Cui}
\affiliation{Universidad Autónoma de Madrid: Madrid, Madrid, ES}
\affiliation{Centro de Investigación Avanzada en Física Fundamental (CIAFF), Universidad Autónoma de Madrid, Madrid~28049, Spain}
\affiliation{Institute for Astronomy, University of Edinburgh, Blackford Hill, EH9 3HJ, Edinburgh, UK}
\email{} 

\author[orcid=0000-0001-8919-7409]{Yun Chen}
\affiliation{National Astronomical Observatories, Chinese Academy of Sciences, Beijing 100101, China}
\affiliation{College of Astronomy and Space Sciences, University of Chinese Academy of Sciences, Beijing, 100049, China}
\email{} 

\author[orcid=0000-0003-2842-9434]{Romeel Dav\'e}
\affiliation{Institute for Astronomy, University of Edinburgh, Blackford Hill, EH9 3HJ, Edinburgh, UK}
\affiliation{Department of Physics and Astronomy, University of the Western Cape, Robert Sobukwe Rd, 7535, Cape Town, South Africa}
\email{} 

\author[orcid=0000-0001-6180-0245]{Katarina Kraljic}
\affiliation{Observatoire Astronomique de Strasbourg, UMR 7550, CNRS, Université de Strasbourg, F-67000 Strasbourg, France}
\email{} 

\begin{abstract}

Using the SIMBA, EAGLE, and IllustrisTNG-100 galaxy formation simulations, we examine the anisotropy of the satellite distribution and its dependencies on central galaxies, host halos, and cosmic filaments. We find that in all simulations the satellite anisotropy is robustly aligned with the halo/central galaxy major axis. This correlation is both redshift- and halo-mass-dependent and also extends to filamentary structures outside the halo to several virial radii. The alignment persists up to $z=1.5$ at high redshifts, and the mass dependence remains down to $M_\mathrm{200c} \approx 10^{11}M_{\odot}$. We identify a clear $3\sigma$ scale-dependent transition in the structural tracers of satellite anisotropy: satellite distributions correlate with central galaxy morphology at small scales ($<0.3R_{\rm 200c}$), are governed by host halo triaxiality at halo scales ($0.3$-$2R_{\rm 200c}$), and align with cosmic filaments beyond $2R_{\rm 200c}$. By tracing satellite trajectories in SIMBA, we uncover the kinematic origin of this transition, demonstrating that satellites prefer halo major-axis aligned regions because their trajectories intersect this axis far more frequently and stay in it for a longer time under the host's gravitational potential. This dynamical processing effectively erases primordial filament-related signals upon accretion ($<2R_{\rm 200c}$), explaining the shift in dominant structural tracers across scales. 
\end{abstract}


\section{Introduction} \label{sec:intro}
In the standard $\Lambda$ cold dark matter ($\Lambda$CDM) cosmological framework, mass assembly occurs hierarchically, with smaller halos forming first and being assembled into larger hosts due to gravity. Large hosts therefore contain numerous subhalos that typically host satellite galaxies.  While a naive expectation might suggest a quasi-spherical or isotropic distribution of these subhalos within the host, both observational surveys and numerical simulations consistently reveal significant direction-dependent anisotropies in the subhalo/satellite distribution \citep{2004MNRAS.348.1236S, 2005PASA...22..184K, 2005ApJ...628L.101B, 2005ApJ...629..219Z, 2008MNRAS.390.1133B}. 
These departures from isotropy are not merely stochastic fluctuations but are correlated with the geometry of the cosmic web and the orientation of the host galaxy's halo. As such, the spatial and kinematic distribution of satellite galaxies provides a critical probe into the processes of galaxy assembly and its connection to the underlying distribution of dark matter.

The investigation of satellite anisotropy in observations mostly focuses on the satellite distribution with the direction of the major and minor axes of the central galaxies. The Holmberg effect \citep{1969ArA.....5..305H, 2004MNRAS.348.1236S} initially suggested a preference for satellites to reside near the minor axes of their hosts. However, using later modern wide-field spectroscopic surveys such as the Sloan Digital Sky Survey (SDSS), an opposite trend was seen.  \citet{2005ApJ...628L.101B} 
showed that satellite samples exhibit statistically significant planar alignments—up to the 99.99\% confidence level—marking a stark departure from isotropy and contradicting the classical Holmberg effect. Furthermore, \citet{2008MNRAS.390.1133B}, \citet{2006MNRAS.369.1293Y}, and \citet{2018ApJ...859..115W} 
demonstrated that the signal of anisotropy is highly dependent on the host's morphology and stellar population, with a primary alignment along the host's major axis particularly occurring in systems with red, early-type central galaxies. This alignment seems to persist across cosmic time: \citet{2011ApJ...731...44N} 
utilized high-resolution Hubble Space Telescope imaging to confirm that such anisotropies are already well-established at intermediate redshifts ($z \sim 0.5$). In agreement with the dependence on central galaxy morphology, \citet{2023MNRAS.525.4685S} find that clusters with more elliptical central galaxies show stronger alignment with both cluster members and the surrounding large-scale structure, with this signal measurable out to $10R_\mathrm{200c}$. Additionally, the spatial positioning connects with other properties of satellite anisotropy signals such as galaxy color, e.g. recent observations have identified ``Anisotropic Satellite Galaxy Quenching" (ASGQ) \citep{2021Natur.594..187M, 2023ApJ...949L..13K, 2025A&A...693A.113Z, 2025MNRAS.537.1542S}. This phenomenon can be caused by the large-scale filaments, which undergo more rapid environmental processing and star-formation quenching compared to those in the less-dense regions of the host halo \citep{2025MNRAS.537.1542S}. In the local field environment, the satellites around the Milky Way and Andromeda galaxies show surprising flattened and co-rotating planes \citep{2012MNRAS.423.1109P, 2014Natur.511..563I}, which can be linked with their large-scale environments and accretion histories \citep{2023ApJ...942...78S, 2024MNRAS.535.3775U, 2024ApJ...965..154G, 2025arXiv250418515M, 2025MNRAS.544.2241G}.

On the simulation side, it has been long predicted that the subhalos (corresponding to satellite galaxies) exhibits intrinsically anisotropic spatial distributions, which are found to preferentially align along the major axes of triaxial dark matter halos. This reflects the filamentary nature of cosmic accretion -- both matter and satellites are funneled into host halos along filaments whose orientation correlates with the halo's principal axes \citep{2005ApJ...629..219Z, 2006EAS....20...41Z, 2007MNRAS.374...16L, 2018MNRAS.473.1562W}. 
Additional evidence comes from orbital studies. \citet{2005PASA...22..184K} found that the apocenters of satellite orbits cluster within a narrow cone ($\sim 40^{\circ}$ opening angle) around the major axis of the host halo. Their analysis linked this configuration not to internal dynamical selection but to the environmental infall patterns that govern how satellites enter the halo. This filament‑driven anisotropy also mirrors the alignments observed in galaxy clusters such as Virgo and Coma, implying a continuity of structure formation physics across scales. Furthermore, \citet{2005ApJ...629L...5L} 
developed an analytic model to demonstrate that the subhalo anisotropy originates from and its strength correlates with tidal fields and hosts' triaxial shapes. 

In this work, we utilize cosmological simulation data from three simulation projects, SIMBA, EAGLE, and TNG100, to understand the origin of the anisotropy distribution of satellite galaxies as well as its connections with halo's and central galaxy's axes.  Specifically, we focus on: (1) the dependence of the alignment trend between halo major axes and the central galaxy major axes on halo mass;  (2) the specific scales where the anisotropy signal of satellite galaxy distributions appears with respect to central galaxies and halos; (3) the evolution of these trends across different redshifts and halo masses; (4) its connection with the large scale environments, i.e. the dependence of the alignment trend between halo major axes and cosmic filament orientations on halo mass and the influence of halos and filaments on the anisotropy of satellite galaxy distributions at various scales. By performing pairwise comparisons of the samples, we clarify the fundamental correlations between satellite galaxy distributions and the orientations of central galaxies, halos, and cosmic filaments. To construct a concrete physical picture for explaining the origin of the anisotropy of satellite galaxies, we lastly investigate the influence of the dark matter distribution at different scales, and track trajectories of satellite galaxies' prior to halo infall.

This paper is organized as follows: Section \ref{sec:data} introduces the data. Section \ref{sec:methods} describes the methods. Sections \ref{sec:SDA_BCG_halo} and \ref{sec:SDA_halo_filament} present our main results and corresponding discussions, while the physical scenarios underlying the aforementioned phenomena are elaborated in Section \ref{sec:physical_insight}. Section \ref{sec:conclusions} summarizes our principal findings and conclusions.

\section{Simulation Data and Sample Selection} \label{sec:data}
Our main data set is from the SIMBA-m100n1024 simulation \citep{2019MNRAS.486.2827D}, with which all high redshift data have been used. Benefiting from its large simulation box size, we can not only retrieve information on the anisotropies of satellite galaxies within halos but also analyze this phenomenon in the large-scale environments beyond halo regions.  As a complement, we also employ and analyze high-resolution simulation data at $z=0.0$ from both the EAGLE-L100N1504 \citep{2015MNRAS.446..521S} and TNG100-1 simulations \citep{2019ComAC...6....2N}. We refer interested readers to the original papers for the details of these three different baryon models. Here, we briefly outline their primary distinctions below; readers interested in exhaustive cross-model comparisons across diverse physical dimensions are directed to  \citet{Vogelsberger2020}, \citet{Dave2020}, and \citet{Habouzit2021}:

\begin{itemize}
\item[(1)]EAGLE and TNG100 adopt the CLOUDY table with metal-line cooling for gas cooling; while the Grackle library \citep{2017MNRAS.466.2217S} is used in SIMBA for gas cooling.

\item[(2)]Their star formation prescriptions differ: EAGLE follows the model from \citet{2008MNRAS.383.1210S}, TNG100 uses the scheme described in \citet{2003MNRAS.339..289S}, and SIMBA implements an H2-based star formation model \citep{2016MNRAS.462.3265D}. 

\item[(3)]The most significant deviations between the three simulations lie in the black hole (BH) model and AGN feedback: EAGLE and TNG use Bondi accretion for BH mass growth, while SIMBA has both Torque-limited and Bondi modes. EAGLE uses thermal (quasar-like) feedback, TNG includes both thermal and kinetic (jet-like) feedback, while SIMBA implements pure kinetic outflows with different velocities to mimic the quasar- and jet-like AGN feedback. 
\end{itemize}

The adoption of high-resolution datasets enables a more detailed and robust analysis of the satellite anisotropies in low-mass halos and at small scales. Relevant information on the simulation datasets employed herein is summarized in Table \ref{tab:params_simulation}. We extracted the physical properties of halos and galaxies from the particle data of these simulations via the 6D Friends-of-Friends (6DFoF) algorithm implemented in the \texttt{CAESAR}\footnote{\tt https://caesar.readthedocs.io/en/latest/} package, followed by the implementation of data screening based on these outputs.

\begin{table}[h!]
    \hspace*{-1.5cm}
    \centering
    \setlength{\tabcolsep}{5pt}  
    \renewcommand{\arraystretch}{1.5}  
    \begin{tabular}{lccccc}
        \toprule
        Name & $z$ & $L_\mathrm{box} \, [\mathrm{Mpc}]$ & $N_\mathrm{DM}$ & $m_\mathrm{gas} \, [M_{\odot}]$ & $m_\mathrm{DM} \, [M_{\odot}]$\\
        \midrule
        SIMBA-m100n1024 & $0.0 - 1.5$ & 147.1 & $1024^3$ & $1.82\times10^{7}$ & $9.60\times10^{7}$\\
        EAGLE-L100N1504 & 0.0 & 100.0 & $1504^3$ & $1.81\times10^{6}$ & $9.70\times10^{6}$\\
        TNG100-1 & 0.0 & 110.7 & $1820^3$ & $1.40\times10^{6}$ & $7.50\times10^{6}$\\
        \bottomrule
    \end{tabular}
    \caption{Table of physical parameters for three simulation projects. From left-to-right the columns show: simulation name suffix; redshift; comoving box length; number of dark matter particles; initial gas particle mass; dark matter particle mass.}
    \label{tab:params_simulation}
\end{table}

For the aforementioned simulation datasets, we performed an initial mass-based selection of halos and their central galaxy samples. We adopt distinct halo mass thresholds for different analyses. For studies on scale-dependent anisotropy, we adopt different halo mass thresholds to guarantee robust resolution of internal structures, tailored to the mass resolution of each simulation suite. Specifically, we select halos with $M_{200\mathrm{c}} \ge 10^{13}\,M_\odot$ for SIMBA, while the lower limit is set to $M_{200\mathrm{c}} \ge 10^{12}\,M_\odot$ for TNG100 and EAGLE. To explore the anisotropy–mass relation, we uniformly extend the lower limit to $M_{200\mathrm{c}} \ge 10^{11}\,M_\odot$. Additionally, all selected halos are required to host central galaxies with over 100 stellar particles. For further refinement of the sample, we required the halos to be dynamically relaxed systems with the central galaxy securely associated with the halo center. To this end, we imposed an additional selection criterion: the offset between the central galaxy and the halo's minimum potential point must be less than $0.05R_{200\rm c}$, where $R_{200\rm c}$ is the equivalent radius that would enclose $M_{200\rm c}$ at an overdensity of 200 times the critical. This criterion is applied to halos from the SIMBA and TNG100 simulations, but we skip it for EAGLE counterparts, as EAGLE snapshot data do not provide particle-derived minimum potential position. It should be noted that the \texttt{CAESAR} package designates the most massive galaxy within a given halo as the central galaxy during the galaxy catalog generation process. This selection, though not perfect for relaxed halos \citep{2017MNRAS.464.2502C, 2020MNRAS.492.6074H, 2021MNRAS.504.5383D, 2022MNRAS.516...26Z}, 
avoids the uncertainties on (1) central galaxy selection; and (2) the influence on major axes estimation from infalling groups. 

When considering both central galaxies and other galaxies in their vicinity, we also impose a mass constraint on these galaxies, requiring their stellar mass to be not less than $100\times m_\mathrm{gas}$. This lower stellar mass limit is primarily motivated by the finite mass resolution of the simulation, which limits the reliability of low-mass subhalos and their associated galaxies. Imposing this threshold ensures that only robustly resolved systems are included in our analysis, avoiding potential biases from poorly resolved objects. Meanwhile, to enhance the reliability of the identified satellite galaxies, the distance between these selected satellite galaxies and their corresponding central galaxies must be greater than twice the half total mass radius of the central galaxies ($2r_\mathrm{c,\, half}$) and less than $10R_{200\rm c}$. It is important to clarify that the satellite galaxies considered in this work are not required to be gravitationally bound, and thus have a broader connotation than satellite galaxies in the general sense. If not specifically mentioned, the aforementioned selection criteria are applied to all data in this paper.

For filament identification, we employed the \texttt{DisPerSE} algorithm \citep{2011MNRAS.414..350S, 2011MNRAS.414..384S} relying on discrete Morse theory \citep{forman2002user}, which was run on the distribution of galaxies. To deal with such a discrete data set, \texttt{DisPerSE} builds on the Delaunay tessellation allowing one to provide a scale-free Delaunay Tessellation Field Estimator \citep{2000A&A...363L..29S} density and reconstruct the local topology. To filter out the topologically less robust features, the commonly adopted persistence threshold $N_\mathrm{\sigma}=3$ was used \citep{2011MNRAS.414..350S, 2011MNRAS.414..384S}.

\section{Methods} \label{sec:methods}

\subsection{Orientation Localization of Central Galaxies, Host Halos and Filaments}
\label{subsec:Orientation_BCGs_halos_webs}

When determining the triaxial ellipsoid orientation of the central galaxy and its host halo, we follow the four-step procedure below:
\begin{itemize}
    \item \textbf{Step 1: Define Particle Sample Boundaries.} First, we specify the spatial range for the particle samples to be processed and extract particles within this range. For the central galaxy, we select stellar particles within a radius of its $2r_\mathrm{c,\, half}$ centered on the central galaxy; for the host halo, we select dark matter particles within the $R_{200\rm c}$ radius.

    \item \textbf{Step 2: Isotropic Random Sampling with Sphere Selection.} Next, we randomly place small sampling spheres isotropically within the aforementioned sample volume, where the radius of each sampling sphere is 1/20 that of the sample volume. A sampling sphere is retained if its internal particle number density is no less than the average number density $\bar{\rho}$ of the sample volume. This process is repeated independently until 5,000 valid sampling spheres are retained, which allows us to characterize the dominant morphology of the central galaxy and its host halo.

    \item \textbf{Step 3: Substructure Removal and Sample Quality Cuts.} We then iterate over all galaxies surrounding the central galaxy without applying any mass constraints. For each surrounding galaxy, we exclude all sampling spheres within twice the half total mass radius centered on that galaxy. This substructure removal mitigates shape measurement biases for both the host halo and the central galaxy, albeit with different necessity levels. For the host halo, embedded substructures skew the measurement of the ellipsoidal principal axes, rendering this step indispensable for accurate halo shape inference; for the central galaxy, we adopt this extra cleaning step only as a conservative check to exclude a portion of ongoing major mergers and stabilize the resulting shape parameters. Halos (and their associated central galaxies) are discarded if the number of remaining sampling spheres falls below 3,500. Only a limited fraction of halos are discarded across the three simulations, with rejection rates of approximately 0.02\% (SIMBA), 0.13\% (EAGLE) and 0.31\% (TNG100).\\.

    \item \textbf{Step 4: Triaxial Ellipsoid Calculation.} Finally, we apply the Principal Component Analysis (\texttt{PCA}) method from the \texttt{sklearn} Python package \citep{scikit-learn} to the remaining sampling spheres. This implementation uses singular value decomposition (SVD) for numerically robust determination of the three principal axes, from which we obtain the axis lengths and orientations characterizing the triaxial ellipsoids of central galaxies and their host halos.

\end{itemize}

Given that the placement of sampling spheres relies on a stochastic process, we repeat the entire procedure 50 times for each central galaxy-halo system. The average triaxial ellipsoid orientation obtained from these 50 iterations is  denoted $\{\vec{e}_\mathrm{major, ave}, \, \vec{e}_\mathrm{median, ave}, \, \vec{e}_\mathrm{minor, ave}\}$. In Appendix \ref{sec:PCA_ITM}, we compare our method with the traditional inertia tensor method (ITM)—where triaxial ellipsoid orientations are determined from the eigenvectors of the particle inertia tensor. Under the assumption of equal particle weights, the principal axes obtained from PCA of the coordinate covariance matrix are mathematically equivalent to those from the ITM. The comparison demonstrates that the orientations derived from the two methods are statistically consistent for halos with $M_\mathrm{200c} \ge 10^{13} M_{\odot}$. The PCA-based method for determining the triaxial ellipsoid orientation has the following useful properties:

\begin{itemize}
\item[(1)] It enables the quantification of the uncertainty in triaxial ellipsoid orientation via multiple stochastic realizations;

\item[(2)] It can be implemented without requiring particle mass information, yielding a more robust measurement of the overall geometry of the structure.

\item[(3)] No assumption is required that the central galaxy or halo profile conforms to a specific functional form.
\end{itemize}

For each individual halo, we extracted the associated large-scale, galaxy-traced filament segments lying between $3R_\mathrm{200c}$ and $5R_\mathrm{200c}$. All these filament segments are topological outputs reconstructed by the \texttt{DisPerSE} algorithm described in Section \ref{sec:data}, traced using galaxies with stellar mass above $10^{8.5}M_\odot$. Each set of interconnected filament segments was identified as a single filament linked to the halo, and the average position vector of the set was defined as the orientation of the filament. Disconnected filament structures are treated independently, so one halo may host multiple filaments with distinct orientations.

\subsection{Statistical Counting of Satellite Galaxies}

To quantify satellite orientation, we implement an orientation-specific counting method anchored to various reference objects, as follows. First, we define target orientations for each structure based on four categories of reference frames:

\begin{itemize} 
\item \textbf{Central Galaxy:} three independent orientations corresponding to the triaxial directions of the central galaxy ellipsoid, with a separate count performed for each direction;

\item \textbf{Host Halo:} three independent orientations corresponding to the triaxial directions of the halo ellipsoid, with one count executed per direction;

\item \textbf{Cosmic Filaments:} all filament directions associated with the target halo, with a count conducted for each individual filament direction;

\item \textbf{Random Orientations:} a set of randomized directions for control purposes, with an independent count carried out for each random direction.
\end{itemize}

To calculate the total number of satellite galaxies along one specified direction, we take the coordinates of the central galaxy as the cone apex and define a double-napped cone region, with each pre-defined target orientation serving as the central axis and a specified conical half-apex angle $\theta$. In this work, we set $\theta=45^\circ$ to ensure a sufficiently large sample size for robust statistical analysis, and as illustrated in Appendix \ref{sec:proj_distribution_galaxy} via the anisotropic distribution of projected satellite galaxies, the anisotropic signal is physically real rather than a spurious feature arising from our selection of $\theta$. For each orientation category (e.g., all major axes of central galaxies across the sample), we first count the number of satellite galaxies matching the orientation for each individual structure and quantify the population of satellite galaxies within the corresponding bicone region; we then sum the counts across all halos for the same orientation type. Therefore, for galaxy counts performed along a specific orientation, the resultant errors originate solely from the uncertainty in the determination of the chosen reference system's orientation.

\section{Contrasting Correlations: Satellite Distribution Anisotropy \\ with Central galaxies Morphologies vs. with Halos Morphologies}
\label{sec:SDA_BCG_halo}

\subsection{Sample Classification}\label{subsec:BCG_halo_sample}

\begin{figure}[htbp]
    \centering
    \begin{minipage}{0.33\textwidth}
        \centering
        \includegraphics[width=\linewidth]{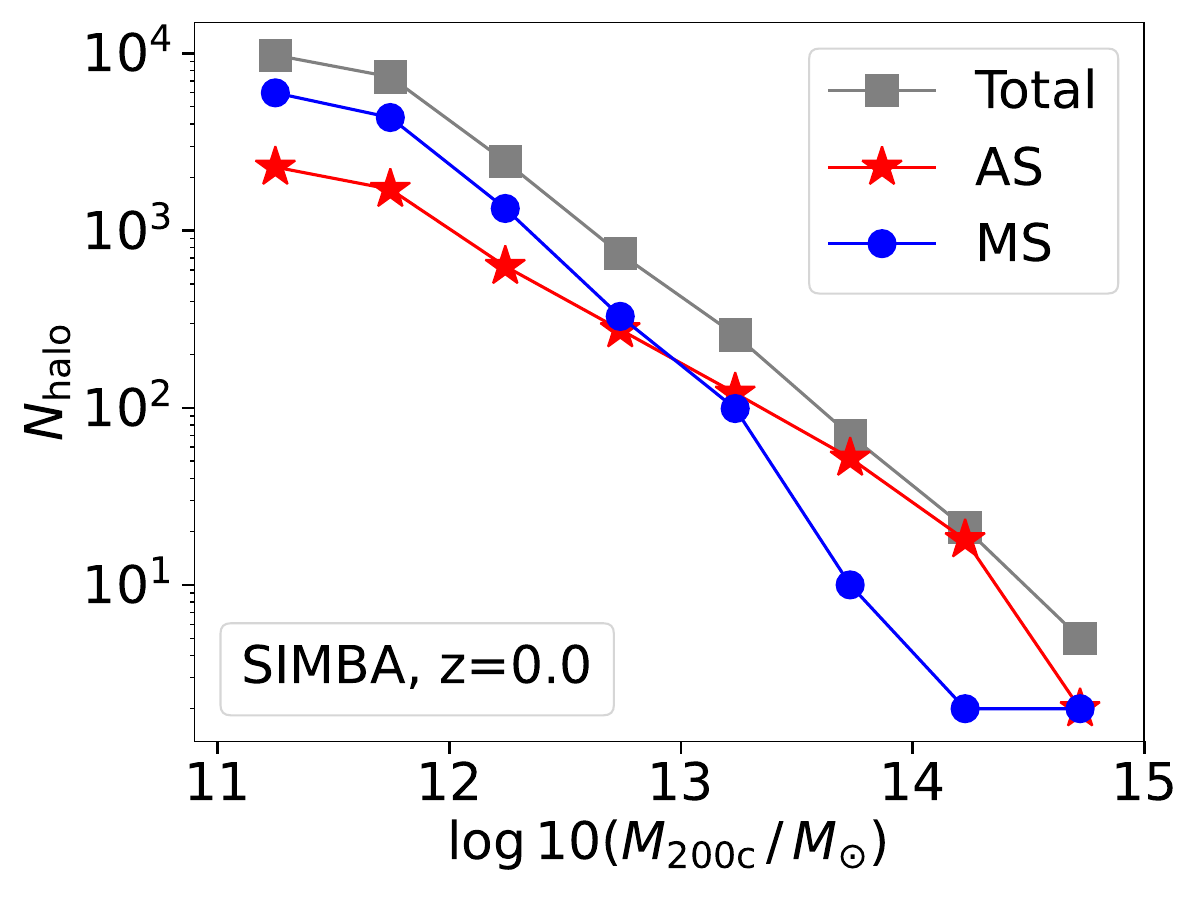}
    \end{minipage}
    \hspace{-5pt}
    \begin{minipage}{0.33\textwidth}
        \centering
        \includegraphics[width=\linewidth]{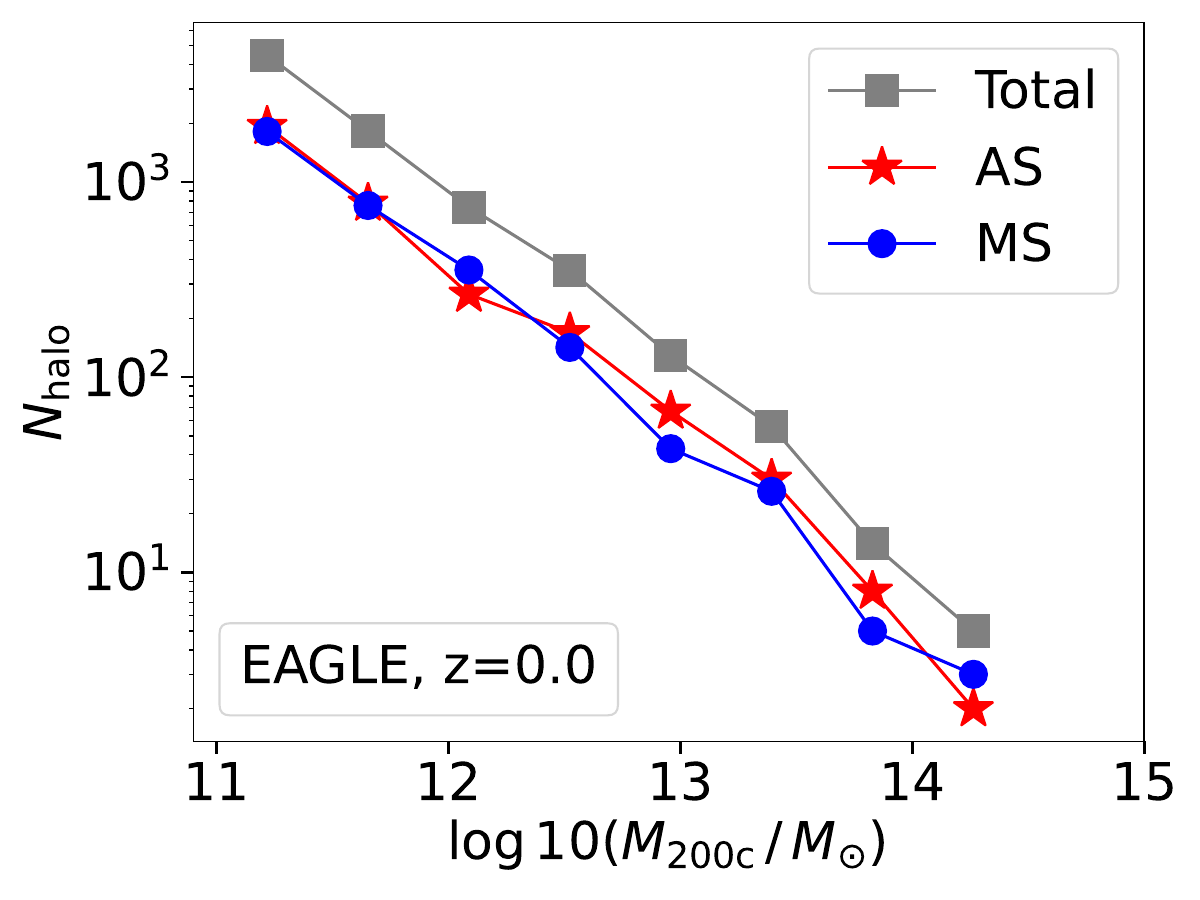}
    \end{minipage}
    \hspace{-5pt}
    \begin{minipage}{0.33\textwidth}
        \centering
        \includegraphics[width=\linewidth]{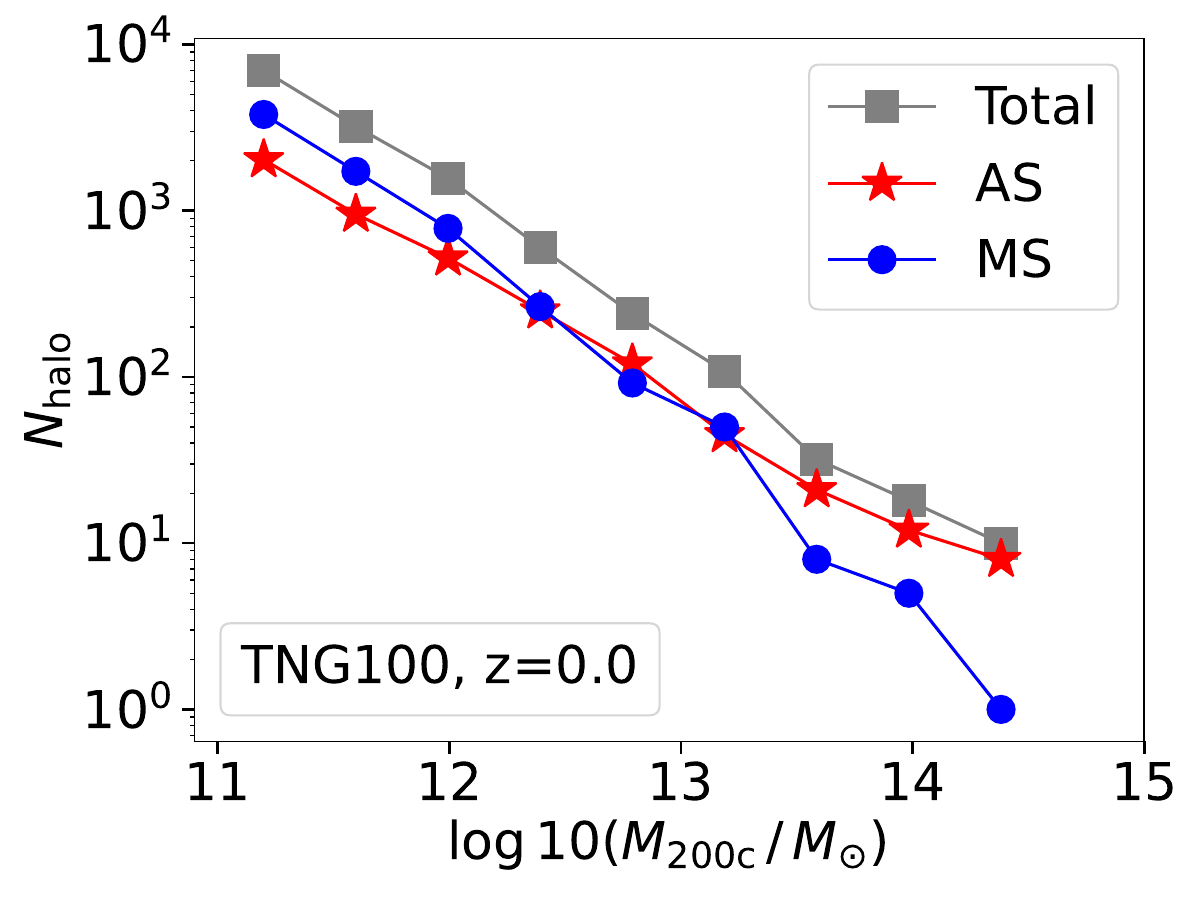}
    \end{minipage}
    \caption{Number of halos as a function of halo mass for the three cosmological simulations: SIMBA, EAGLE, and TNG100. A deficit of massive halos can be seen in the MS relative to the AS.}
    \label{fig:simulation_mass_distribution}
\end{figure}

In a halo, the major axis orientation of the central galaxy does not always align with that of its host halo \citep{2016MNRAS.460.3772S, 2020MNRAS.496.2591O}. Based on this, we classify the systems into two groups: those for which the angle between the major axes of the central galaxy and its host halo is less than $40^\circ$ are assigned to the aligned sample (AS); those for which the angle exceeds $50^\circ$ are assigned to the misaligned sample (MS). Furthermore, we compute the halo number-halo mass relations for the total sample, the AS, and the MS across three cosmological simulations at $z=0$, with the results presented in Figure \ref{fig:simulation_mass_distribution}. We find that massive halos are less abundant in the MS compared to the AS, and this trend is particularly prominent in the SIMBA and TNG100 simulations. Note that low-mass halos may lack satellite galaxies that satisfy our mass selection criteria. Accordingly, the numbers of halos effectively contributing to the subsequent analyses are listed in Tables \ref{tab:data_information_spatial} and \ref{tab:data_information_M200c}.

\begin{figure}[htbp]
    \centering
    \includegraphics[width=0.50\textwidth]{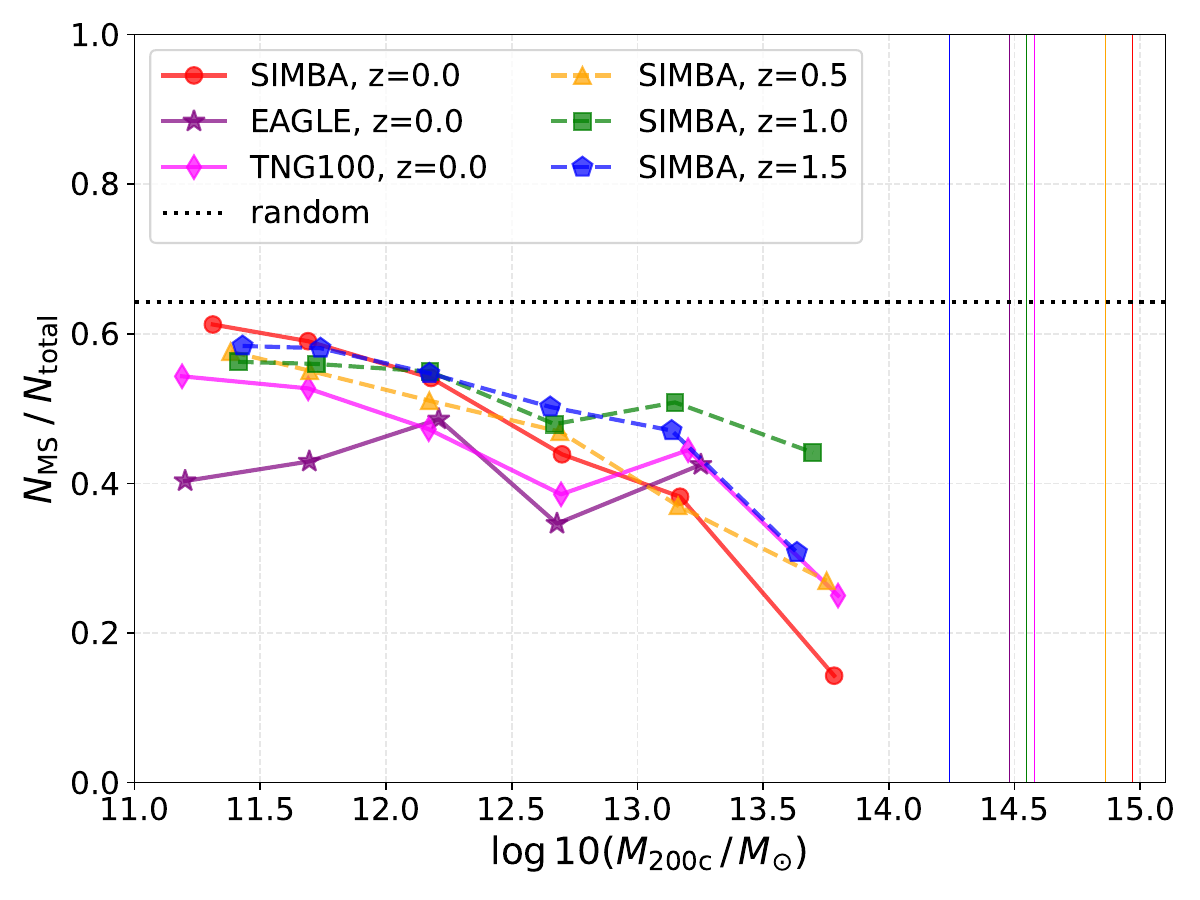}
    \caption{Misaligned fraction of central galaxy and halo major axes vs. halo $M_{\rm 200c}$ across simulations and redshifts. The vertical thin lines correspond to the maximum $M_\mathrm{200c}$ of the halo data in the same color. The misaligned fractions corresponding to the datasets consistently decrease with increasing halo mass.}
    \label{fig:halo_BCG_misalign_rate}
\end{figure}

We further explore the dependence of the misalignment probability on halo $M_{200\rm c}$. Figure \ref{fig:halo_BCG_misalign_rate} shows the relationship between the ratio of misaligned samples to the total samples and halo $M_{200\rm c}$ from the simulation data used in this work. At $z=0$, except that the proportion of misaligned samples in low halo mass derived from the EAGLE data is significantly lower than that from SIMBA and TNG, all datasets exhibit the same trend: the higher the halo mass, the lower the probability of it being in a misaligned state. This conclusion is also in agreement with those of \citet{2016IAUS..308..448K}, \citet{2024A&A...688A..40R}, and \citet{2025A&A...703A..67M}, who use different simulation data. On the other hand, using SIMBA data at $z=0.5$, $1.0$ and $1.5$, we confirm that this conclusion remains robustly valid, albeit that the misalignment rates tend to increase with redshift. We further find that even for low-mass halos with $M_{200\rm c}$ slightly above $10^{11}M_{\odot}$, misaligned cases are still lower than the probability of misalignment for random orientations $(P=0.64)$. This indicates that, statistically speaking, the major axis orientation of the central galaxy tends to be aligned with its host halo's major axis.

\subsection{Anisotropies of Satellite Galaxies across Spatial Scales and Cosmic Time}\label{subsec:Nx_Nz_R_halo_BCG}

\begin{figure}[htbp]
    \centering
    \begin{minipage}{0.33\textwidth}
        \centering
        \includegraphics[width=\linewidth]{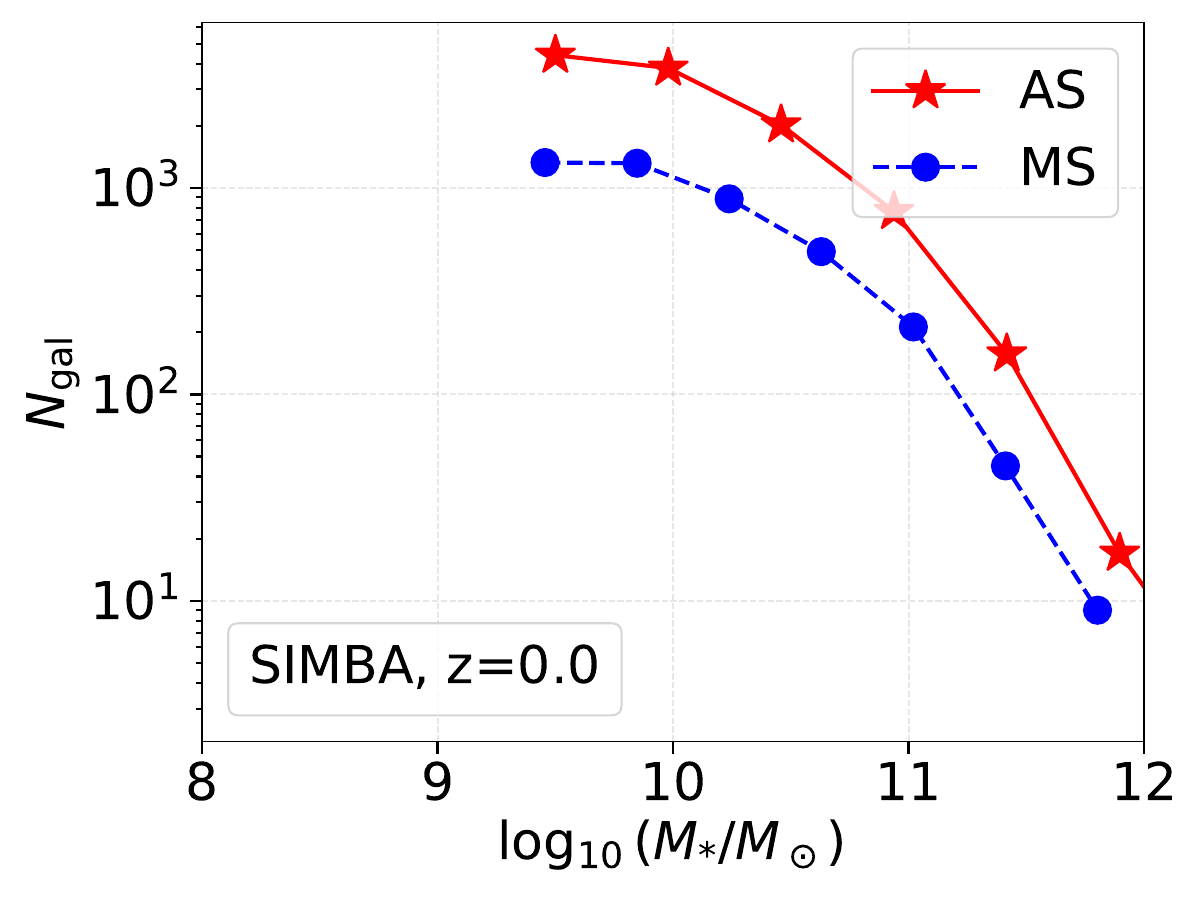}
    \end{minipage}
    \hspace{-3pt}
    \begin{minipage}{0.33\textwidth}
        \centering
        \includegraphics[width=\linewidth]{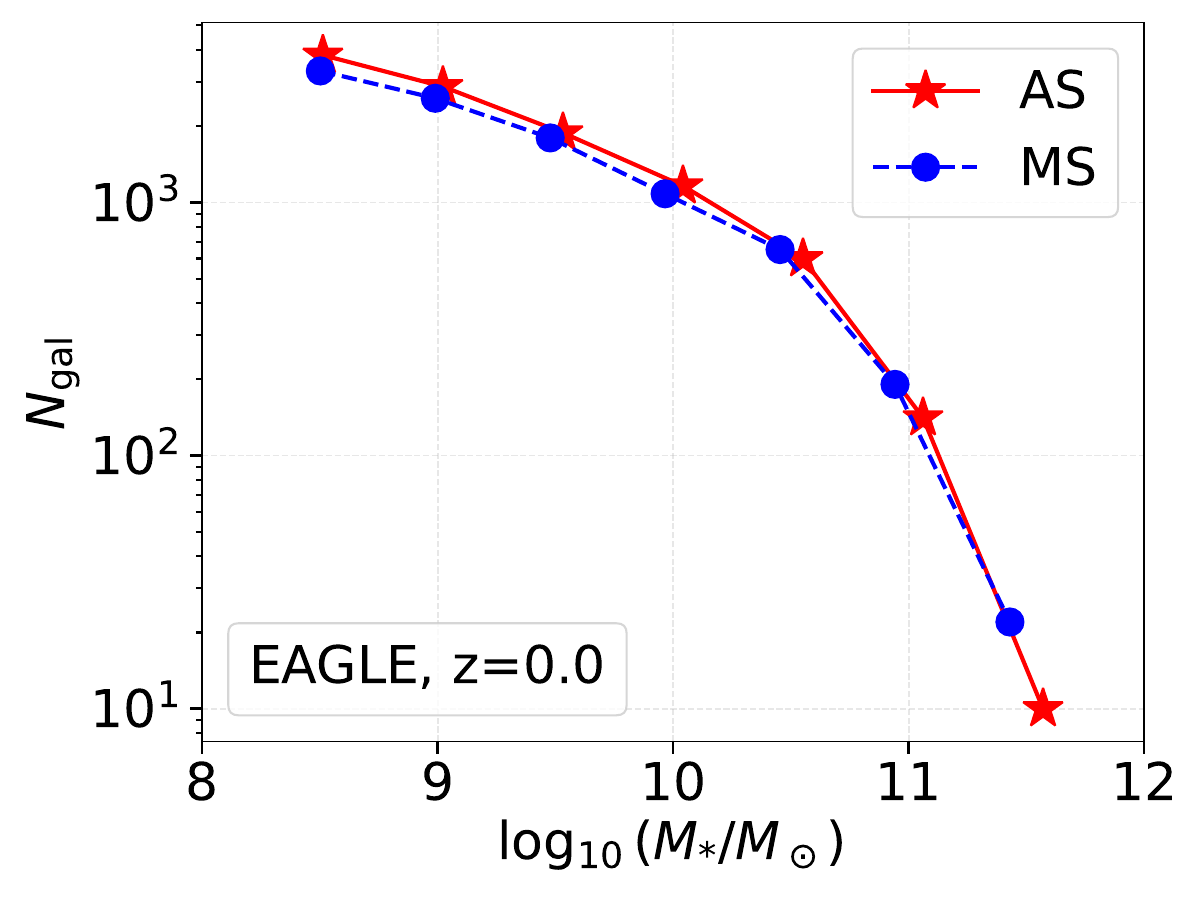}
    \end{minipage}
    \hspace{-3pt}
    \begin{minipage}{0.33\textwidth}
        \centering
        \includegraphics[width=\linewidth]{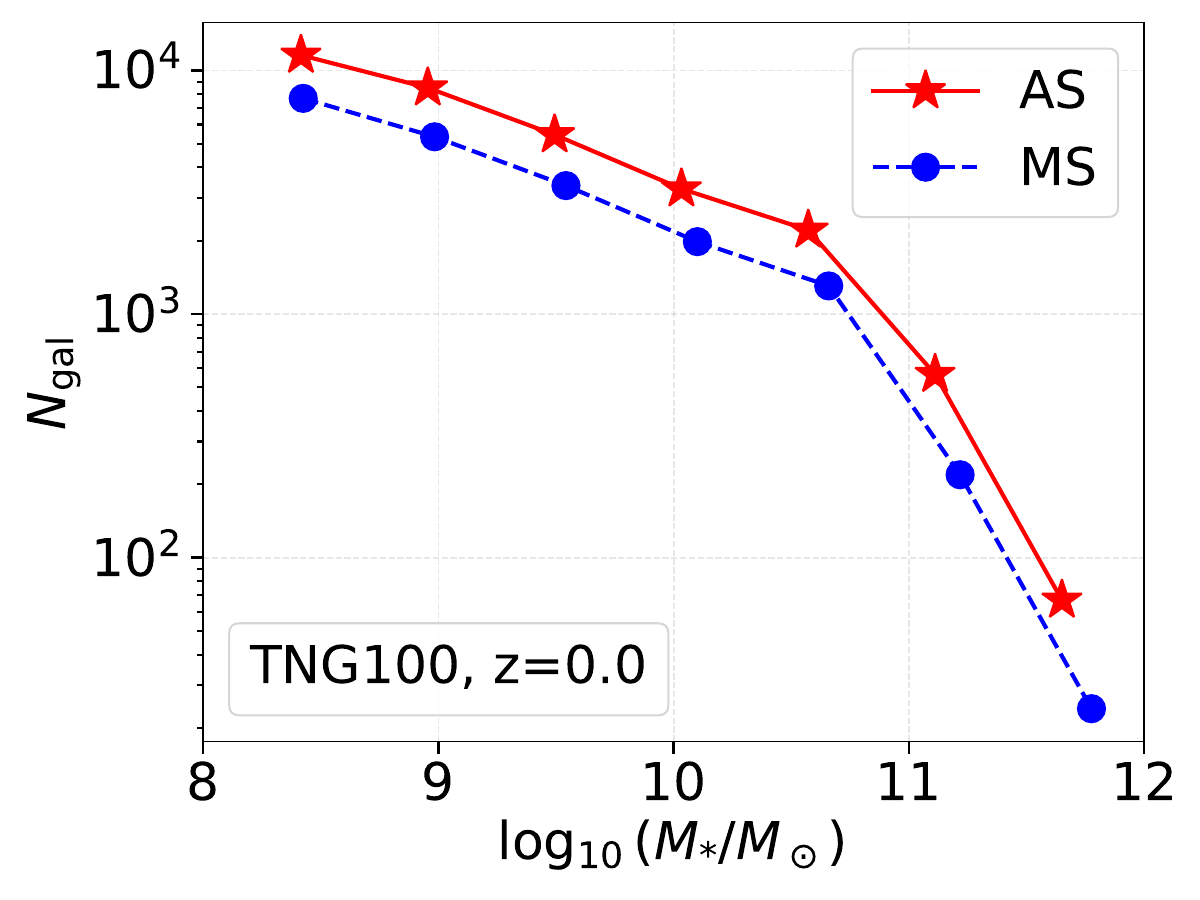}
    \end{minipage}
    \caption{Number of galaxies as a function of stellar mass. We adopt halo mass cuts of $M_\mathrm{200c}\ge 10^{13}\,M_\odot$ (SIMBA) and $M_\mathrm{200c}\ge 10^{12}\,M_\odot$ (EAGLE, TNG100), alongside a radial cut $2r_\mathrm{c,\, half}<R<10R_\mathrm{200c}$ for satellite galaxies. Galaxy counts show similar trends for AS and MS across all stellar masses.}
    \label{fig:sate_starmass_10R200c_distribution}
\end{figure}

When analyzing the anisotropy signal of satellite galaxy distributions, simulation baryon models, spatial resolution and volume can play important roles. While the baryon models control the simulated galaxy properties, the smoothing scale determines whether small-scale satellite galaxy distributions can be accurately resolved, and the simulation box volume restricts the maximum scale accessible for large-scale structure studies which determines the maximum halo mass and statistics. Therefore, a comprehensive cross-analysis of diverse simulation datasets enables robust cross-validation, which is vital for strengthening the reliability of the results. To account for the effects of the simulation resolution, we impose different halo masses constraints for the three simulations. Specifically, we adopt a cutoff of $M_{200\rm c}\ge10^{13}M_{\odot}$ for SIMBA simulations, whereas lower thresholds of $M_{200\rm c}\ge10^{12}M_{\odot}$ are applied to EAGLE and TNG100 datasets \footnote{According to the classical definition, the dark matter halo mass range for galaxy groups is $10^{12} M_{\odot} < M_{200\rm c} < 10^{14} M_{\odot}$, and for galaxy clusters it is $10^{14} M_{\odot} < M_{200\rm c} < 10^{15} M_{\odot}$. Therefore, based on our selection criterion, the objects we identify include both galaxy clusters and galaxy groups.}. This difference arises because low-mass halos are better captured by higher resolution simulations. For satellite galaxies, we only retain those located in the radial range $2r_\mathrm{c,\,half} < R < 10R_\mathrm{200c}$ for our analysis. The final selection criteria for the simulation data mentioned in Section \ref{sec:data}, as well as the information on the aligned and misaligned samples obtained, are presented in Table \ref{tab:data_information_spatial}. The corresponding relations between satellite galaxy number and stellar mass are displayed in Figure \ref{fig:sate_starmass_10R200c_distribution}, where we find no statistically significant differences in satellite distributions between the AS and MS.

\begin{table}[h!]
    \centering
    \setlength{\tabcolsep}{5pt}  
    \renewcommand{\arraystretch}{1.5}  
    \begin{tabular}{lccccc}
        \toprule
Name & $z$ & \, $N_\mathrm{c}$ (AS, MS) \, & \, $N_\mathrm{s}$ (AS, MS) \, & $M_\mathrm{200c,\, min} \, [M_{\odot}]$ & $M_\mathrm{*s,\, min} \, [M_{\odot}]$\\
        \midrule
        SIMBA-m100n1024 & 0.0 & 187, 107 & 11168, 4285 & $1\times10^{13}$ & $1.82\times10^{9}$\\
        SIMBA-m100n1024 & 0.5 & 112, 100 & 7356, 4939 & $1\times10^{13}$ & $1.82\times10^{9}$\\
        SIMBA-m100n1024 & 1.0 & 48, 80 & 2395, 3885 & $1\times10^{13}$ & $1.82\times10^{9}$\\
        SIMBA-m100n1024 & 1.5 & 28, 53 & 1377, 1941 & $1\times10^{13}$ & $1.82\times10^{9}$\\
        EAGLE-L100N1504 & 0.0 & 411, 424 & 10474, 9623 & $1\times10^{12}$ & $1.81\times10^{8}$\\
        TNG100-1 & 0.0 & 671, 727 & 31605, 19924 & $1\times10^{12}$ & $1.40\times10^{8}$\\
        \bottomrule
    \end{tabular}
    \caption{Final selection criteria and aligned/misaligned sample information for the simulation suites, employed in the analysis of anisotropy across different spatial scales. From left to right the columns show: simulation name suffix; redshift; number of central galaxies (AS, MS); number of satellite galaxies (AS, MS); minimum halo mass; minimum satellite galaxy stellar mass. Satellite galaxies are selected within $2r_\mathrm{c,\, half}<R<10R_\mathrm{200c}$.}
    \label{tab:data_information_spatial}
\end{table}

\begin{figure}[h!]
\includegraphics[width=0.5\textwidth]{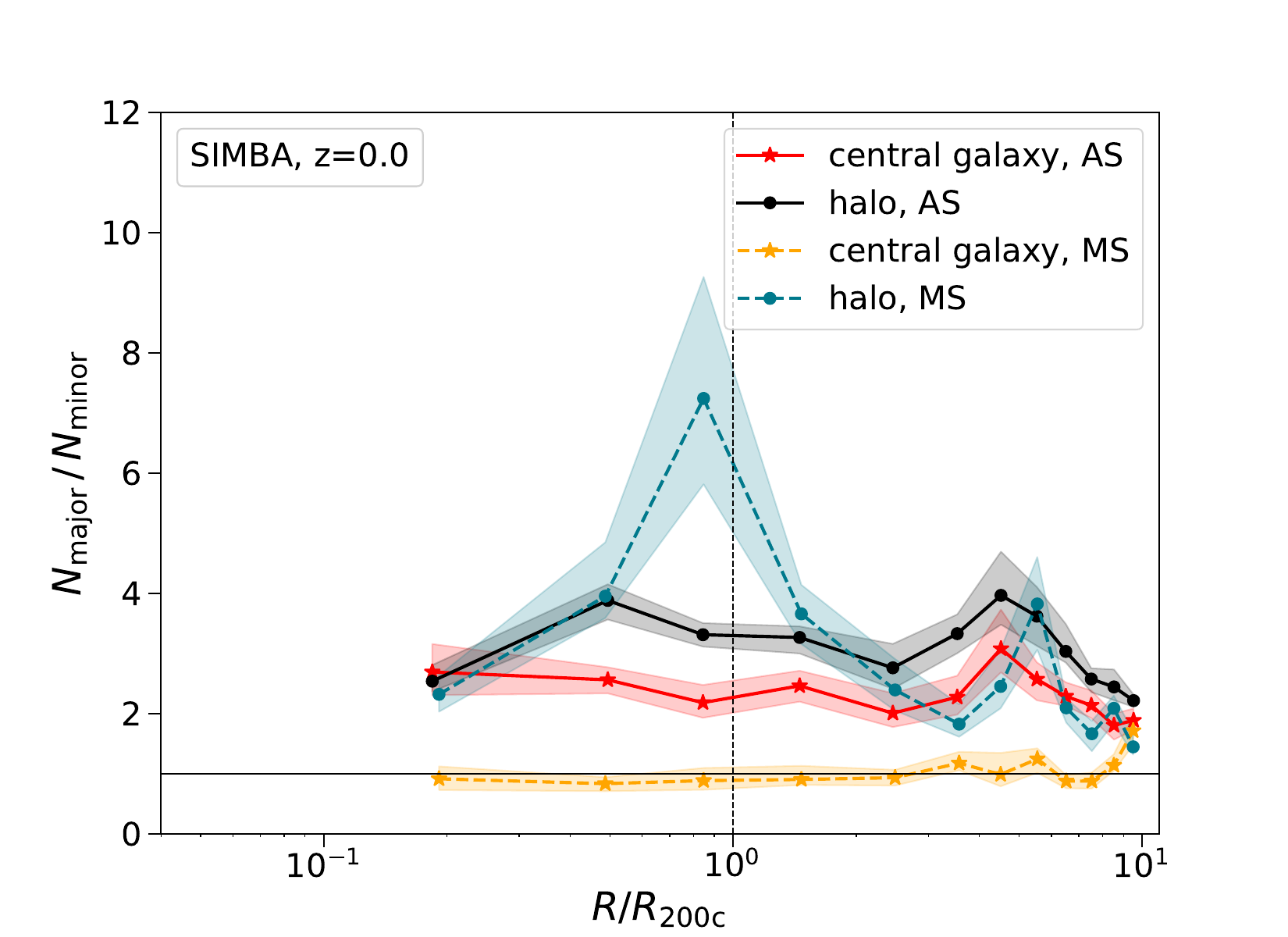}
\hspace{-25pt}
\includegraphics[width=0.5\textwidth]{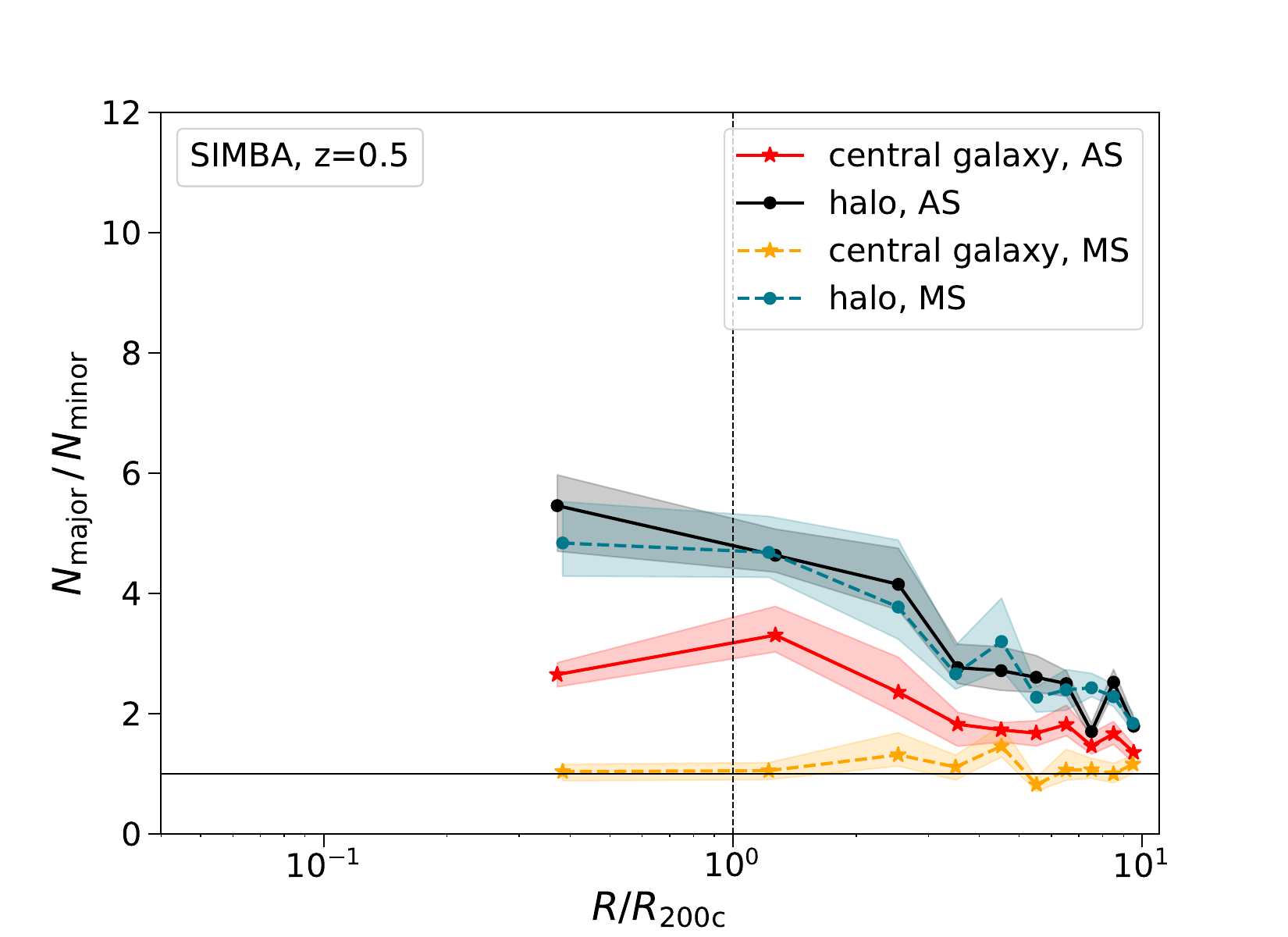}

\vspace{-2.5pt}

\includegraphics[width=0.5\textwidth]{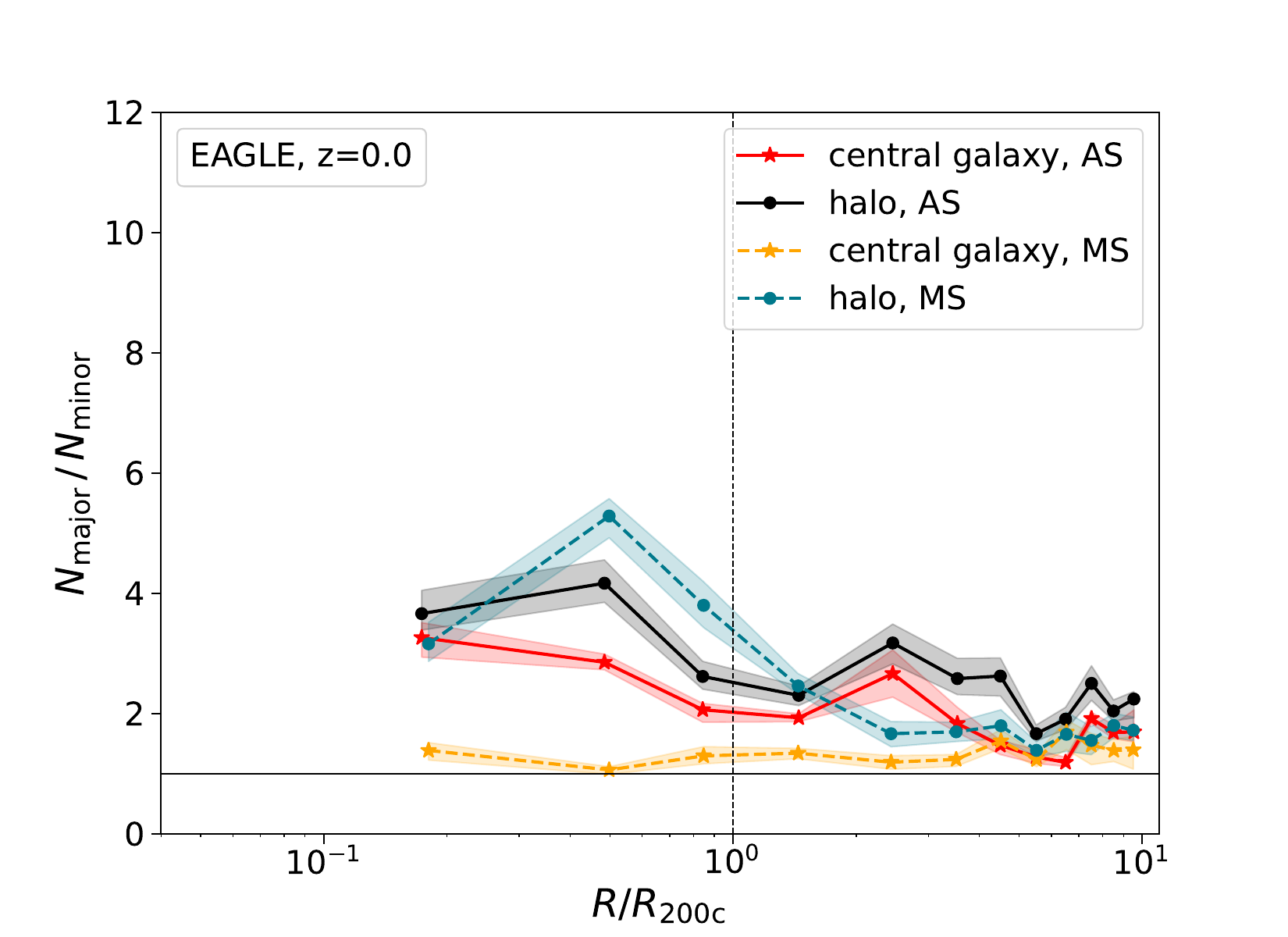}
\hspace{-25pt}
\includegraphics[width=0.5\textwidth]{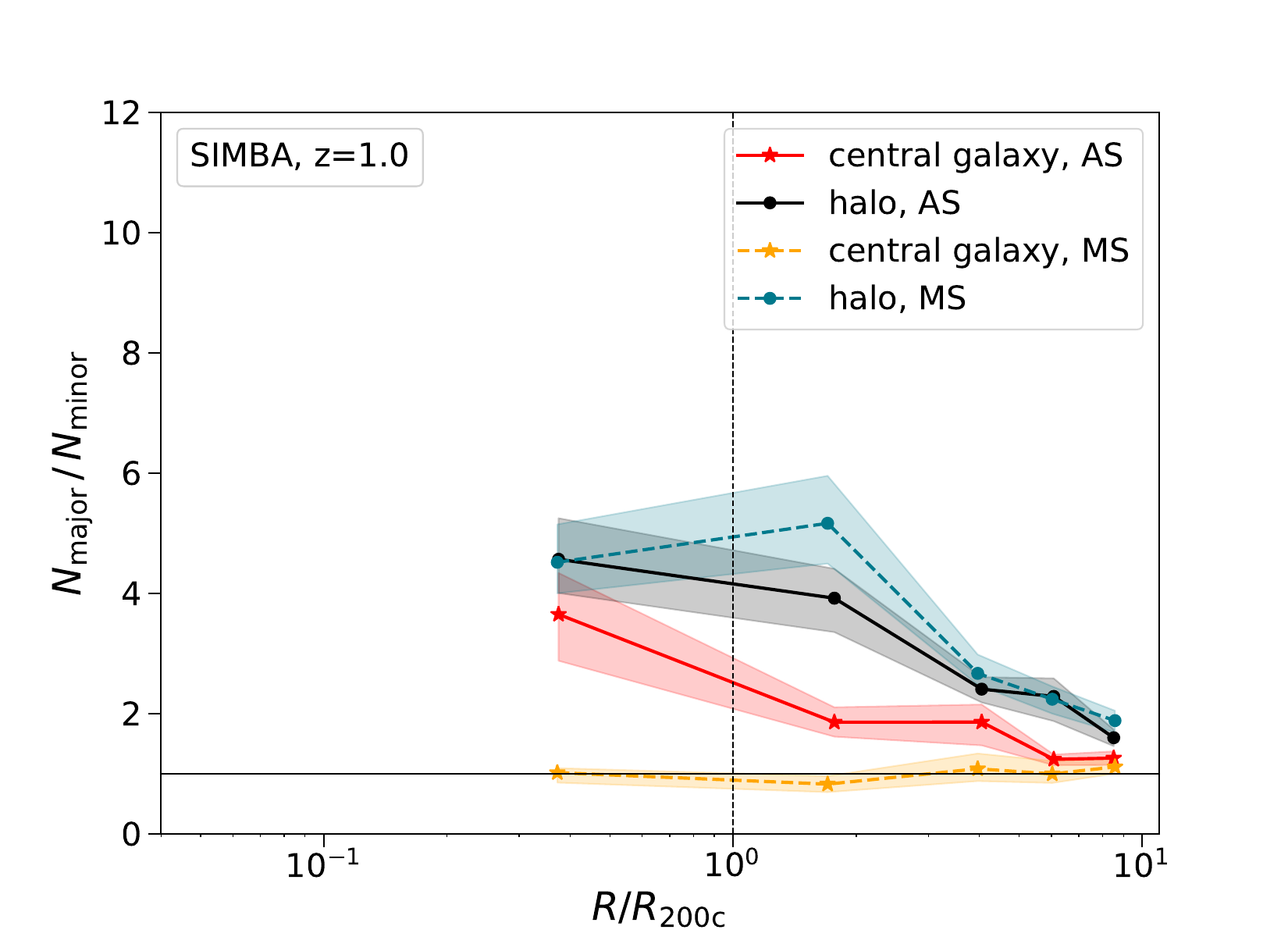}

\vspace{-2.5pt}

\includegraphics[width=0.5\textwidth]{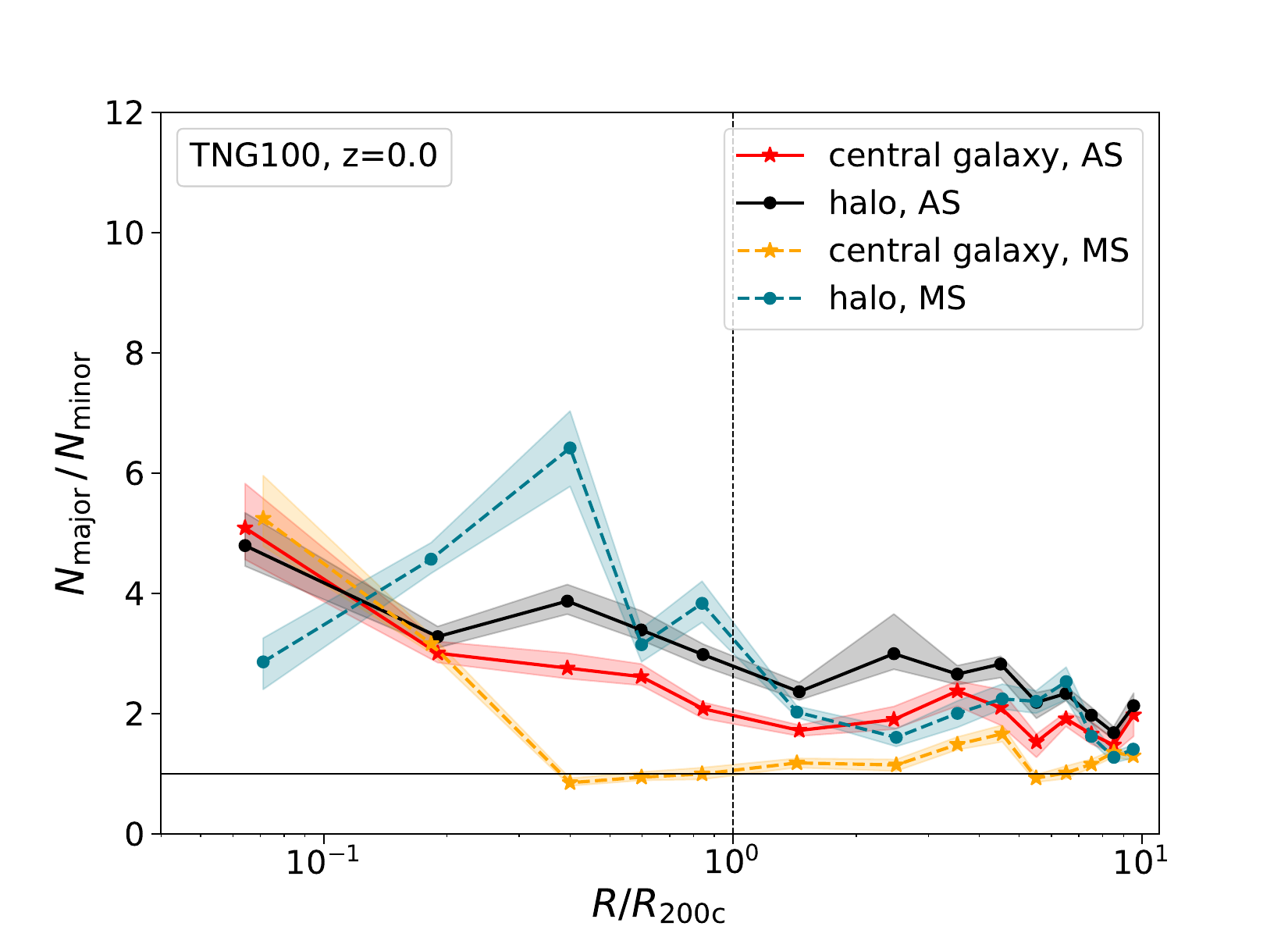}
\hspace{-25pt}
\includegraphics[width=0.5\textwidth]{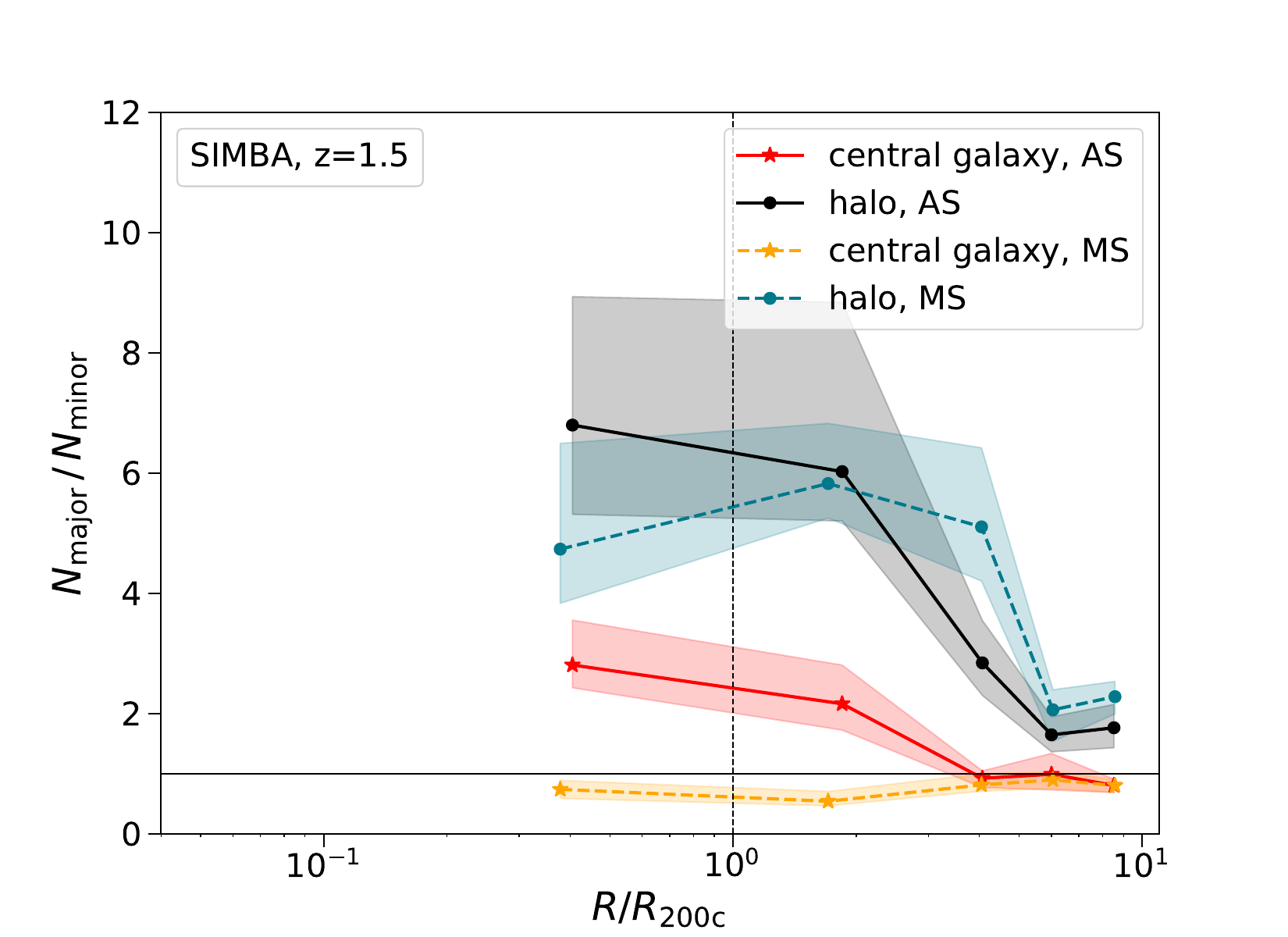}

\caption{Major-to-minor axis satellite galaxy count ratios for halos and central galaxies in the AS and MS as a function of radius. \textit{Left column}: results at $z=0.0$ from the SIMBA, EAGLE, and TNG100 simulations. \textit{Right column}: results at $z=0.5,1.0,1.5$ from SIMBA. The horizontal black solid line corresponds to a count ratio of 1. Shaded bands denote 3$\sigma$ confidence intervals, quantifying the anisotropy of satellite distributions relative to the triaxial coordinate frames of host halos and central galaxies. Halo-related anisotropy dominates at halo scales and above, whereas central galaxy signals are most prominent on small scales.} 
\label{fig:Nx_Nz_R_halo_BCG_z}
\end{figure}

We first analyze the SIMBA simulations, computing the major-to-minor axis satellite count ratios for halos and central galaxies in both the AS and MS at 4 different redshifts as shown in Figure \ref{fig:Nx_Nz_R_halo_BCG_z}. For the AS, we find that within the triaxial coordinate frames of both central galaxies and halos, the number of satellite galaxies along the major axis exceeds that along the minor axis by a factor of $\sim 3$; this anisotropy is detected at the 3$\sigma$ confidence level (central $99.73\%$ sample percentile interval), demonstrating that the anisotropic signal extends to scales larger than individual halos — a result consistent with the work of \citet{2022MNRAS.514.1077R} using SDSS DR16. Notably, although the major axes of central galaxies and halos are closely aligned in the AS, the directional discrepancy in satellite counts (major axis vs. minor axis) is significantly stronger for halos than for central galaxies -- a result indicating that satellite distribution anisotropy is dominated by halo morphology rather than central galaxy morphology, this conclusion is consistent with that of some previous works \citep{2013ApJ...779..160Z, 2016MNRAS.460.3772S}. At high redshift, this anisotropy signal is stronger towards halo center and becomes weaker at large radii, but still at around $\sim 2$, except the result from central galaxies direction at the highest redshift 1.5. However, at $z=0$, this radial dependence seems to be much weaker with both lines generally flat. Note that we cannot make a valid comparison of the inner small-radius anisotropy between redshifts. Due to the smaller sample size at high redshift, we employ broader radial bins than for $z=0$, which pushes the leftmost data point outward.

In contrast, continuing our analysis of the SIMBA data, for central galaxies in the MS, an excess of galaxies on the major-axis relative to the minor axis is almost flat at a value of 1, which means that some halos have their central galaxy major axis orientations not aligned with the assembly of satellite galaxies, i.e. they tend to point randomly at all the redshifts. Strikingly, when using the major-axis orientations of halos in the MS, anisotropic signals remain statistically significant, closely resembling the halo-based results from AS across all redshifts. Notably, for the MS, the amplitude of the halo-based anisotropy weakens on small scales, which may arise from differences in the dark matter distribution within different radial ranges of halos, as discussed in detail in Section \ref{subsec:R_Morphologies}. Furthermore, to verify that the AS/MS classification criterion exerts no influence on the conclusions drawn from Figure \ref{fig:Nx_Nz_R_halo_BCG_z}, we elaborate on the results derived from other criteria in Appendix \ref{sec:AS_MS_classification}. This means that the correlation between satellite galaxy anisotropy distribution and halo morphology is stable for all samples and at all redshifts. Thus, we should be very cautious to use the central galaxy's orientation to indicate satellite assembly positions.

In the left column of Figure \ref{fig:Nx_Nz_R_halo_BCG_z}, we further test data from the EAGLE and TNG100 simulations at $z=0$. Despite applying a lower mass threshold for halo selection in these two simulations, both yield results qualitatively consistent with those from SIMBA. Furthermore, the high resolution and sufficient data volume of the TNG100 simulation enable us to derive more detailed satellite galaxy anisotropy information within $0.3R_{\rm 200c}$. We find that within such a small region, the central galaxy-associated anisotropy signal in the AS gradually exceeds that of halos -- a trend indicating a strong correlation between central galaxy morphology and satellite distribution in the central halo regions, with behavior consistent with the central galaxy-associated satellite anisotropy observed in the MS at small scales. This means that the satellite galaxies are strongly affected by the central galaxy instead of halo at the central region.

\clearpage

\subsection{Anisotropies of Satellite Galaxies across Halo Masses and Cosmic Time}\label{subsec:Nx_Nz_M_halo_BCG}

\begin{figure}[htbp]
    \centering
    \begin{minipage}{0.33\textwidth}
        \centering
        \includegraphics[width=\linewidth]{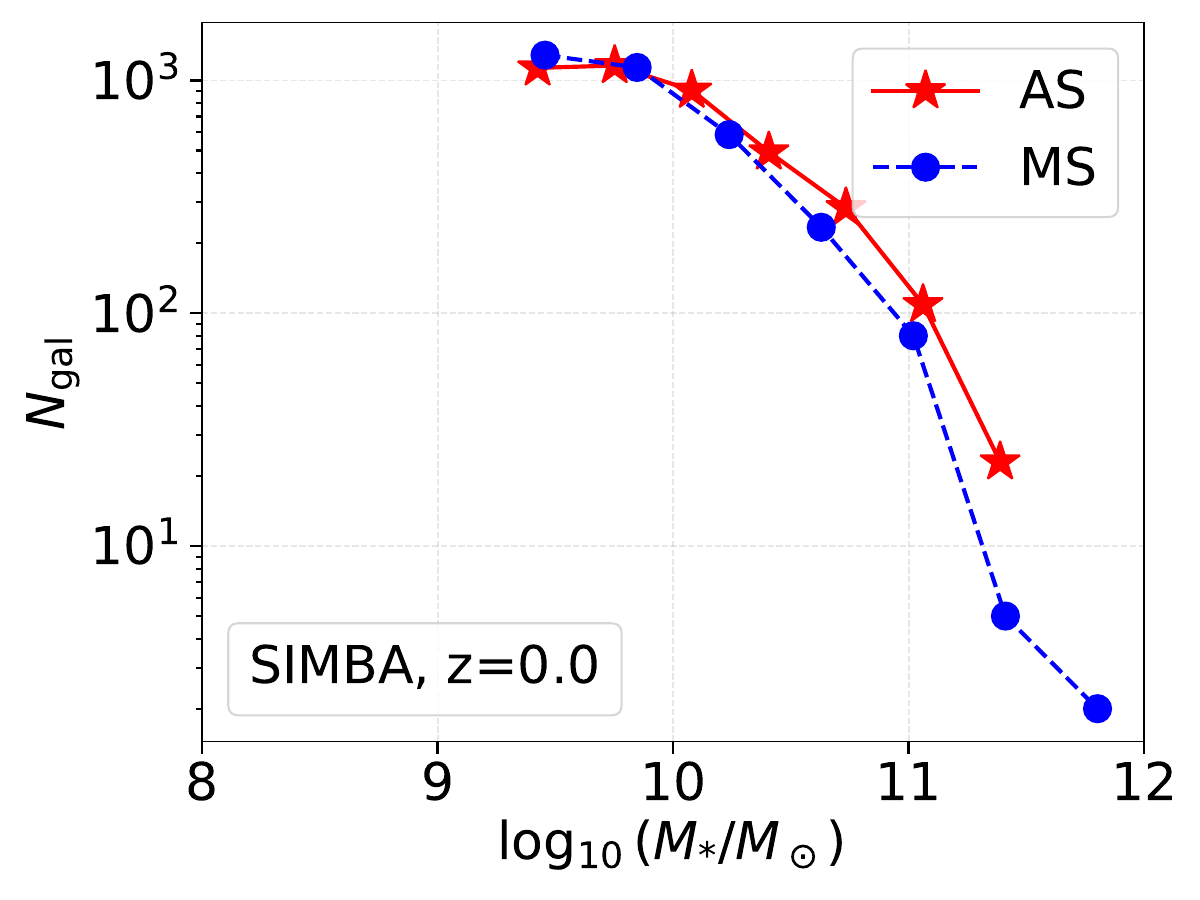}
    \end{minipage}
    \hspace{-3pt}
    \begin{minipage}{0.33\textwidth}
        \centering
        \includegraphics[width=\linewidth]{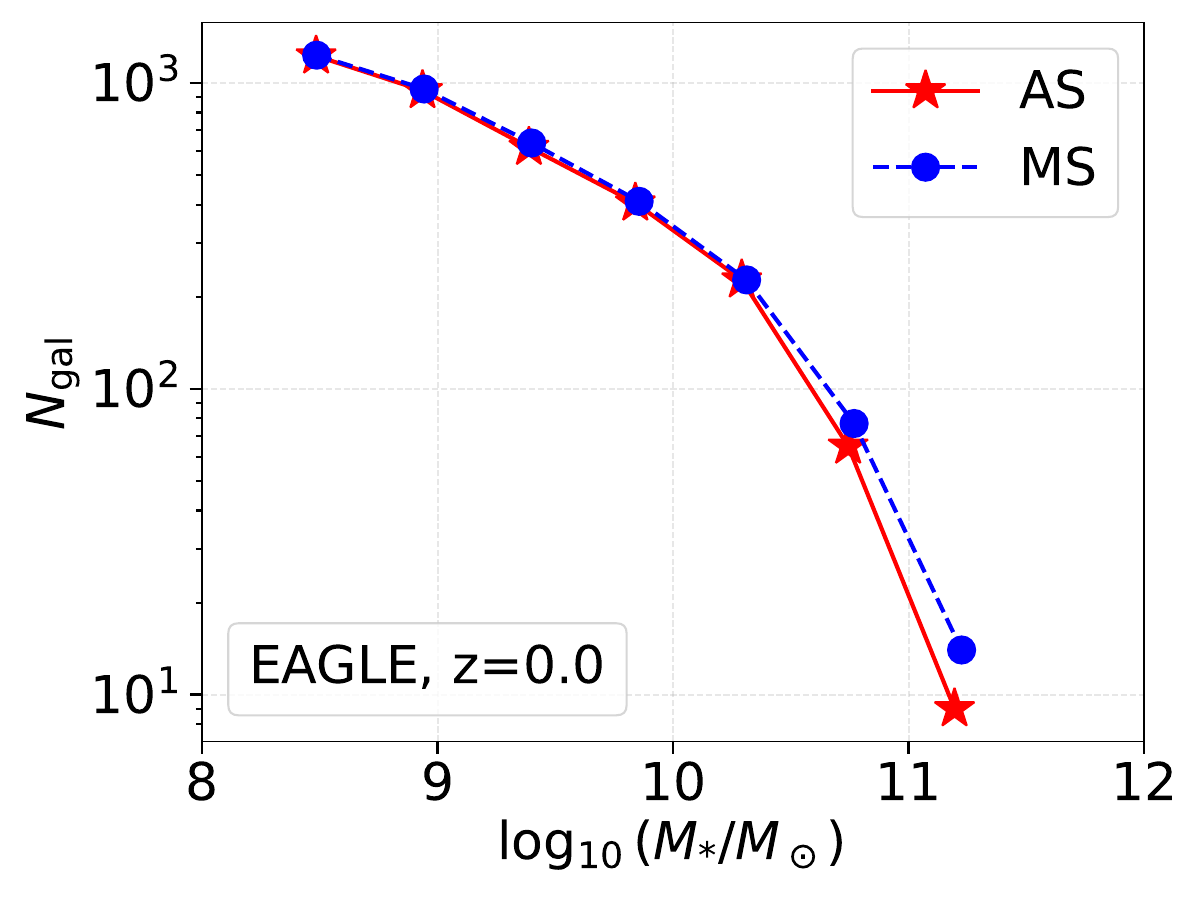}
    \end{minipage}
    \hspace{-3pt}
    \begin{minipage}{0.33\textwidth}
        \centering
        \includegraphics[width=\linewidth]{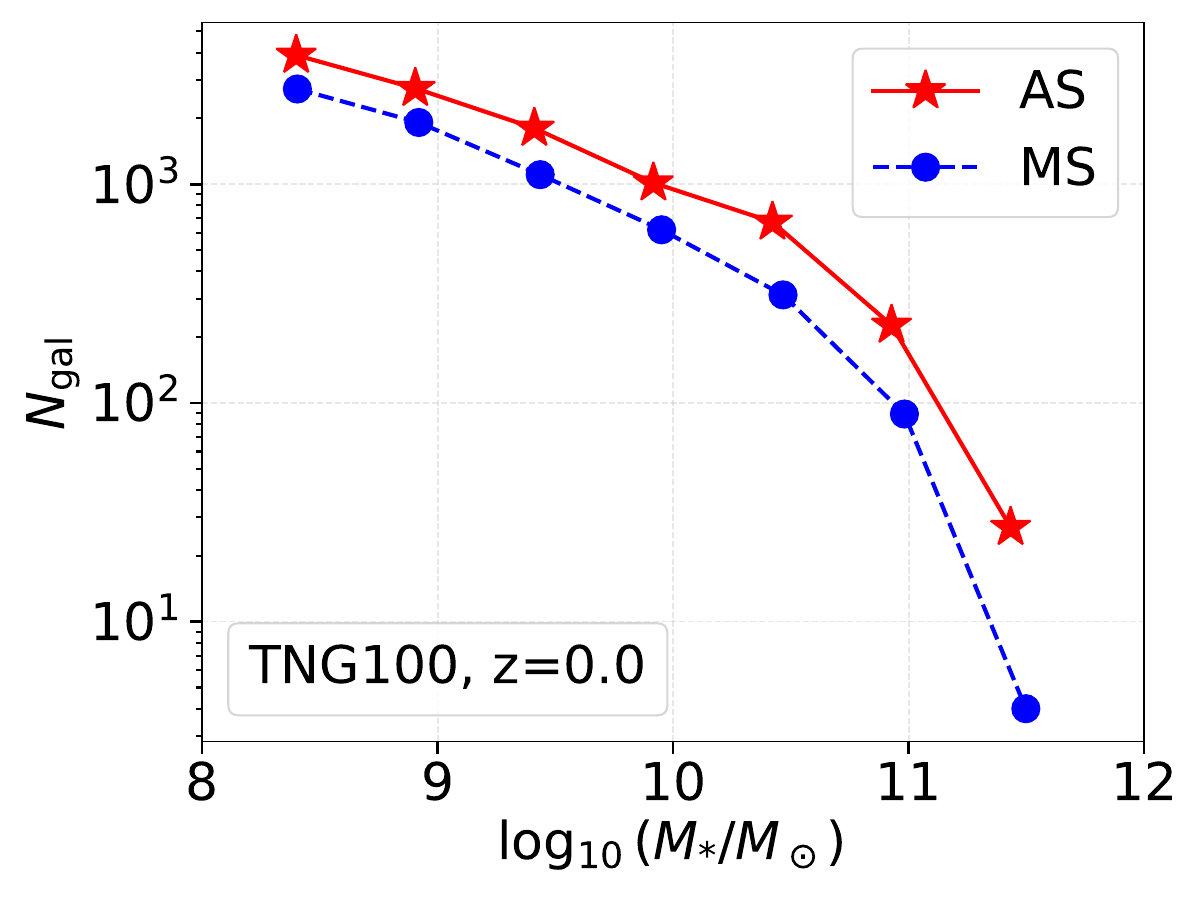}
    \end{minipage}
    \caption{Number of galaxies as a function of stellar mass. We adopt halo mass cuts of \(M_\mathrm{200c}\ge 10^{11}\,M_\odot\) (SIMBA, EAGLE, TNG100), alongside a radial cut \(2r_\mathrm{c,\, half}<R<3R_\mathrm{200c}\) applied to satellite galaxies. The AS and MS yield nearly identical galaxy count trends at all stellar masses.}
    \label{fig:sate_starmass_3R200c_distribution}
\end{figure}

To analyze the correlation between satellite galaxy anisotropy and the mass of the corresponding host halos, we adopt the identical satellite galaxy statistical procedure described in Section \ref{subsec:Nx_Nz_R_halo_BCG}. The key differences in this section are twofold: we adjust the lower mass threshold of halo $M_{\rm 200c}$ to $10^{11}\ M_\odot$ and partition the halo $M_{\rm 200c}$ into multiple bins based on this revised criterion; we restrict our measurements to the directional differences in satellite galaxy counts within the radial range of $2r_{\rm c,\, half}$ to $3R_{\rm 200c}$ inside halos. The final selection criteria and sample information corresponding to different simulation datasets are summarized in Table \ref{tab:data_information_M200c}. Figure \ref{fig:sate_starmass_3R200c_distribution} shows the relation between satellite count and stellar mass for the samples. Though these samples differ from those in Section \ref{subsec:Nx_Nz_R_halo_BCG}, the AS and MS distributions remain broadly consistent, in line with the results from Figure \ref{fig:sate_starmass_10R200c_distribution}.

\begin{table}[h!]
    \centering
    \setlength{\tabcolsep}{5pt}  
    \renewcommand{\arraystretch}{1.5}  
    \begin{tabular}{lccccc}
        \toprule
Name & $z$ & \, $N_\mathrm{c}$ (AS, MS) \, & \, $N_\mathrm{s}$ (AS, MS) \, & $M_\mathrm{200c,\, min} \, [M_{\odot}]$ & $M_\mathrm{*s,\, min} \, [M_{\odot}]$\\
        \midrule
        SIMBA-m100n1024 & 0.0 & 831, 1455 & 4108, 3327 & $1\times10^{11}$ & $1.82\times10^{9}$\\
        SIMBA-m100n1024 & 0.5 & 739, 1491 & 2758, 3295 & $1\times10^{11}$ & $1.82\times10^{9}$\\
        SIMBA-m100n1024 & 1.0 & 571, 1297 & 1342, 2868 & $1\times10^{11}$ & $1.82\times10^{9}$\\
        SIMBA-m100n1024 & 1.5 & 427, 967 & 933, 1903 & $1\times10^{11}$ & $1.82\times10^{9}$\\
        EAGLE-L100N1504 & 0.0 & 632, 748 & 3491, 3551 & $1\times10^{11}$ & $1.81\times10^{8}$\\
        TNG100-1 & 0.0 & 1034, 967 & 10384, 6772 & $1\times10^{11}$ & $1.40\times10^{8}$\\
        \bottomrule
    \end{tabular}
    \caption{Final selection criteria and aligned/misaligned sample information for the simulation suites, employed in the analysis of anisotropy across different halo mass. Columns follow the same nomenclature as Table \ref{tab:data_information_spatial}. Satellite galaxies are selected within $2r_\mathrm{c,\, half}<R<3R_\mathrm{200c}$.}
    \label{tab:data_information_M200c}
\end{table}

\begin{figure}[h!]
\includegraphics[width=0.5\textwidth]{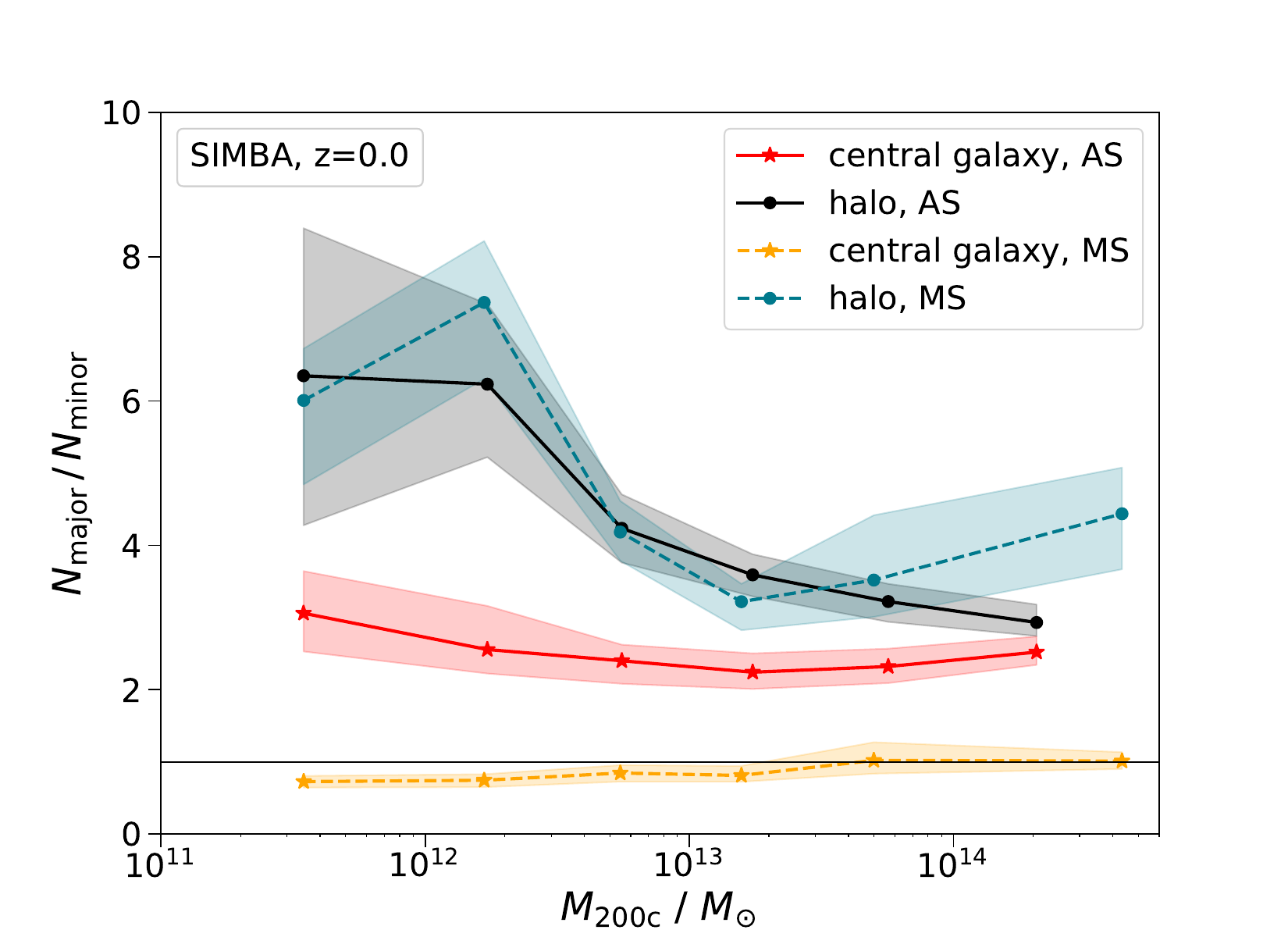}
\hspace{-25pt}
\includegraphics[width=0.5\textwidth]{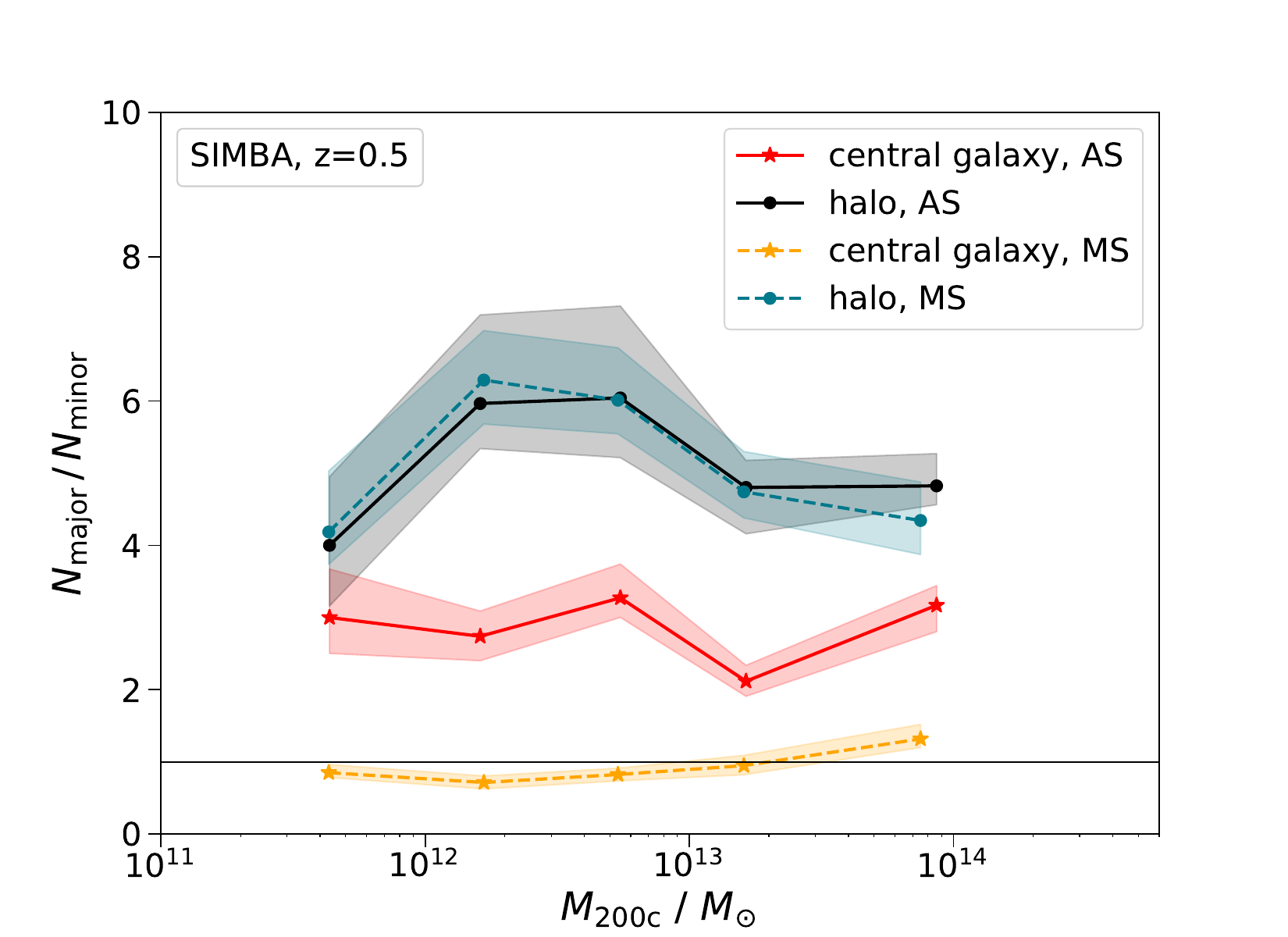}

\vspace{-2.5pt}

\includegraphics[width=0.5\textwidth]{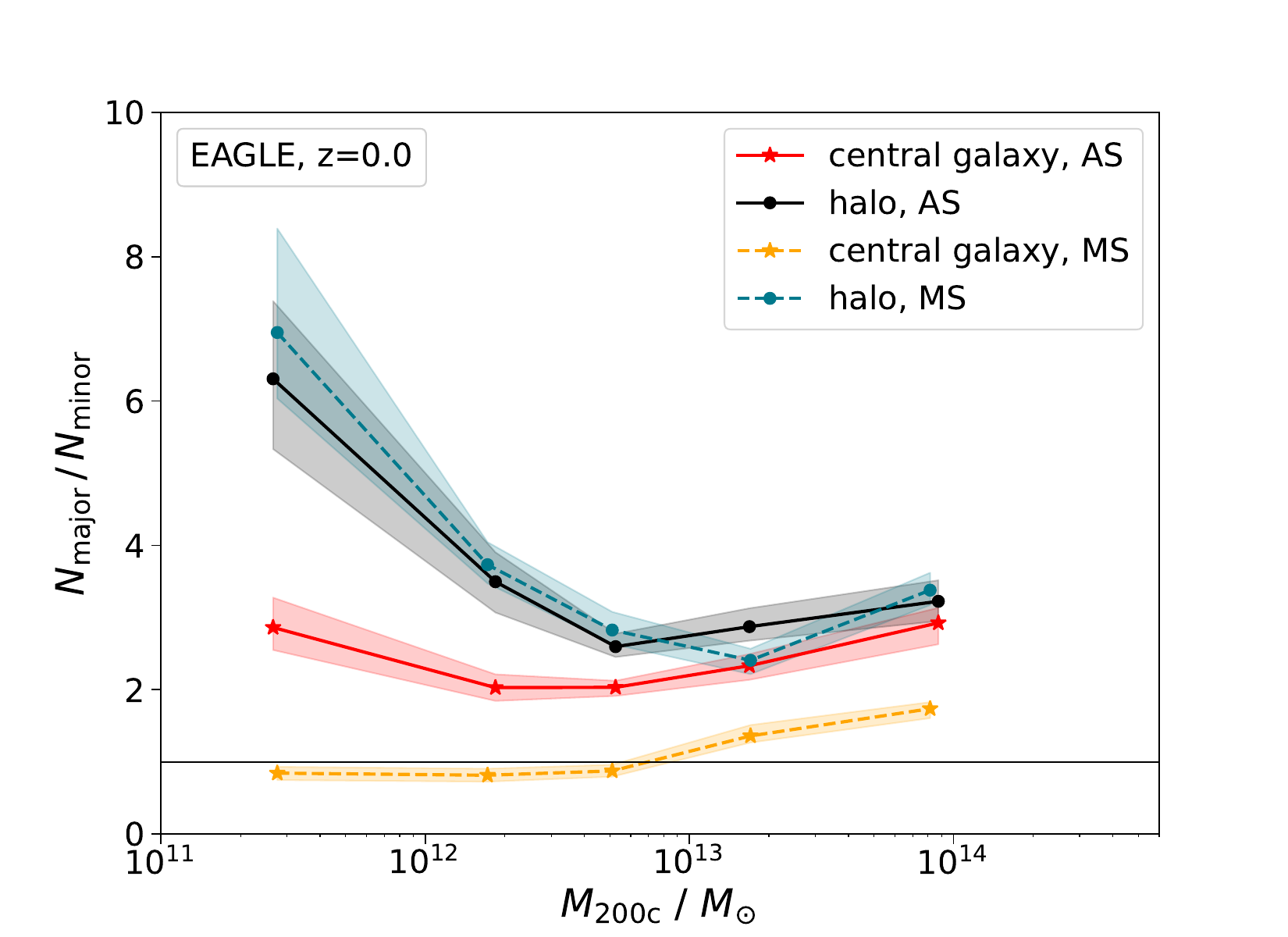}
\hspace{-25pt}
\includegraphics[width=0.5\textwidth]{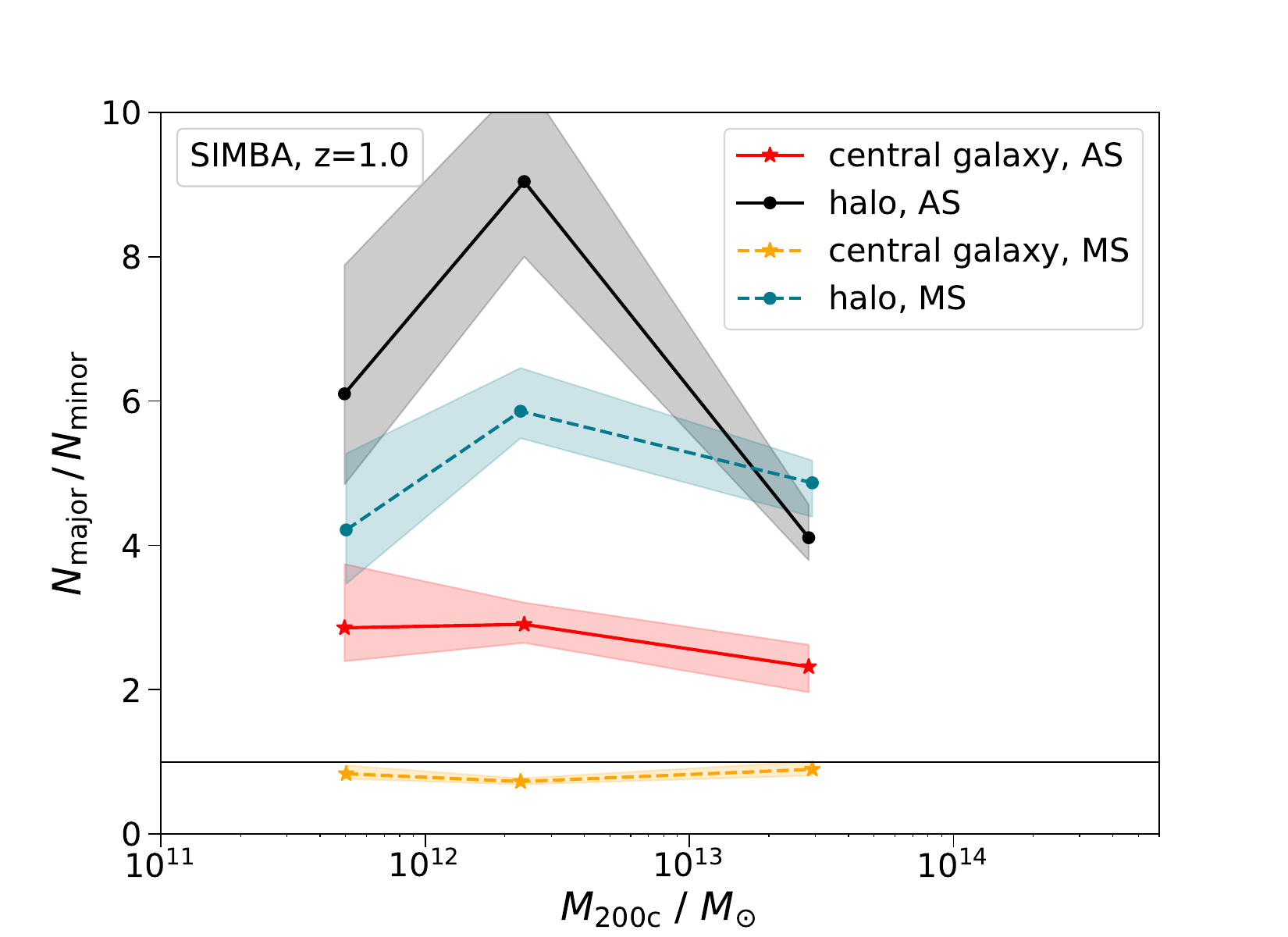}

\vspace{-2.5pt}

\includegraphics[width=0.5\textwidth]{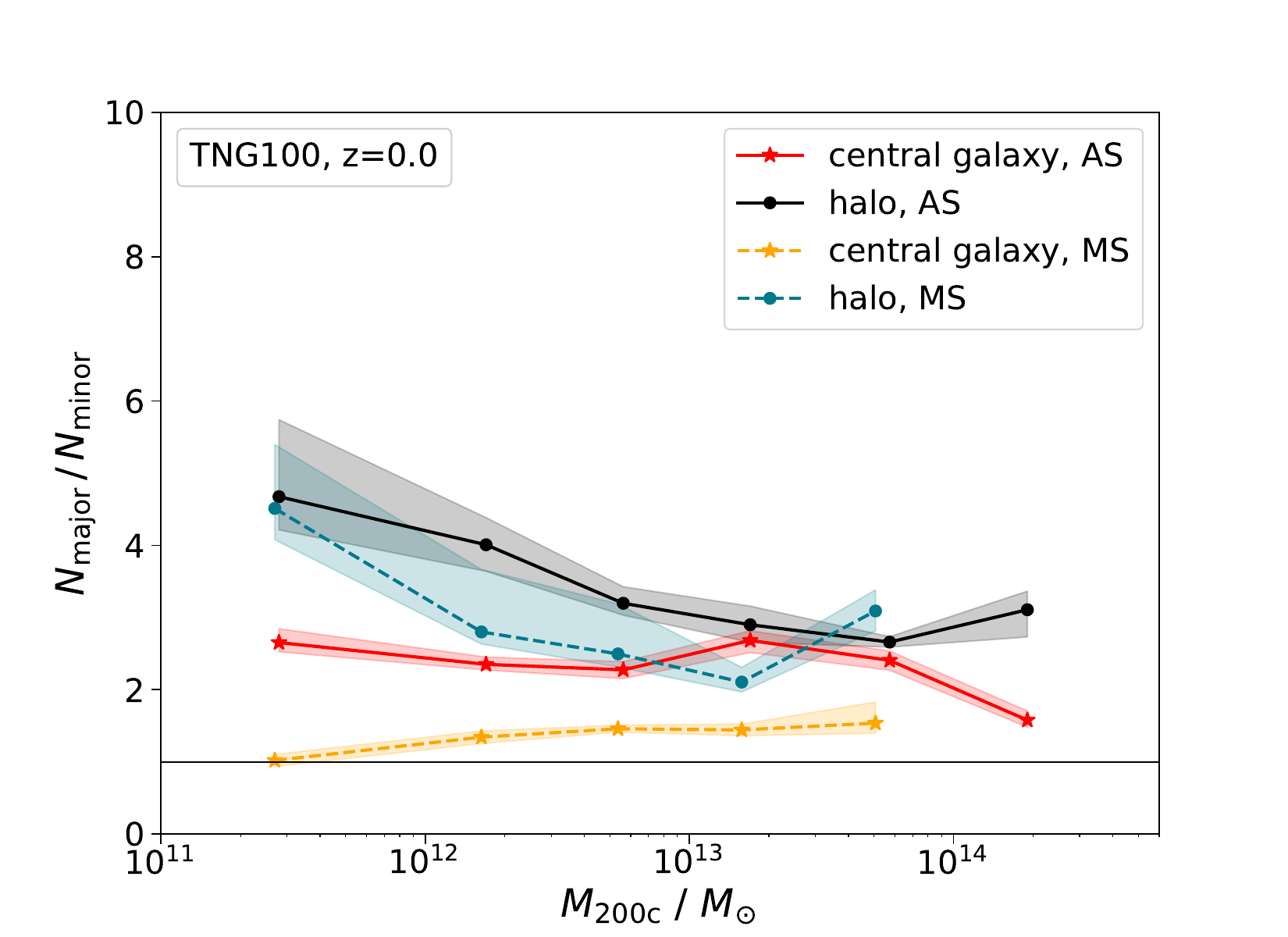}
\hspace{-25pt}
\includegraphics[width=0.5\textwidth]{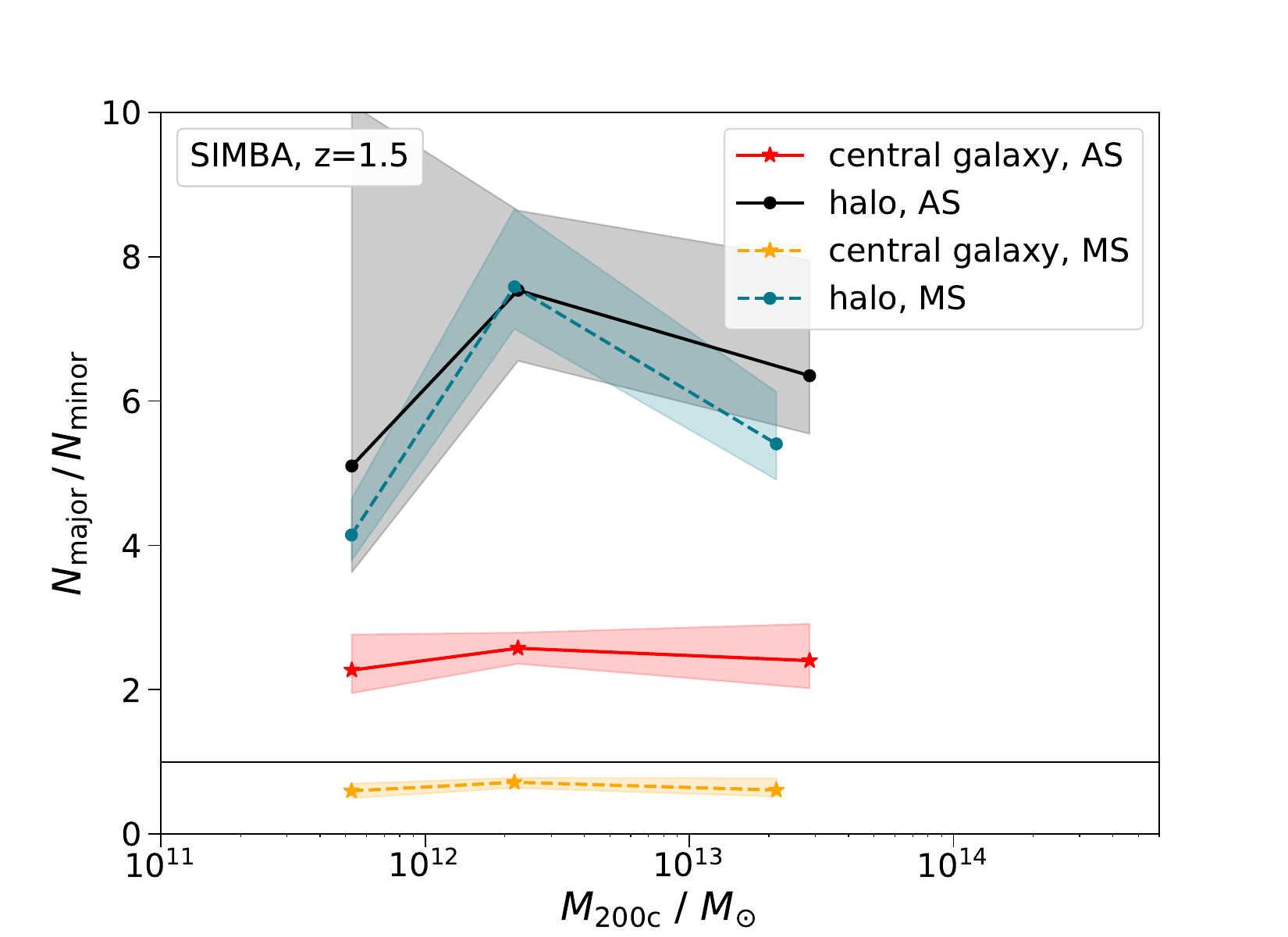}
\caption{Major-to-minor axis satellite galaxy count ratios for halos and central galaxies in the AS and MS as a function of halo mass. \textit{Left column}: results at $z=0.0$ from the SIMBA, EAGLE, and TNG100 simulations. \textit{Right column}: results at $z=0.5,1.0,1.5$ from SIMBA. The curve color coding, and visual conventions are identical to those in Figure \ref{fig:Nx_Nz_R_halo_BCG_z}.} Halo-related satellite anisotropy remains the strongest across all mass bins, followed by central galaxy-associated anisotropy in the AS, while MS central galaxies show no significant signal.
\label{fig:Nx_Nz_M_halo_BCG_z}
\end{figure}

We first present the relationship between satellite galaxy anisotropy and halo mass using SIMBA data at $z=0$ in Figure \ref{fig:Nx_Nz_M_halo_BCG_z}. It can be observed that for samples across different mass bins, the halo-associated anisotropy ratio is above 2, statistically significant at the 3$\sigma$ confidence level, and stronger than the central galaxy-associated counterpart; this behavior is consistent for both the AS and MS results. On the other hand, the central galaxy-associated results for the MS show almost no overall anisotropy signal. Finally, all the above conclusions are fully validated by data at redshifts $z=0.5,\, 1.0,\, 1.5$, demonstrating their redshift independence. This finding is completely consistent with the results presented in Section \ref{subsec:Nx_Nz_R_halo_BCG}, further reinforcing the conclusion that the anisotropy of satellite galaxy distributions can be biased using the central galaxy morphology but is clearly correlated/consistent with halo morphology. Another notable phenomenon is that the strength of the halo-associated anisotropy signal decreases with increasing halo mass throughout the high-mass range. Though massive halos are intrinsically triaxial \citep{2006MNRAS.367.1781A}, one plausible explanation is that large-scale tidal fields from filaments and walls suppress anisotropy by altering satellite distributions, thereby weakening the correlation between signal and halo morphology. In contrast, the strength of the central galaxy-associated anisotropy signal in the AS shows almost no halo-mass dependencies at all the investigated redshifts.

To ensure the robustness of our findings, we cross-validate the SIMBA-specific results using the higher mass resolution EAGLE and TNG100 simulations. The left column of Figure \ref{fig:Nx_Nz_M_halo_BCG_z} also presents the same relationship between the anisotropy of satellite galaxy distributions and halo mass, but derived from the EAGLE and TNG100 datasets. We find that the halo-associated anisotropy signal is substantially stronger than its central galaxy-associated counterpart, with the corresponding axis ratio generally greater than 2. This signal is also statistically significant at the 3$\sigma$ confidence level in both simulations. On the other hand, the ratios from both AS and MS using halo morphology information show halo mass dependence with low mass halos tend to have larger ratios, while the central galaxy-associated signal from the AS result tends to have a lower ratio and not to correlate with halo mass -- very similar conclusions are drawn from the SIMBA simulation. As shown in Figure \ref{fig:halo_BCG_misalign_rate}, the alignment between massive halos and the major axes of their central galaxies is more consistent. This favorable alignment strengthens our expectation of a significant detection in observational data, even though the satellite anisotropy signal for central galaxies in the AS is slightly weaker than that of halos in the AS, with a typical ratio of about 3. Such behavior might naturally explains why previous studies identified stronger central galaxy-associated satellite anisotropies in massive halos and for red central galaxies \citep{2013ApJ...779..160Z, 2018ApJ...859..115W, 2022MNRAS.514.1077R, 2025A&A...699A.215R}. By contrast, the mass dependence of anisotropy for halo-associated satellites is more complex, meaning that the strength of this anisotropy does not vary monotonically with mass, as demonstrated in \citet{2025A&A...699A.215R}. However, unlike the SIMBA and EAGLE results, the central galaxy-associated anisotropy signal for the MS from the TNG100 simulation seems to be detectable across the full mass range, albeit with lower amplitude. Combined with the relationship between anisotropy signal and scale discussed in Section \ref{subsec:Nx_Nz_R_halo_BCG}, which infers that the central galaxy-associated signal is dominated by satellite galaxies on small scales, we have a full picture of the satellite anisotropy distributions across different scales, halo properties, as well as the effects of baryon models.

\begin{figure}[h!]
    \centering
    \includegraphics[width=0.5\textwidth]{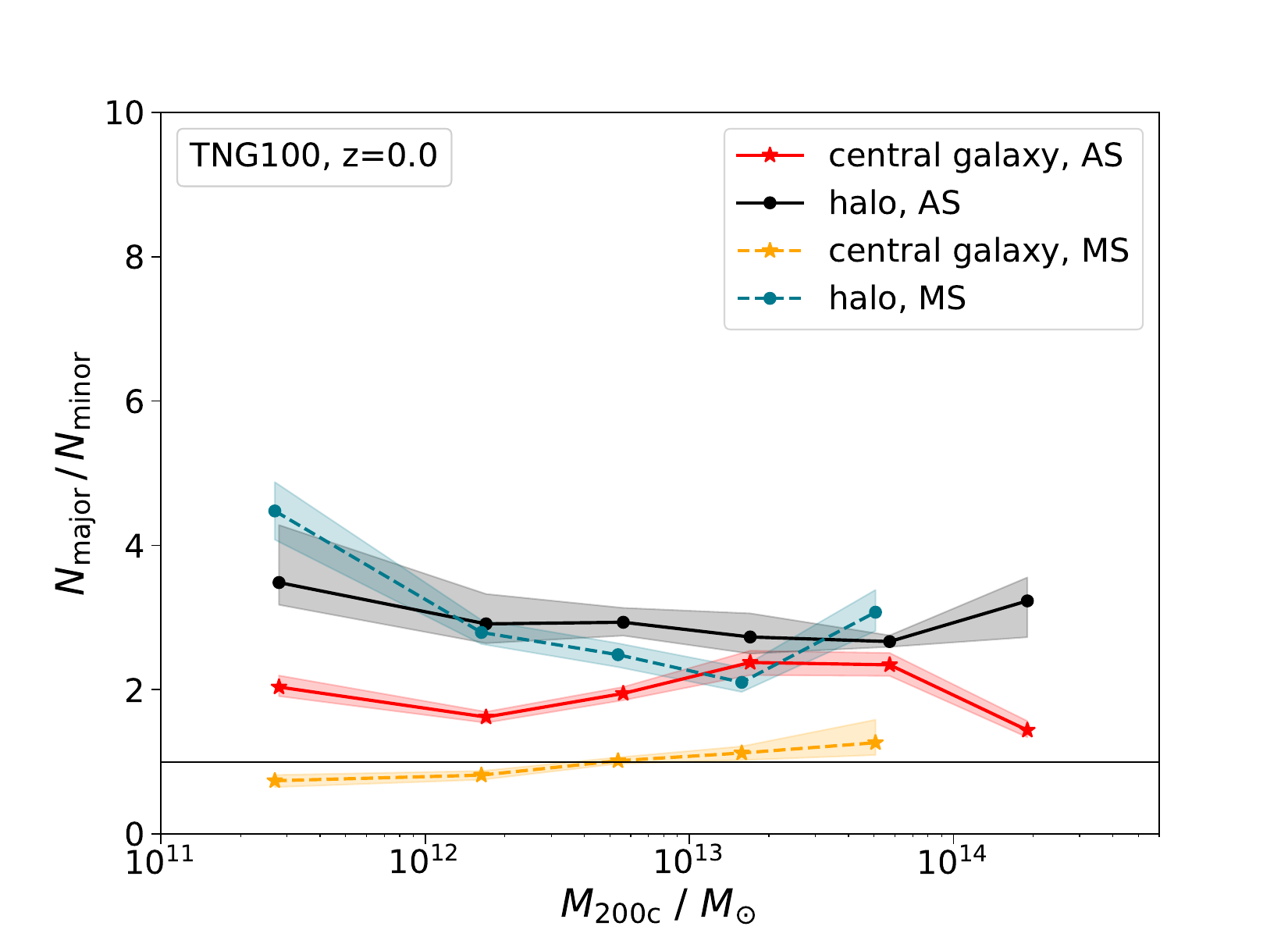}
    \caption{Major-to-minor axis satellite galaxy count ratios for halos and central galaxies in the AS and MS at $z=0$ from the TNG100 simulation, plotted as a function of halo mass with satellite galaxies within $0-0.3R_{\rm 200c}$ excluded from the sample. All axis definitions, curve color coding, and visual conventions are identical to those in Figure \ref{fig:Nx_Nz_M_halo_BCG_z}. Halo-related anisotropy does not strongly depend on small-scale galaxy distributions, whereas central galaxy-associated anisotropy in the MS is highly sensitive to small-scale structure.}
\label{fig:EAGLE_TNG100_Nx_Nz_M_R_0.3_3_halo_BCG}
\end{figure}

To verify this conclusion, we only select satellite galaxies within the radial range of $r_{\rm s}/R_{\rm 200c}\in(0.3,\, 3.0)$, where $r_{\rm s}$ denotes the radial coordinate of satellite galaxies. By re-analyzing the distribution of this subset of satellite galaxies, we can eliminate the contribution of small-scale satellite galaxies to the anisotropy signal, and the corresponding results are presented in Figure \ref{fig:EAGLE_TNG100_Nx_Nz_M_R_0.3_3_halo_BCG}. As expected, after eliminating the contribution of small-scale galaxies, the central galaxy-associated anisotropy signal in the MS weakens significantly, with the overall anisotropy signal becoming extremely faint. This result undoubtedly confirms that the central galaxy-dominated anisotropy signal is predominantly confined to small-scale regions. Meanwhile, the halo-associated anisotropy ratio in both the AS and MS is still consistently above 2, and remains statistically significant at the 3$\sigma$ confidence level, demonstrating that the halo-dominated anisotropy signal can persist across cluster-scale ranges. The aforementioned conclusions provide robust cross-validation with the findings presented in Section \ref{subsec:Nx_Nz_R_halo_BCG}.

\section{Contrasting Correlations: Satellite Distribution Anisotropy \\ with Halos Morphologies vs. with Filaments Morphologies}\label{sec:SDA_halo_filament}

\subsection{Sample Classification}\label{subsec:halo_web_sample}

Similar to the classification of the sample into AS and MS based on the alignment between the halos' and the central galaxies' major axes in Section \ref{subsec:BCG_halo_sample}, we can also divide the samples into AS and MS when considering the alignment between the halo major axes and filament orientations. However, a single halo is likely to be linked to multiple filaments, a scenario that is particularly common for massive halos. For the sake of conservatism, we therefore consider all filaments associated with each halo: a halo is classified as AS if there exists at least one filament in its vicinity whose orientation forms an angle of less than $40^\circ$ with the halo's major axis; conversely, it is classified as MS if the orientation of all filaments surrounding the halo forms an angle of more than $50^\circ$ with the halo's major axis. Unless otherwise specified, all references to the AS and MS in this chapter refer to classifications based on the orientational misalignment between halos and filaments.

\begin{figure}[htbp]
    \centering
    \includegraphics[width=0.5\textwidth]{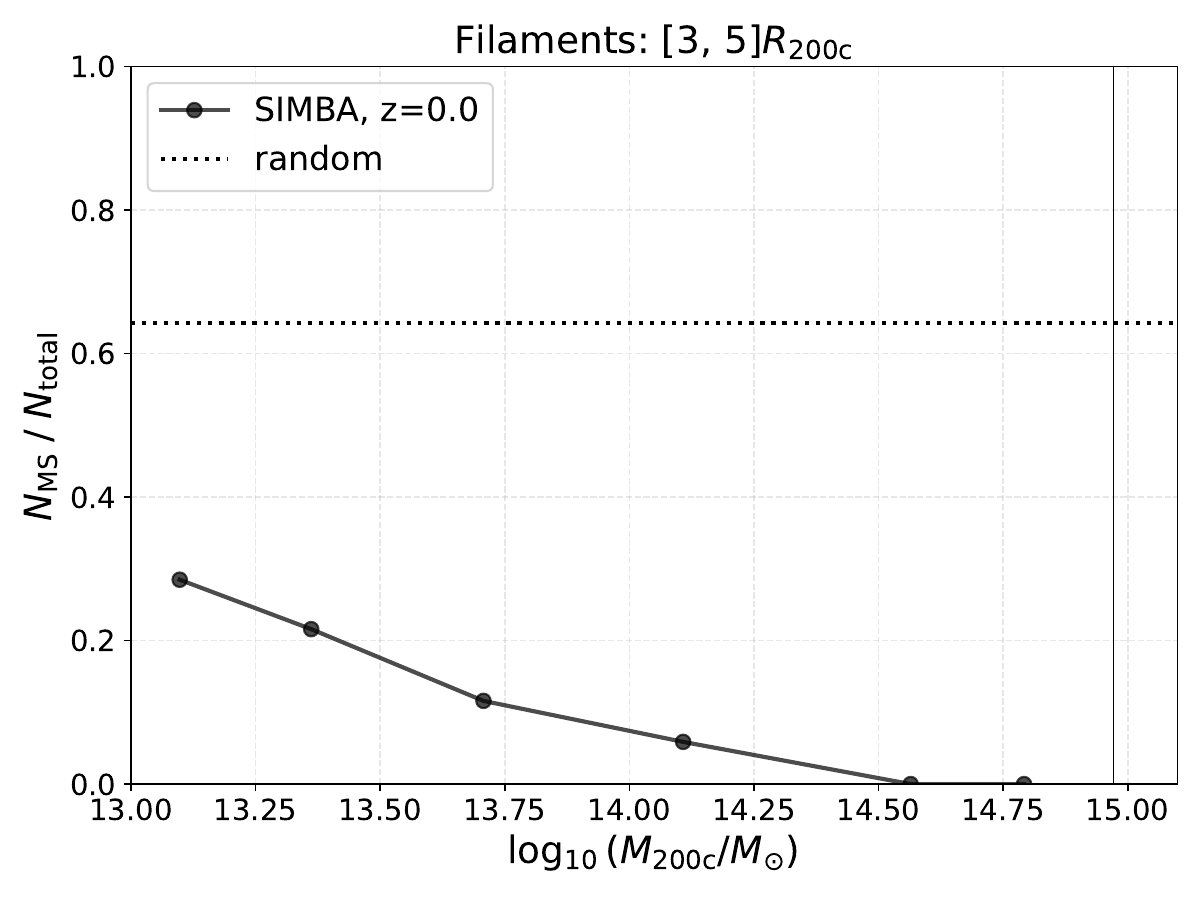}
    \caption{Misaligned fraction of halo major axes and filament orientations vs. halo $M_{\rm 200c}$ derived from SIMBA simulation at $z=0$. The vertical thin lines correspond to the maximum $M_\mathrm{200c}$ of the halo data. The misaligned fractions decrease with increasing halo mass.}\label{fig:SIMBA_halo_web_misalign_rate_z_0}
\end{figure}

Using SIMBA data at $z=0$, we select a subset of halos with $M_{\rm 200c}\ge10^{13}\ M_\odot$ for our analysis. In this process, we obtain a sample of $211$ halos with $13533$ satellite galaxies for the AS, and $73$ halos with $2546$ satellite galaxies for the MS. In Figure \ref{fig:SIMBA_halo_web_misalign_rate_z_0}, we present the fraction of MS halos relative to the total sample in different halo mass bins. Similarly to Figure \ref{fig:halo_BCG_misalign_rate}, this fraction decreases monotonically with increasing halo mass, which indicates that more massive halos have a higher probability of hosting at least one filament aligned with their major axes. This is definitely related to our criterion for AS -- a higher possibility of alignment with a larger connectivity. On the other hand, the misaligned fractions observed within the current mass range are significantly lower than those corresponding to random orientations, which indicates a substantial correlation between filament orientations and halo major axes. However, the persistent level when using \texttt{DisPerSE} to identify filaments is also responsible for this systematic change -- a higher value will reduce the number of filaments, thus increasing the possibility of misalignment. Nevertheless, we are interested in the relative differences, which should be qualitatively robust.

\subsection{Spatial Anisotropies of Satellite Galaxy Distributions}

\begin{figure}[h!]
    \centering
    \begin{minipage}[b]{0.52\textwidth}
        \centering
        \includegraphics[width=\textwidth]{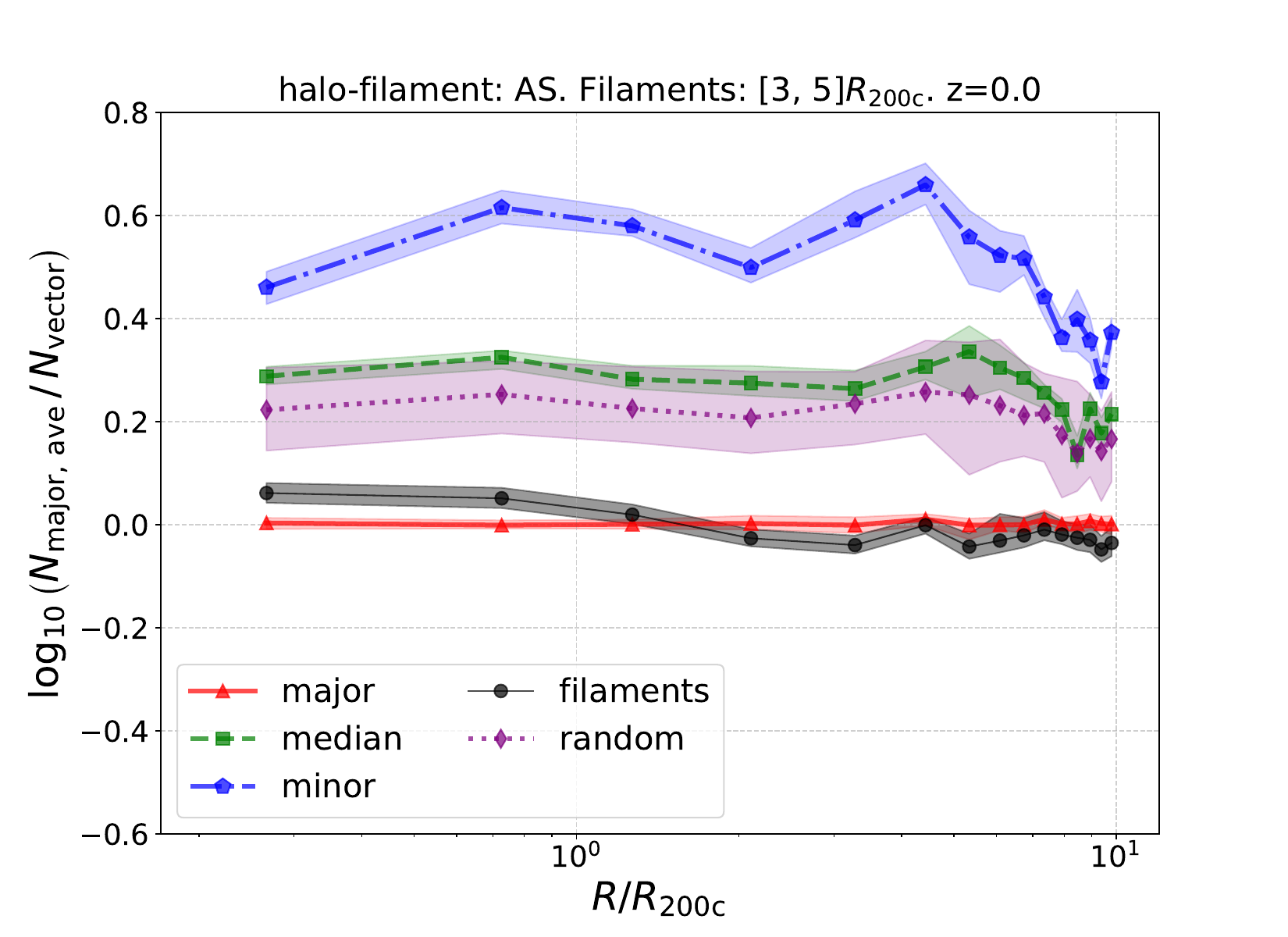}
        \small 
    \end{minipage} \hspace{-30pt}
    \begin{minipage}[b]{0.52\textwidth}
        \centering
        \includegraphics[width=\textwidth]{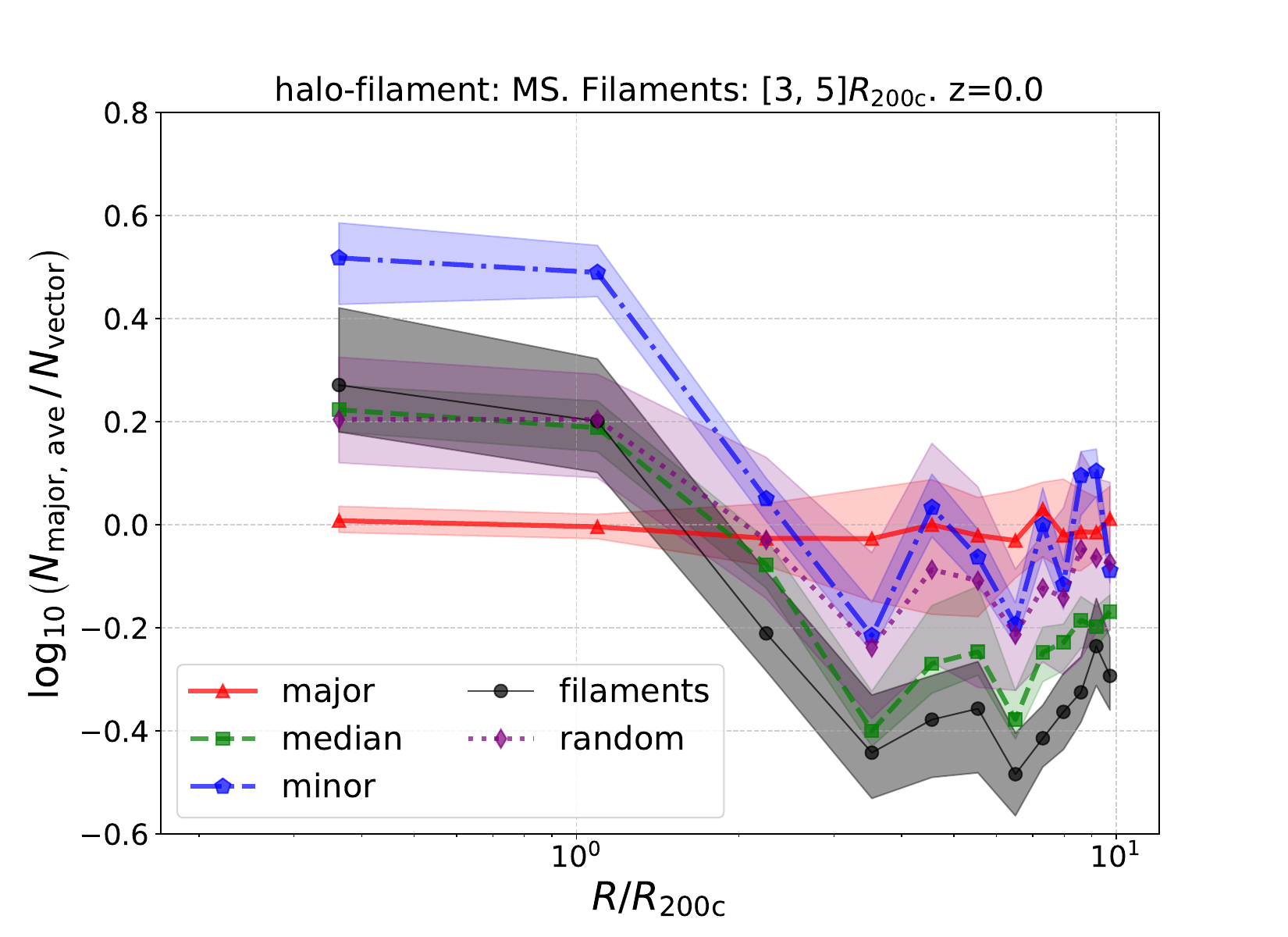}
        \small
    \end{minipage}
    \caption{The base-10 logarithm of the ratio of satellite galaxy counts along the averaged major halo axis to those along other orientations, plotted as a function of radial distance from the halo center. The shaded bands around the curves denote the $3\sigma$ confidence intervals of the measured ratios. Within $2R_\mathrm{200c}$, satellite galaxies aligned with the minor axis show the lowest count across all orientations, while those along the major axis yield the highest count. Beyond $2R_\mathrm{200c}$, the filament orientation corresponds to the maximum satellite number instead.}
\label{fig:SIMBA_Nx_Nvec_R_halo_web}
\end{figure}

In this subsection, we focus on the relationship between the number ratios of satellite galaxies as a function of the radial distance, considering three directional cases: the triaxial directions of halos, filament orientations, and random directions. For the analysis of filament orientations, we only consider filaments in the AS whose orientations form an angle of less than $40^\circ$ with the connected halos' major axes, whereas for the MS, we take into account all associated filaments. To fully characterize the differences in satellite galaxy counts across different filament orientations, we design the following procedure: each halo is randomly assigned one of its linked filaments; we then compile and statistically analyze the satellite galaxy data corresponding to these filaments across all halos. Finally, we repeat this procedure 1000 times, and the statistical scatter induced by the differences in satellite galaxy counts across various filament orientations is reflected in the error bands shown in Figure \ref{fig:SIMBA_Nx_Nvec_R_halo_web}. Additionally, unlike the treatment in Section \ref{sec:SDA_BCG_halo}, here we incorporate filament and random orientations for cross-category comparison. To enable a unified comparison across categories corresponding to different orientations, we take $N_\mathrm{major,ave}$ as an indirect comparative quantity, where $N_\mathrm{major,ave}$ is defined as the number of galaxies counted along the major-axis direction, obtained by averaging 50 repeated orientation determinations.

From the left panel of Figure \ref{fig:SIMBA_Nx_Nvec_R_halo_web}, $\log_{10}\left(N_\mathrm{major, \, ave}/N_\mathrm{minor}\right)$ exceeds $\log_{10}\left(N_\mathrm{major, \, ave}/N_\mathrm{random}\right)$ by approximately 0.2-0.4. This implies that the number of satellite galaxies along the minor axis is correspondingly 1.5-2.5 times lower than in the random orientation. By the same reasoning, the number along the median axis is comparable to the random case, while the numbers along both the filament orientation and halo major axis are significantly higher (by a factor of $\sim 2$) than in the random orientation. In addition, within the radial range of approximately $<2R_\mathrm{200c}$, the number of satellite galaxies along the halo major axis is about 1.15 times that along the filament orientation. Although the two values are close, this difference is still statistically significant at the 3$\sigma$ confidence level. However, at larger scales, the number of satellite galaxies along the filament orientation becomes slightly higher than that along the halo major axis. These results indicate that the anisotropy of satellite galaxy distributions on halo scales is significantly correlated with the triaxial geometry of halos; this correlation, while outside of halos, is shaped by the orientation of large-scale filaments. This finding also corroborates the reliability of the filament structures that we extracted, and such a trend is physically consistent with the enhanced satellite abundance within filamentary environments reported in \citet{2026ApJ...998..251M}. 

When examining the right panel of Figure \ref{fig:SIMBA_Nx_Nvec_R_halo_web}, the contrast between the anisotropic distribution of satellite galaxies with halo morphology and that with filament orientations becomes much more pronounced. We can see that within $\sim 2R_\mathrm{200c}$, the number of satellite galaxies along the minor axis remains significantly lower than that along the major axis—a signature of anisotropy intrinsic to halos. However, unlike the left panel, this anisotropy fades rapidly at larger scales, eventually becoming almost undetectable -- very similar to the random direction. On the other hand, analysis of satellite galaxy counts along filament and median axis orientations shows that within $\sim 2R_\mathrm{200c}$, these counts are still lower than those along the halo major axis -- close to random direction. However, at larger scales, both $N_\mathrm{filaments}$ and $N_\mathrm{median}$ are 1.5–2.5 times $N_\mathrm{random}$. This outcome arises because satellite galaxy distributions on large scales trace filamentary structures. For the MS, the filamentary directions align preferentially with the median axes; such filaments are substantially misaligned with halo major axes, and consequently, halo major axes no longer correlate with satellite galaxy distributions at large radii. In summary, the anisotropy signal associated with halos is restricted to the halo scale, whereas filaments exert a stronger influence on the anisotropy signal at larger scales, which may support the satellite galaxies' supply along both halo major axes for the AS case and median axes for the MS case. This can explain the finding that satellite galaxies prefer the direction of cosmic filaments, as demonstrated by \citet{2015MNRAS.450.2727T} and \citet{2020ApJ...900..129W}.

\section{Physical Insight: Dynamical Behaviors of Satellite Galaxies}
\label{sec:physical_insight}

\subsection{Radial Dependence of Central Galaxies and Halos Morphologies}
\label{subsec:R_Morphologies}

The misalignment between the central galaxy major axis and the halo major axis essentially indicates a substantial discrepancy between central galaxy morphology and halo morphology. At the halo center, the stellar distribution, i.e. the central galaxy morphology, is primarily determined by the central subhalo's gravitational potential. Within the halo and outside the central region, the dark matter distribution, which controls the distribution of satellite galaxies, is governed by structure formation processes mainly contributed by external filaments. We suspect that both the number and orientation of filaments can affect how halos accrete material, which in turn reshapes the dark matter distribution within halos. Over time, this may even swap the halo's major and median axes, leading to the observed misalignment between filaments and the halo major axis. Meanwhile, the dynamics of central regions also contributes to the distribution of inner satellite galaxies. This consequently leads to discrepancies between satellite anisotropy distributions and the different orientations of dark matter halos at different radii. 

\begin{figure}[htbp]
    \centering
    \begin{minipage}[b]{0.45\textwidth}
        \centering
        \includegraphics[width=\textwidth]{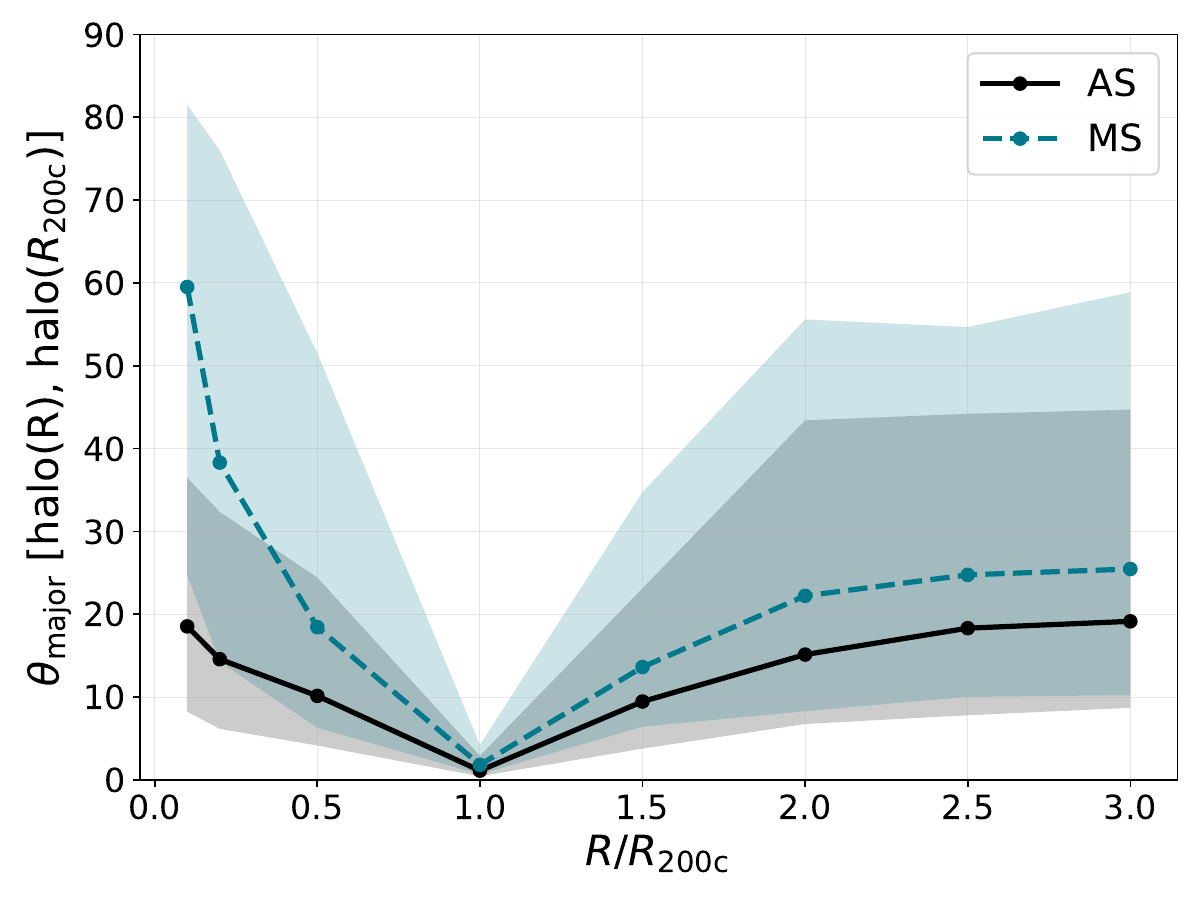}
        \small 
    \end{minipage} \hspace{0pt}
    \begin{minipage}[b]{0.45\textwidth}
        \centering
        \includegraphics[width=\textwidth]{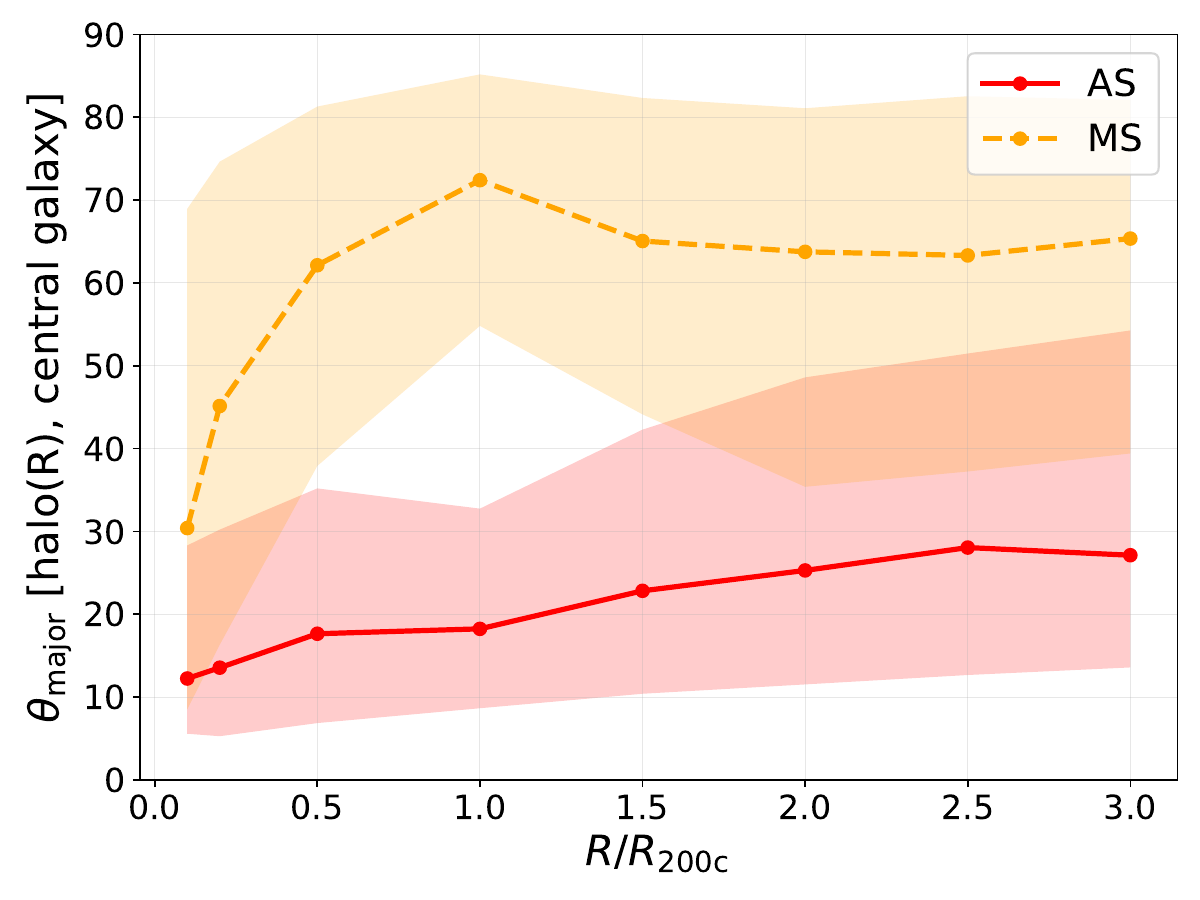}
        \small
    \end{minipage}
    \caption{Left panel: The angle between the halo major axis derived within different radial apertures and that defined within $R_{\rm 200c}$. Right panel: The angle between the halo major axis derived within different radial apertures and the central galaxy major axis. Curves show the median angle, and shaded bands correspond to the $1\sigma$ confidence interval. For the MS, the morphology of small-scale halos exhibits a significant discrepancy from that of their outer counterparts, whereas for both the AS and MS, the orientation of the central galaxy major axis is more consistent with that of the halo major axis in the central region.}
    \label{fig:SIMBA_theta_major_halo_R_z}
\end{figure}

To better understand this picture, Figure \ref{fig:SIMBA_theta_major_halo_R_z} shows the angle changes of the halo major axes (left panel) and between central galaxy and halo major axes (right panel) as a function of radius for our halo sample with $M_\mathrm{200c} \ge 10^{13}M_{\odot}$. We split our galaxies into the AS and MS, adopting the same definitions established in Section \ref{sec:SDA_BCG_halo} based on the angle between central galaxies and their host halo major axes. As shown in the left panel of Figure \ref{fig:SIMBA_theta_major_halo_R_z}, for halos in the AS, the orientation of the dark matter major axis within different radial apertures shows no significant deviation from that defined within $R_\mathrm{200c}$, even at small scales, which is because the major axes of the central galaxy and halo in the AS are inherently well-aligned. However, the scenario is entirely different for the MS: the orientation of the dark matter major axis at smaller scales deviates significantly from that within $R_\mathrm{200c}$ -- a great influence due to the central galaxy potential, whereas this discrepancy weakens substantially at larger scales. This phenomenon of halo misalignment at different radii is also discussed in \citet{2014ApJ...786....8W}. Undoubtedly, this indicates that for samples where the central galaxy and halo major axes are misaligned, the morphology of the halo at small radii differs more significantly from that at larger scales. Regarding the cause of this misalignment, we suspect that the orientation changes of the central galaxy’s principal axes play a major role, which is discussed in Appendix \ref{sec:omega_major_minor} in detail.

For the AS in the right panel, the central galaxy major axis is well-aligned with the halo major axis derived at different radial apertures, which demonstrates the consistency of halo morphology across different radial scales for this sample. For the MS, however, the central galaxy major axis is more closely aligned with the halo major axis at small scales -- the influence coming from its potential, but exhibits a significant misalignment at large scales. This indicates that regardless of the sample type, the orientation of the central galaxy major axis provides a better tracer of the halo orientation at the central region, which is consistent with previous work \citep{2006MNRAS.369.1293Y, 2016IAUS..308..448K, 2023MNRAS.521.5483R}.

The above results demonstrate that the orientation of the central galaxy's major axis can only serve as a robust tracer of the major axis of the dark matter/satellite galaxies' distribution in the central region. While the dynamics of satellite galaxies is governed by the gravitational potential within their host halos, its spatial distribution can characterize the dark matter distribution across a broad range of scales, spanning both small and large regimes. Combined with the conclusions drawn in Section \ref{sec:SDA_BCG_halo} and Section \ref{sec:SDA_halo_filament}, this explains why the anisotropy of satellite galaxies correlates strongly with central galaxy morphology on small scales, exhibits a clear association with halo morphology on halo scales, and displays a pronounced connection with filamentary structures on larger scales.

\subsection{Radial Dependence of Satellite Galaxy Kinematics via Tracking}

Satellite galaxies are orbiting within their dark matter halos, instead of falling towards the halo center in a direct line. This brings the question of how the satellite's anisotropies originated. To quantitatively answer this question, for which it is expected that the anisotropy of satellite galaxy distributions arises from their anisotropic infall in the gravitational field, we analyzed the motions of satellite galaxies via tracking along the central galaxy axis and halo axis in the AS and MS, respectively. In this subsection, the satellite galaxies are selected from halos with $M_\mathrm{200c} \ge 10^{13}M_{\odot}$. During this analysis, we traced the past trajectories of satellite galaxies located along the major (minor) axis at $z=0$, until the redshift reached $1.0$ or the radial coordinate of the galaxy exceeded $2R_\mathrm{200c}$. For any given galaxy along the major (minor) axis, to eliminate the influence of sample selection on statistical results, we only considered the trajectory of the galaxy before its final entry into the major (minor) axis of the host halo at the corresponding redshift. The total tracked time is denoted as $\Delta t_\mathrm{total}$. We define $\Delta t_\mathrm{major}$ and $\Delta t_\mathrm{minor}$ as the total durations during which the satellite lies within a $45^{\circ}$ cone aligned with the major and minor principal axes, respectively. By analyzing the fractions of $\Delta t_\mathrm{major}$ and $\Delta t_\mathrm{minor}$ relative to $\Delta t_\mathrm{total}$, we can determine whether the past motions of satellite galaxies exhibit any tendency correlated with central galaxy or halo morphology. For the i-th galaxy, we compute $\Delta t_\mathrm{i, \, major}$, $\Delta t_\mathrm{i, \, minor}$ and $\Delta t_\mathrm{i, \, total}$, from which we obtain the key statistical quantities:
\begin{eqnarray}
\langle \Delta t_\mathrm{major} / \Delta t_\mathrm{total} \rangle = \frac{1}{N_\mathrm{galaxy}} \sum^{N_\mathrm{galaxy}}_{i=0}\frac{\Delta t_\mathrm{i, \, major}}{\Delta t_\mathrm{i, \, total}}, \\
\langle \Delta t_\mathrm{minor} / \Delta t_\mathrm{total} \rangle = \frac{1}{N_\mathrm{galaxy}} \sum^{N_\mathrm{galaxy}}_{i=0}\frac{\Delta t_\mathrm{i, \, minor}}{\Delta t_\mathrm{i, \, total}}.
\end{eqnarray}

\begin{figure}[h!]
    \centering
    \begin{minipage}[b]{0.4\textwidth}
        \centering
        \includegraphics[width=\textwidth]{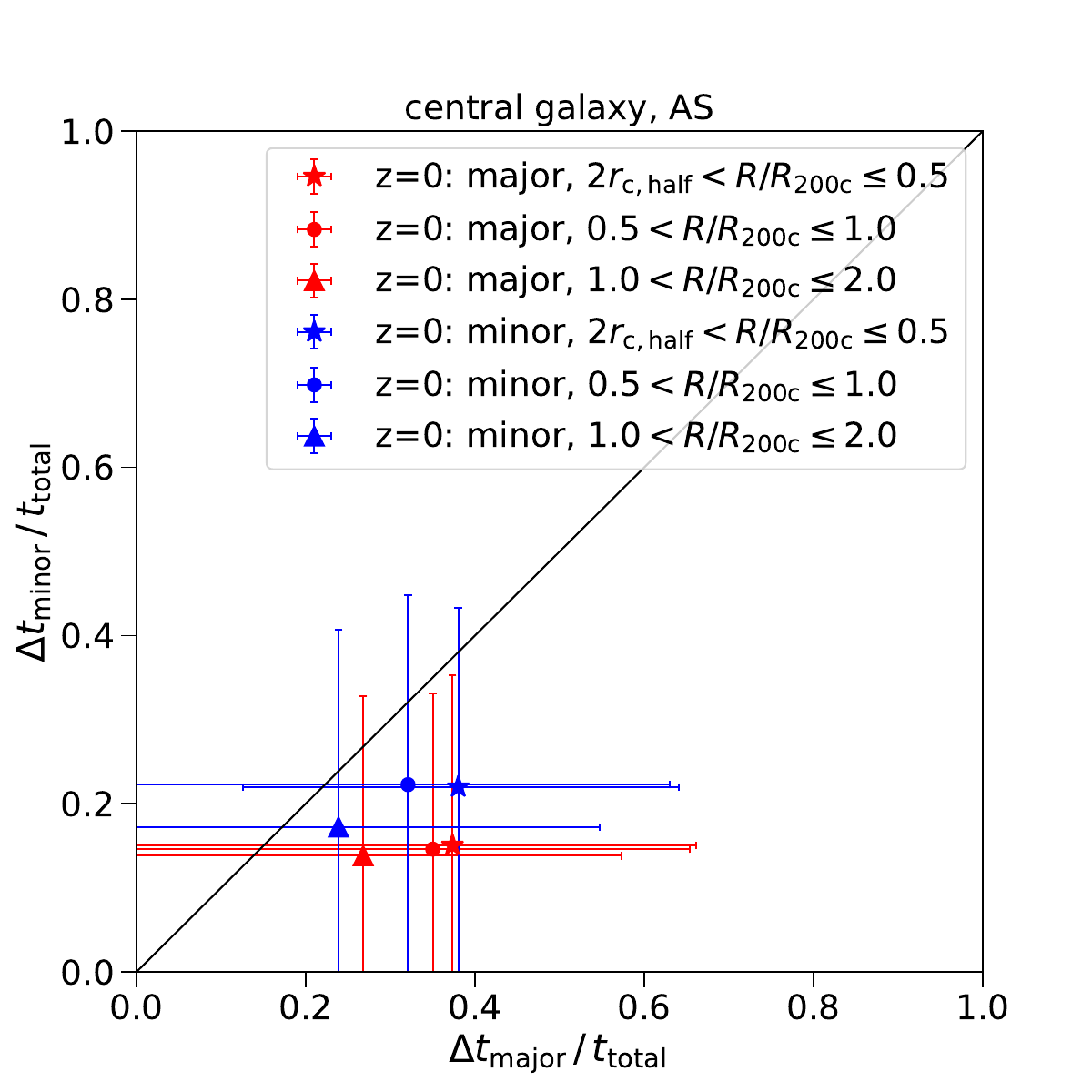}
        \small 
    \end{minipage} \hspace{-20pt}
    \begin{minipage}[b]{0.4\textwidth}
        \centering
        \includegraphics[width=\textwidth]{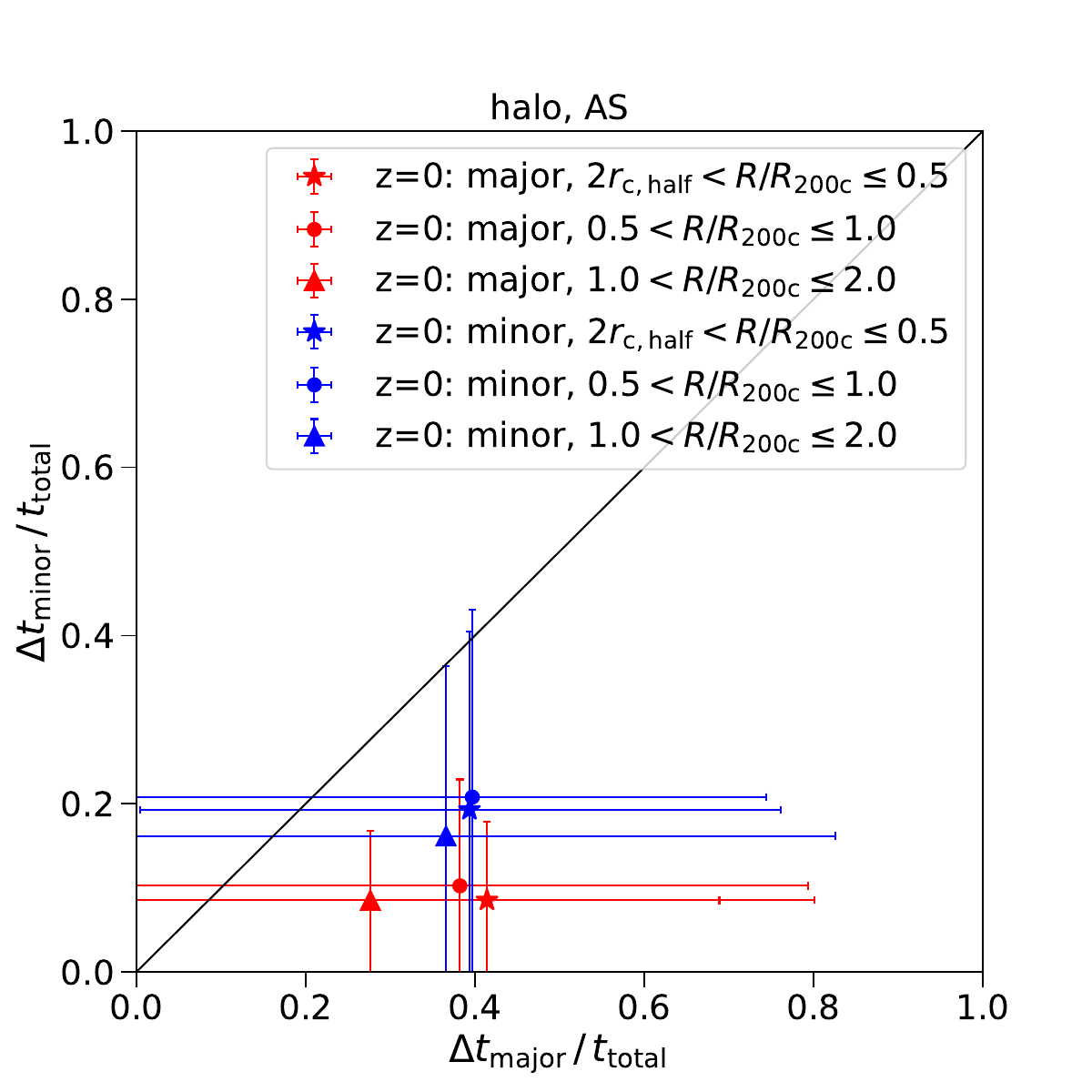}
        \small
    \end{minipage}
    
    \vspace{-4pt}
    
    \begin{minipage}[b]{0.4\textwidth}
        \centering
        \includegraphics[width=\textwidth]{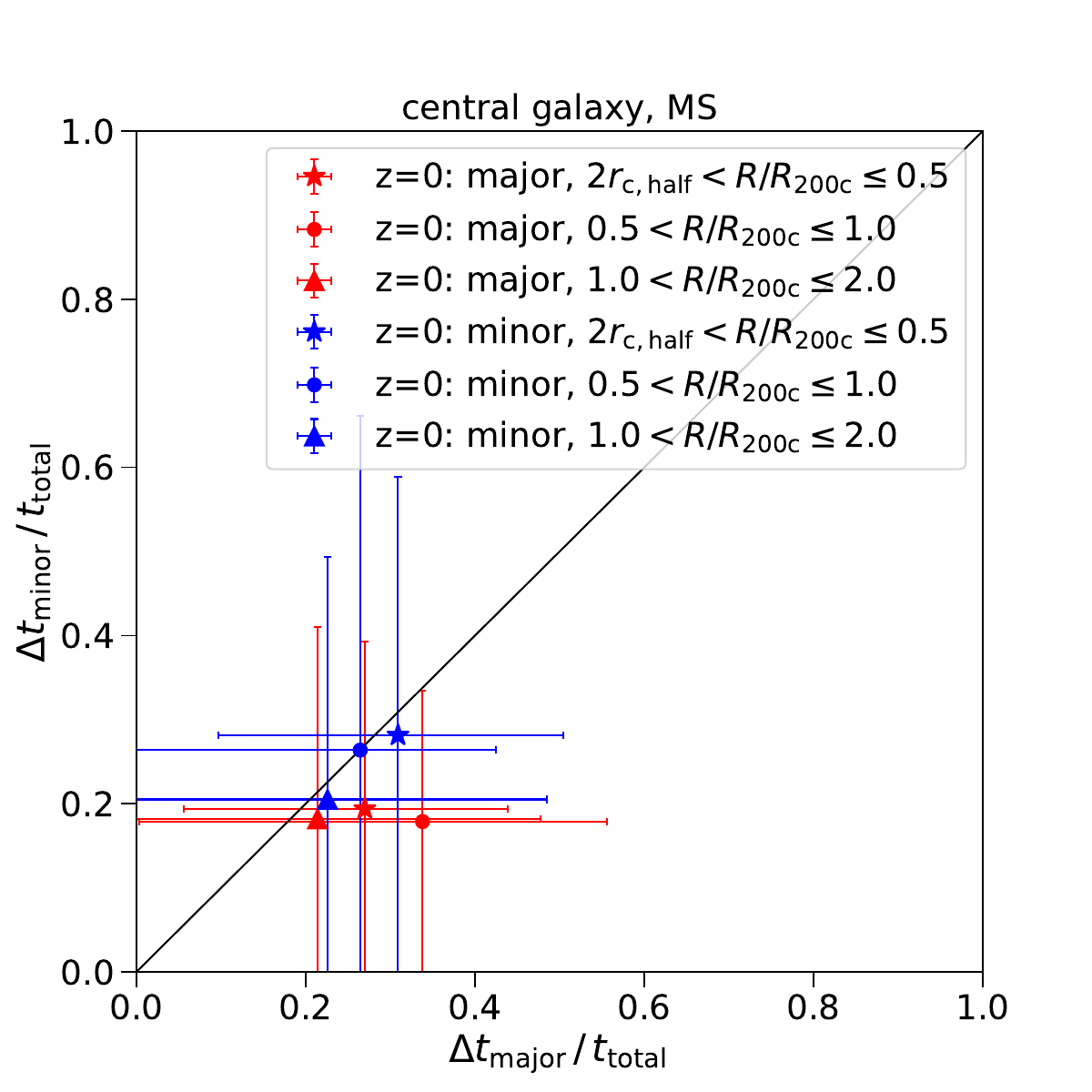}
        \small 
    \end{minipage} \hspace{-20pt}
    \begin{minipage}[b]{0.4\textwidth}
        \centering
        \includegraphics[width=\textwidth]{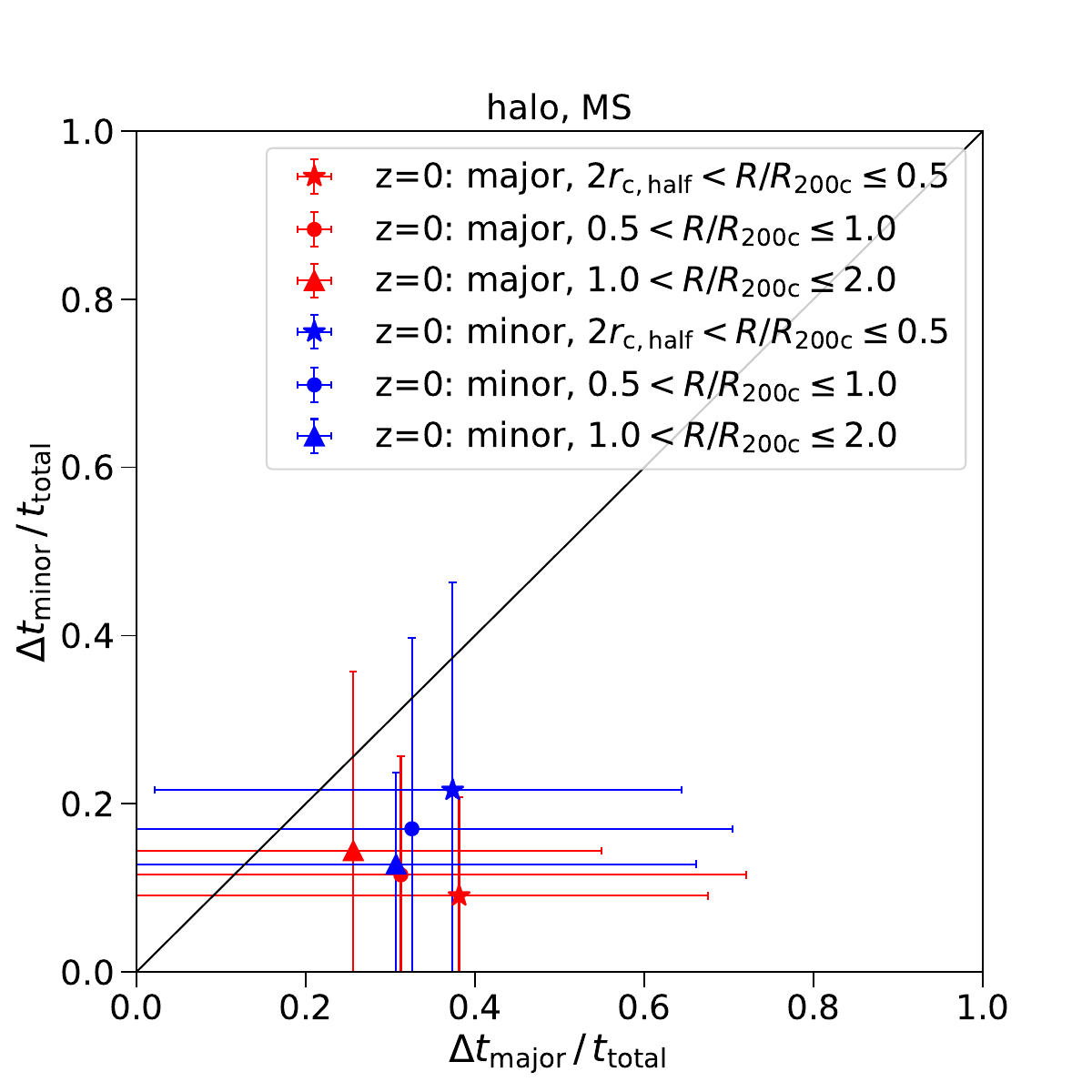}
        \small
    \end{minipage}
    \caption{Kinematic preferences of satellite galaxies in AS and MS, quantified via the residence time ratios $\Delta t_\mathrm{major}/\Delta t_\mathrm{total}$ and $\Delta t_\mathrm{minor}/\Delta t_\mathrm{total}$ across radial bins. Data points represent the mean values, with error bars corresponding to the 16th-84th percentiles of the sample distribution. AS galaxies show consistent major-axis trajectory preference, stronger for centrally concentrated objects; MS galaxies exhibit no such bias when sorted in central galaxy triaxial frame ($\Delta t_\mathrm{major} \approx \Delta t_\mathrm{minor}$), but recover major-axis preference when sorted in halo triaxial frame.}
\label{fig:motion_trend}
\end{figure}

In Figure \ref{fig:motion_trend}, we present the kinematic preferences of satellite galaxies in different orientations across various radial bins. During the tracking, we find that the satellite galaxies will spend a certain time on both minor and major axes regardless of their final positions. For galaxies in both AS and MS, we obtain a more prominent $\langle \Delta t_\mathrm{major}/\Delta t_\mathrm{total} \rangle$ relative to $\langle \Delta t_\mathrm{minor}/\Delta t_\mathrm{total} \rangle$ when using halo axes instead of central galaxy axes, demonstrating that the orbital anisotropy of satellite galaxies is more tightly linked to the host halo rather than the central galaxy, consistent with earlier findings. Consistently, for galaxies in the AS, the mean value of $\langle \Delta t_\mathrm{major} / \Delta t_\mathrm{total} \rangle$ is higher than $\langle \Delta t_\mathrm{minor} / \Delta t_\mathrm{total} \rangle$ for satellite galaxies staying on either major or minor axis at $z=0$. This result indicates that galaxies were more likely to appear along the major axis in the past or have spent a longer time at the major axis, rather than the minor axis. Moreover, galaxies closer to the center at $z=0$ exhibit more pronounced kinematic preferences in the past than galaxies at outer radius. This is because the motions of such satellite galaxies, being more strongly influenced by the gravitational potential of the halo, sufficiently reflect the spatial anisotropy of the potential. Centrally concentrated satellite galaxies generally accrete into the cluster at earlier cosmic times and thus undergo a longer evolutionary timescale. Consistently, \citet{2015ApJ...813....6K} found that the alignment preference of subhalos along the major axis of their host halo becomes more pronounced after accretion. This physical picture may qualitatively explain the prominent anisotropy signal observed in the spatial distribution of red satellite galaxies in previous studies \citep{2006MNRAS.369.1293Y, 2016IAUS..308..448K, 2020ApJ...893...87T, 2022MNRAS.514.1077R}.

However, when galaxies are considered in the MS, those along the central galaxy major and minor axes show almost none of the aforementioned preferences. Instead, these galaxies all exhibit a trend where $\langle \Delta t_\mathrm{major} / \Delta t_\mathrm{total} \rangle \approx \langle \Delta t_\mathrm{minor} / \Delta t_\mathrm{total} \rangle$. This expected result dynamically confirms that the orientations of these misaligned central galaxies point to randomized directions. Therefore, the major axis of central galaxy is a biased tracer for the satellite galaxies' assembly or halo gravitational potential distribution. When we analyze galaxies along the halo major and minor axes, we once again observe a strong preference for larger $\langle \Delta t_\mathrm{major} / \Delta t_\mathrm{total} \rangle$. This behavior is analogous to that seen in the AS, including the fact that galaxies closer to the center display more prominent kinematic preferences in their motions.

\begin{figure}[htbp]
    \centering
    \includegraphics[width=0.8\textwidth]{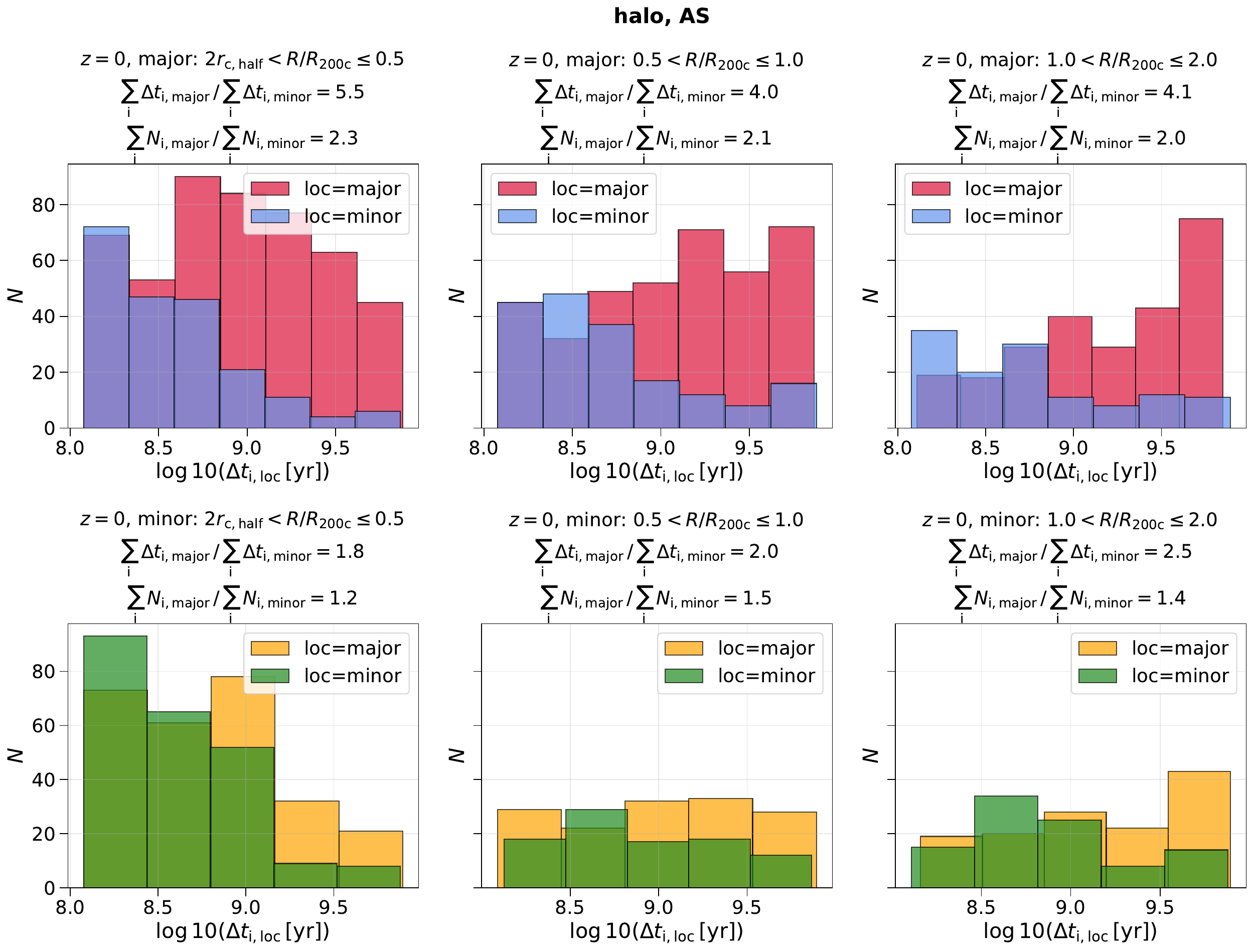}\\[3pt]
    \includegraphics[width=0.8\textwidth]{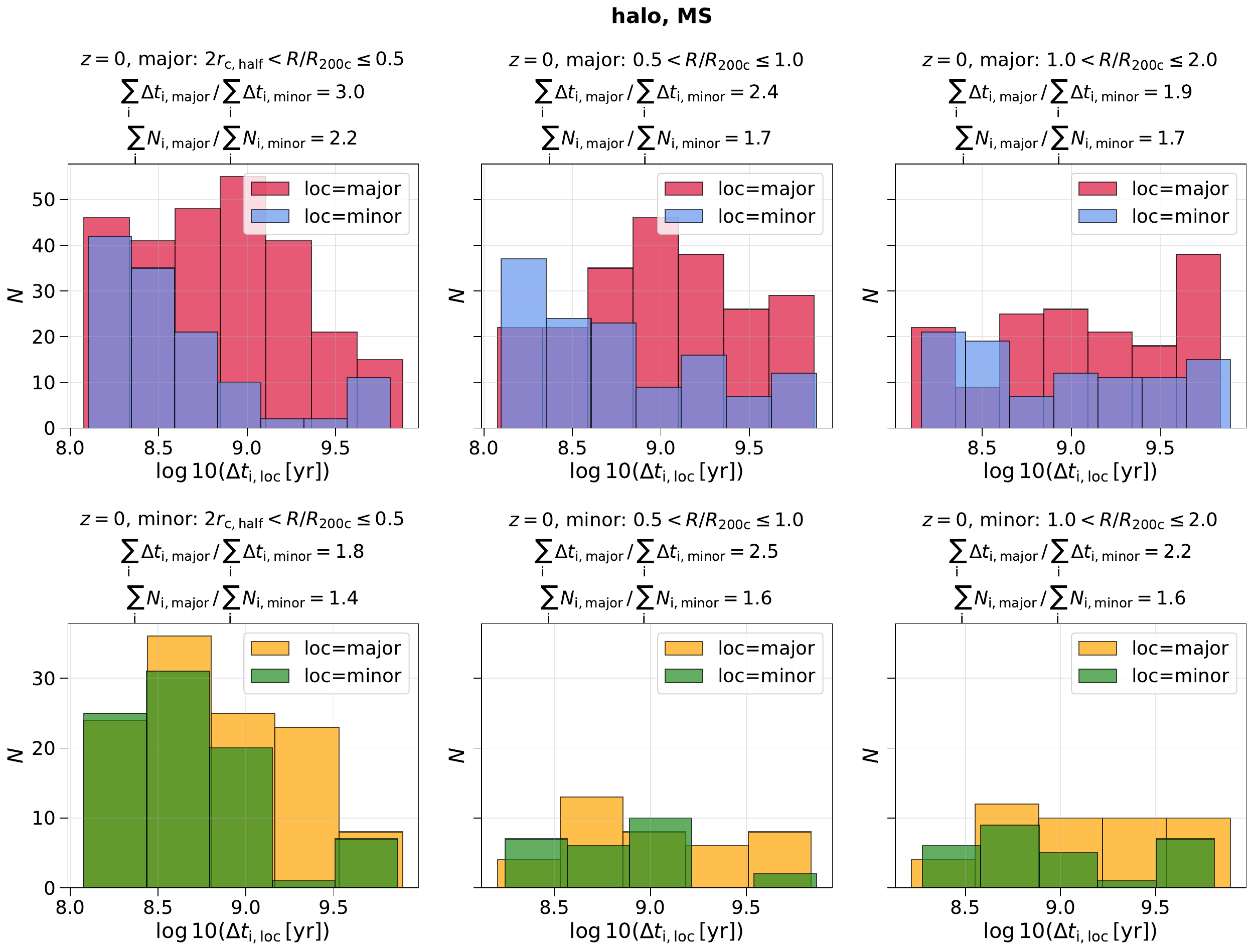}
    \caption{Distribution of the time durations for satellite galaxies crossing the major and minor axes in the triaxial halo ellipsoid coordinate system. 
    \textit{Top}: halos and satellites from the AS.
    \textit{Bottom}: halos and satellites from the MS.
    Satellites in both samples preferentially cross the major axis more frequently and with longer time durations.}
    \label{fig:dm_axis_dt}
\end{figure}

The previous figure only shows relative results for qualitative conclusions. To further quantify this dynamical information, we study all satellite galaxies' axis-crossing times, i.e. the time it spent in the major and minor axes of host halos. The satellite galaxy samples of AS and MS used here are identical to those shown in the top-left and bottom-right panels of Figure \ref{fig:motion_trend}, respectively, since these samples exhibit the strongest anisotropic signal. 
Figure \ref{fig:dm_axis_dt} present the histogram of these crossing times, for which these satellites are needed to cross the major or minor axis of the host halo. The horizontal axes $\Delta t_\mathrm{i, major}$ and $\Delta t_\mathrm{i, minor}$ represent the time interval for a single galaxy to cross the cone region aligned with the halo’s major and minor axis, respectively. $\sum\limits_\mathrm{i} \Delta t_\mathrm{i, major}$ and $\sum\limits_\mathrm{i} \Delta t_\mathrm{i, minor}$ are the total accumulated crossing time over all galaxies for major-axis and minor-axis passages. $\sum\limits_\mathrm{i} N_\mathrm{i, major}$ and $\sum\limits_\mathrm{i} N_\mathrm{i, minor}$ denote the total number of major-axis and minor-axis crossings summed across all galaxies, which we use to compare the overall crossing frequency for the two axes. The first and second rows show satellite galaxies from the AS staying on the major and minor axes at $z = 0$, respectively, with each row containing three panels for the three radius bins. The third and fourth rows show the corresponding major-axis and minor-axis galaxies from the MS. 
As shown in Figure \ref{fig:dm_axis_dt}, for satellites in both AS and MS, regardless of whether they lie on the major or minor axes of the halo at $z=0$, we find $\sum\limits_\mathrm{i} N_\mathrm{i, major} / \sum\limits_\mathrm{i} N_\mathrm{i, minor} > 1$. This shows that the number of major-axis crossings is greater than that of minor-axis crossings, and the trend is much more apparent in the first and third rows than in the second and fourth rows, respectively. Furthermore, based on orbit tracing for individual satellites, we measure the time it takes for each galaxy to cross the conical aperture defined around the major or minor axis. We find that satellites crossing the major axis typically have longer crossing times, whereas those associated with the minor axis are dominated by shorter crossing times. This result indicates that satellite orbits tend to intersect the major axis of the host halo more frequently and for longer staying times.

In summary, the anisotropy signal of satellite galaxy distributions is manifested not only in their preferential spatial configuration at a given epoch but also in the directional preference of their past trajectories. This trajectory preference is also highly correlated with halo morphology, which is the physical reason for making them stay longer along its major axis. This pronounced kinematic influence of halos on satellite galaxies also explains why a strong anisotropy signal can be detected in higher-redshift datasets. As such, this satellite galaxy anisotropy phenomenon is not only spatially stable, but also dynamically stable. In addition, although satellite galaxies tend to cluster along filamentary structures on scales larger than those of halos, their azimuthal angles undergo substantial alterations under the influence of halo gravitational potentials after they accrete into halos, thereby erasing the original filament-related signals. This conclusion is consistent with the finding in Section \ref{sec:SDA_halo_filament} that the anisotropy signal within halos is not dominated by filaments.

\clearpage

\section{Conclusions} \label{sec:conclusions}

Using SIMBA simulation data, together with EAGLE and TNG100 data at $z=0$, we probe the physical origins of the anisotropy of satellite galaxies in this paper. To perform a thorough study, we first analyzed the probability of misalignment between the central galaxy major axis and the halo major axis, as well as the anisotropy signals of satellite galaxies along the respective major and minor axes of both central galaxies and halos. The anisotropy distribution is further investigated across different halo masses, radial bins and redshifts. In addition, we used SIMBA data at $z=0$ to investigate the correlation between the halo major axis and filament orientations. Therefore, we can have a full picture of the anisotropy signals of satellite galaxies across different spatial scales. Finally, we track the satellite galaxy's evolution history to understand the formation of the anisotropy signal from a kinematic point of view. Through extensive systematic investigations and comparisons, we derive the following key conclusions:

\begin{itemize}

\item \textbf{central galaxy-halo misalignment vs. halo mass}

In the SIMBA and TNG100 simulations, the misalignment probability between the major axes of central galaxies and their host halos correlates negatively with $M_{200\mathrm{c}}$, meaning more massive halos exhibit stronger axis alignment. We also analyzed high-redshift data from SIMBA up to $z=1.5$, and this alignment preference weakens at higher redshifts. We suspect this arises from frequent merging events at higher redshift, which can strongly perturb the morphologies of both central galaxies and their host halos, thus inhibiting the alignment of their major axes. This trend holds even for low-mass halos slightly above $M_\mathrm{200c}=10^{11}M_{\odot}$.

\item \textbf{Halo-filament misalignment vs. halo mass}

For halos with $M_\mathrm{200c}\ge10^{13}M_{\odot}$, the misalignment rate between the host halo major axis and cosmic filament orientations also decreases with increasing $M_\mathrm{200c}$, with massive halos being far more likely to have at least one aligned filament and this rate being significantly lower than that of random orientations, which demonstrates a statistically significant alignment between the cosmic filament direction and the halo major axis.

\item \textbf{Stability of satellite anisotropy signal}

The anisotropy signal of satellite galaxy distributions is statistically significant at the 3$\sigma$ level and persists across spatial scales from $2r_\mathrm{c,half}$ to $10R_\mathrm{200c}$, with robustness confirmed across all simulations with different baryon models, resolutions and volumes. Using the SIMBA dataset, we further find that this anisotropy signal extends up to $z=1.5$, which demonstrates the reliability of our conclusions.

\item \textbf{Scale-dependent structural tracers of satellite anisotropy}

Satellite galaxy anisotropy is governed by different structural components on varying spatial scales, and this conclusion is verified by multiple simulation datasets at the 3$\sigma$ confidence level. According to TNG100, central galaxy morphology dominates the anisotropy at small scales ($<0.3R_{200\mathrm{c}}$). Combined with SIMBA, TNG100 and EAGLE, we find that the triaxial geometry of host halos takes precedence at intermediate halo scales ($0.3R_{200\mathrm{c}} \sim 2R_{200\mathrm{c}}$). For large scales ($>2R_{200\mathrm{c}}$), results from SIMBA indicate that orientations of galaxy-traced filaments exert a stronger influence than host halo morphology.

\item \textbf{Satellite anisotropy vs. halo mass threshold}

For halos with $M_\mathrm{200c}\ge10^{11}M_{\odot}$, the anisotropy ratio of satellite galaxy distributions is consistently above 2 and shows a significant 3$\sigma$ signal strongly correlated with the halo major axis across all mass bins, albeit slightly lower at higher mass bins; this result is validated by all simulation datasets. A slightly lower but consistent anisotropy signal can also be correlated to the central galaxy major axis but only to the AS.

\item \textbf{Kinematic origin of satellite anisotropy}

By tracing the progenitors of $z=0$ satellite galaxies, we characterize their kinematic preferences through the analysis of residence time ratios: the fraction of time spent in proximity to the halo major axis ($\langle \Delta t_\mathrm{major}/\Delta t_\mathrm{total} \rangle$) versus the minor axis ($\langle \Delta t_\mathrm{minor}/\Delta t_\mathrm{total} \rangle$). This analysis reveals the physical origin of the satellite anisotropy: satellites preferentially have a higher number of encounters with the halo major axis and stay in it for a longer time. This reveals that satellite anisotropy exhibits both spatial and dynamical stability. However, this stable distribution seems not being contributed by the same number of galaxies. This kinematic evidence suggests that the satellite anisotropy observed at halo scales may be a direct manifestation of the host halo's gravitational potential. Notably, even in the early universe, the halo gravitational potential appears sufficiently mature to sustain such anisotropy via this dynamical mechanism, which could explain the persistence of these signals in higher-redshift datasets up to $z=1.5$.

\item \textbf{Non-radial motions of satellites}

By analyzing the trajectories of satellites, we find that satellite motions within the halo are not purely radial; instead, substantial azimuthal variations driven by the host halo's gravitational potential post-accretion may suppress the original filament-related anisotropy signals within $2R_\mathrm{200c}$.

\end{itemize}

Through our systematic analysis of the relationship between satellite galaxy distribution anisotropy and different host structures, we have obtained a clear physical picture for the origin of such anisotropies. These results advance our understanding of satellite distributions across systems ranging from the Local Group with mass around $10^{12}\,M_\odot$ \citep{2022MNRAS.513.2385C, 2023MNRAS.521.4863S, 2025A&A...698A.178M} to massive galaxy groups. In studies focusing on the density profiles of dark matter halos, satellite distributions and the orientations of central galaxies can be used to better infer the underlying dark matter distribution. Furthermore, our findings provide valuable insights for modeling the gravitational potential of lens objects in gravitational lensing studies. 

All constraints on satellite anisotropy presented here rely solely on numerical simulations. Follow-up work could extend this investigation to observational galaxy survey data, which may compare the alignment offsets between central galaxies and their host halo major axes and further verify conclusions regarding satellite anisotropy. In addition, this paper mainly explores how satellite anisotropy depends on spatial scale and halo mass. Future studies can incorporate internal satellite properties such as stellar mass and color to characterize the distinct spatial distribution patterns of different satellite populations.

\appendix

\section{PCA vs. Inertia Tensor Method for Major and Minor Axis Determination}
\label{sec:PCA_ITM}
\begin{figure}[htbp]
\centering
\begin{minipage}[b]{0.47\textwidth}
    \centering
    \includegraphics[width=\textwidth]{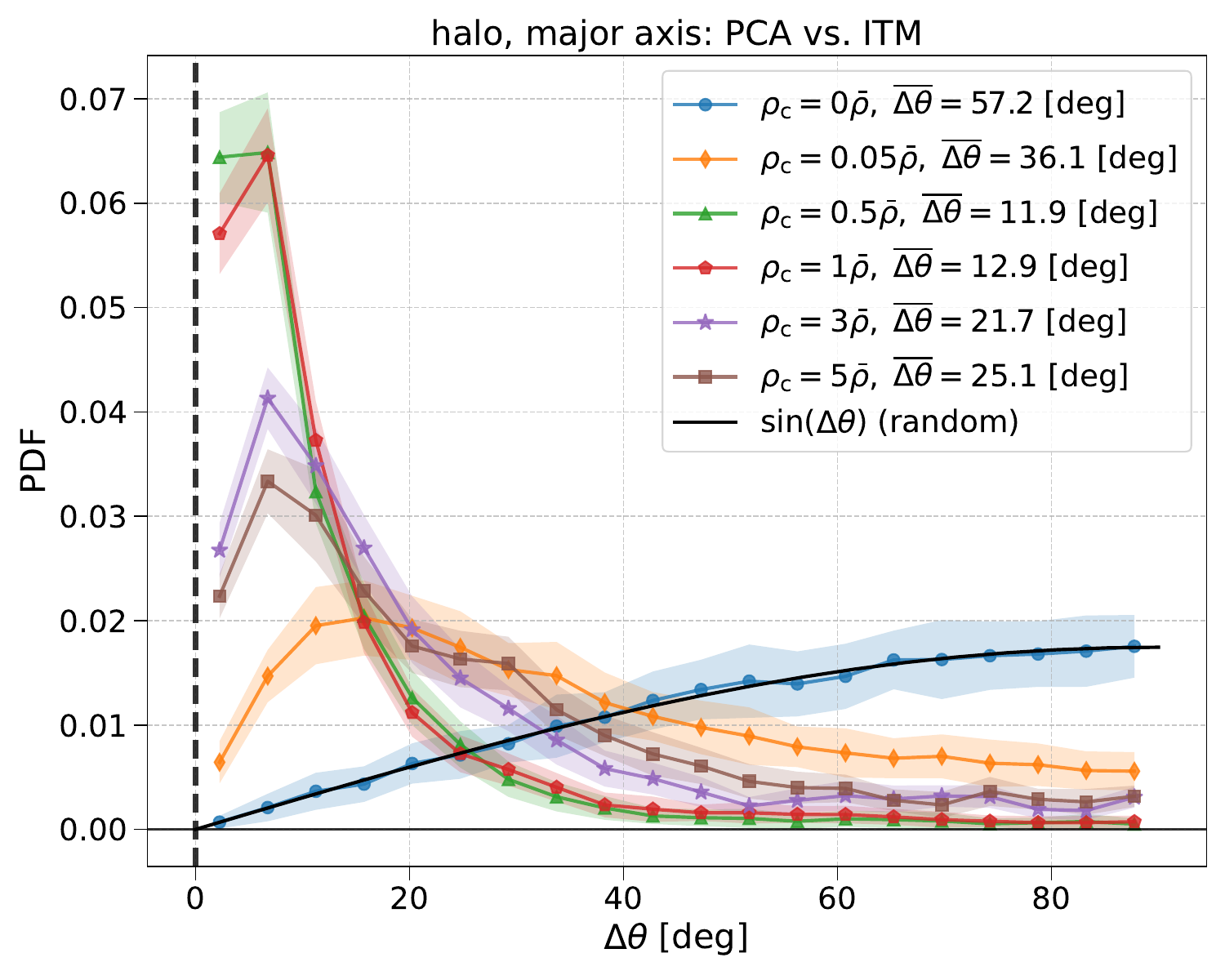}
    \small
\end{minipage}
\hspace{0pt}
\begin{minipage}[b]{0.47\textwidth}
    \centering
    \includegraphics[width=\textwidth]{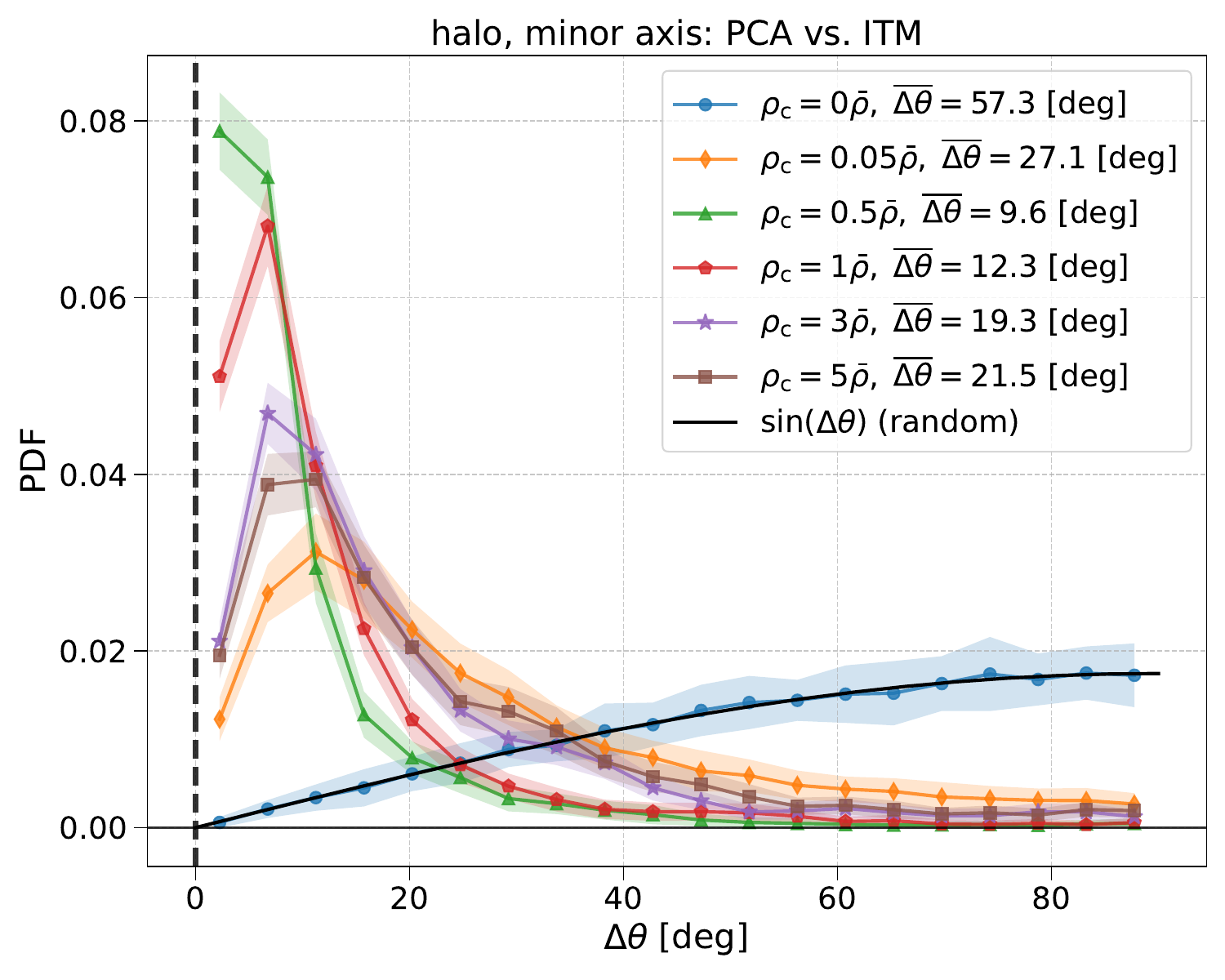}
    \small
\end{minipage}

\vspace{2mm} 

\begin{minipage}[b]{0.47\textwidth}
    \centering
    \includegraphics[width=\textwidth]{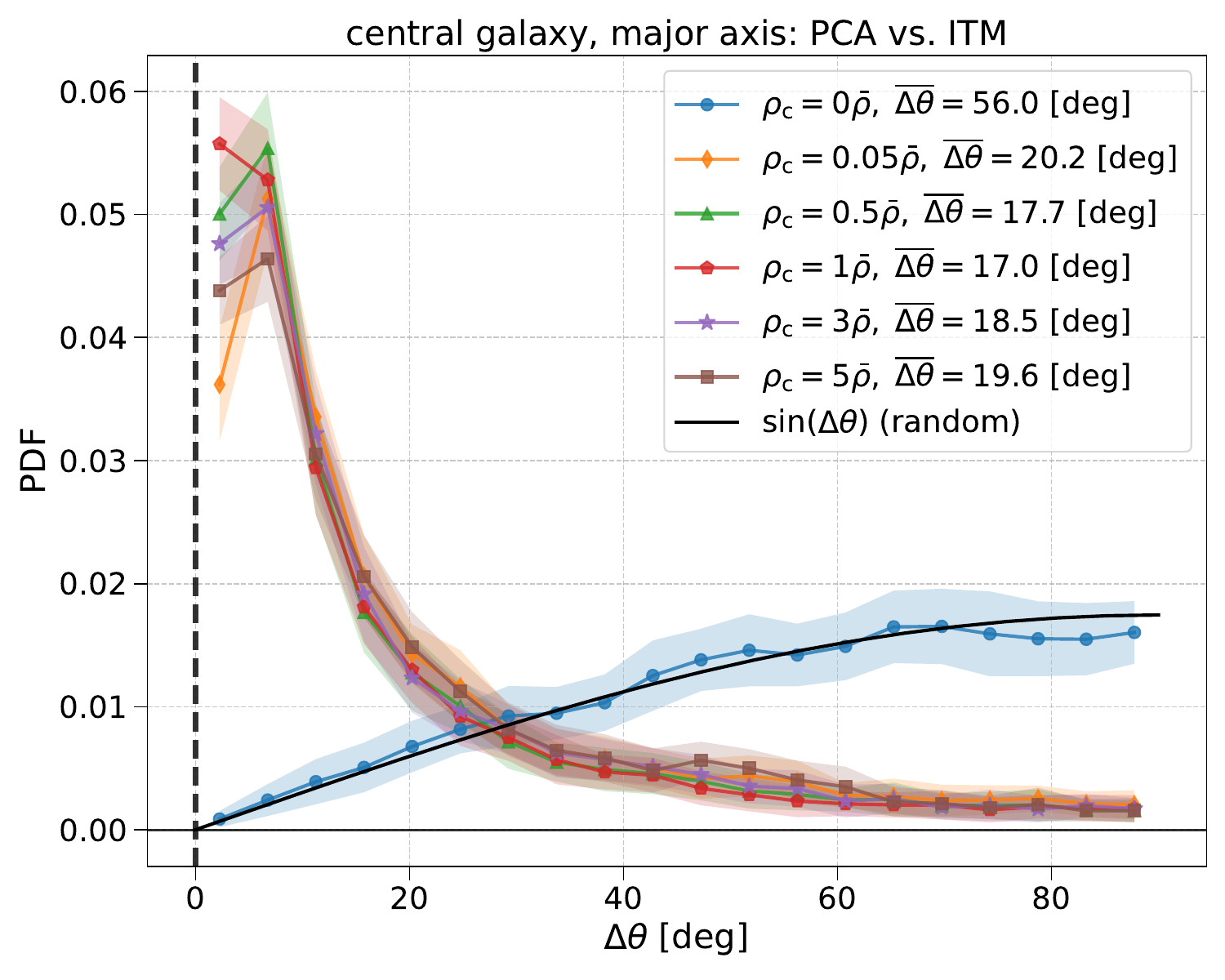} 
    \small
\end{minipage}
\hspace{0pt}
\begin{minipage}[b]{0.47\textwidth}
    \centering
    \includegraphics[width=\textwidth]{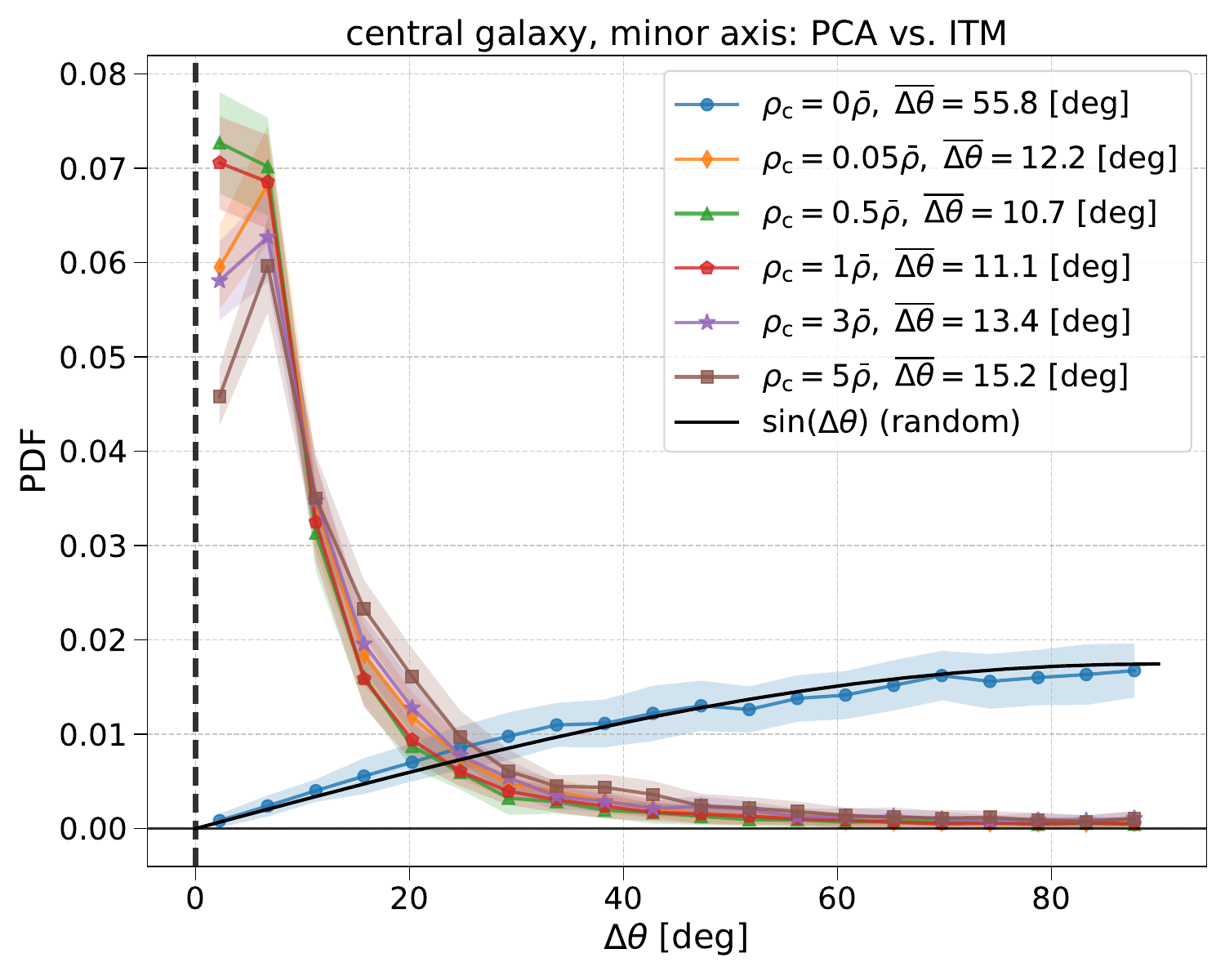} 
    \small
\end{minipage}

\caption{Probability density functions (PDFs) of alignment angles, together with the mean misalignment angles $\overline{\Delta\theta}$ indicated in the legend, are derived via ITM, and via PCA at different particle number density thresholds $\rho_\mathrm{c}$. The upper and lower rows show results for halos and central galaxies, respectively; the left and right columns correspond to major axes and minor axes. Solid lines show the mean values, and shaded regions indicate the $1\sigma$  standard deviation. The sample includes halos with $M_\mathrm{200c}\ge10^{13}M_{\odot}$ and their central galaxies from the SIMBA simulation at $z=0$. The black solid line represents the angle distribution from random orientations. The PCA results for $\rho_\mathrm{c}=0.5\bar{\rho}$ and $1\bar{\rho}$  show the smallest differences from the ITM.}
\label{fig:PCA_vs_ITM}
\end{figure}

To compare the differences in determining the major- and minor-axis orientations of the same ellipsoid between the PCA method introduced in Section \ref{subsec:Orientation_BCGs_halos_webs} and the conventional ITM, we use halo and central galaxy samples from the SIMBA simulation with $M_\mathrm{200c}\ge10^{13}M_{\odot}$. We quantify the discrepancy between the two methods using both the distribution of the misalignment angle $\Delta \theta$ and its mean value $\overline{\Delta\theta}$, as shown in Figure \ref{fig:PCA_vs_ITM}. To illustrate the dependence on the density threshold in the PCA method, we adopt a set of values $\rho_\mathrm{c}=[0,0.05,0.5,1,3, 5]\bar{\rho}$ for comparison.

From Figure \ref{fig:PCA_vs_ITM}, we find that for halos, the results at $\rho_\mathrm{c}=0.5\bar{\rho}$ and $1\bar{\rho}$ are more consistent with those from ITM, and noticeable variations exist among different $\rho_\mathrm{c}$. This indicates that the iso-density contours at different radii within halos exhibit distinct ellipsoidal morphologies, which likely shares the same physical origin as the radial dependence of halo morphologies discussed in Section \ref{subsec:R_Morphologies}. For central galaxies, by contrast, all non-zero $\rho_\mathrm{c}$ values yield consistent results that also agree with ITM. This demonstrates that the iso-density contours of central galaxies at different densities maintain a coherent major- and minor-axis orientation.

Finally, for both halos and central galaxies, the distribution at $\rho_\mathrm{c}=0$ is fully consistent with a random orientation, implying that the seed spheres are placed randomly without tracing the underlying particle distribution, in good agreement with our expectation.

\section{Projected Distribution of Satellite Galaxies}
\label{sec:proj_distribution_galaxy}
\begin{figure}[h!]
    \centering
    \begin{minipage}[b]{0.30\textwidth}
        \centering
        \includegraphics[width=\textwidth]{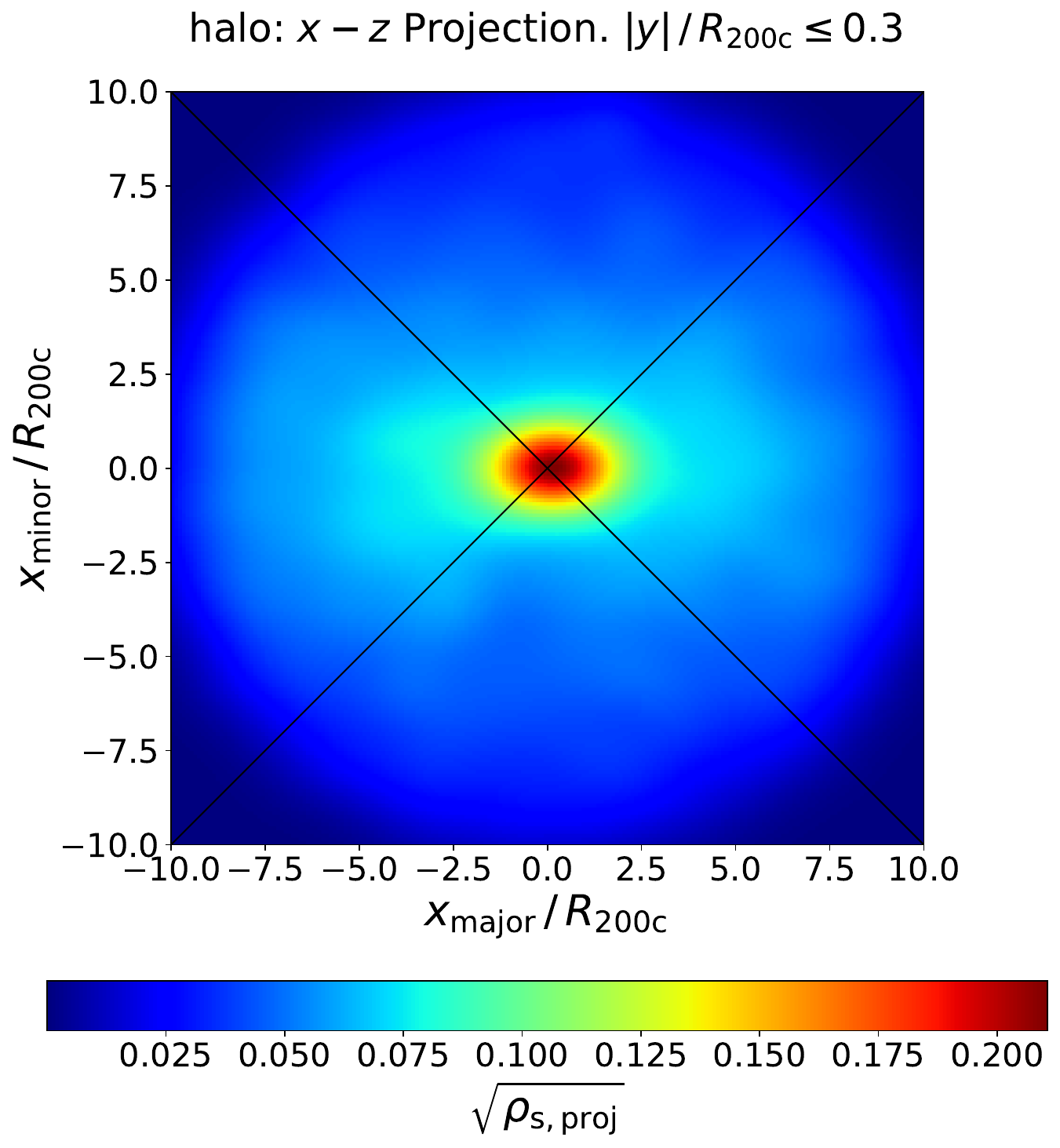}
    \end{minipage}
    \hspace{0pt}
    \begin{minipage}[b]{0.30\textwidth}
        \centering
        \includegraphics[width=\textwidth]{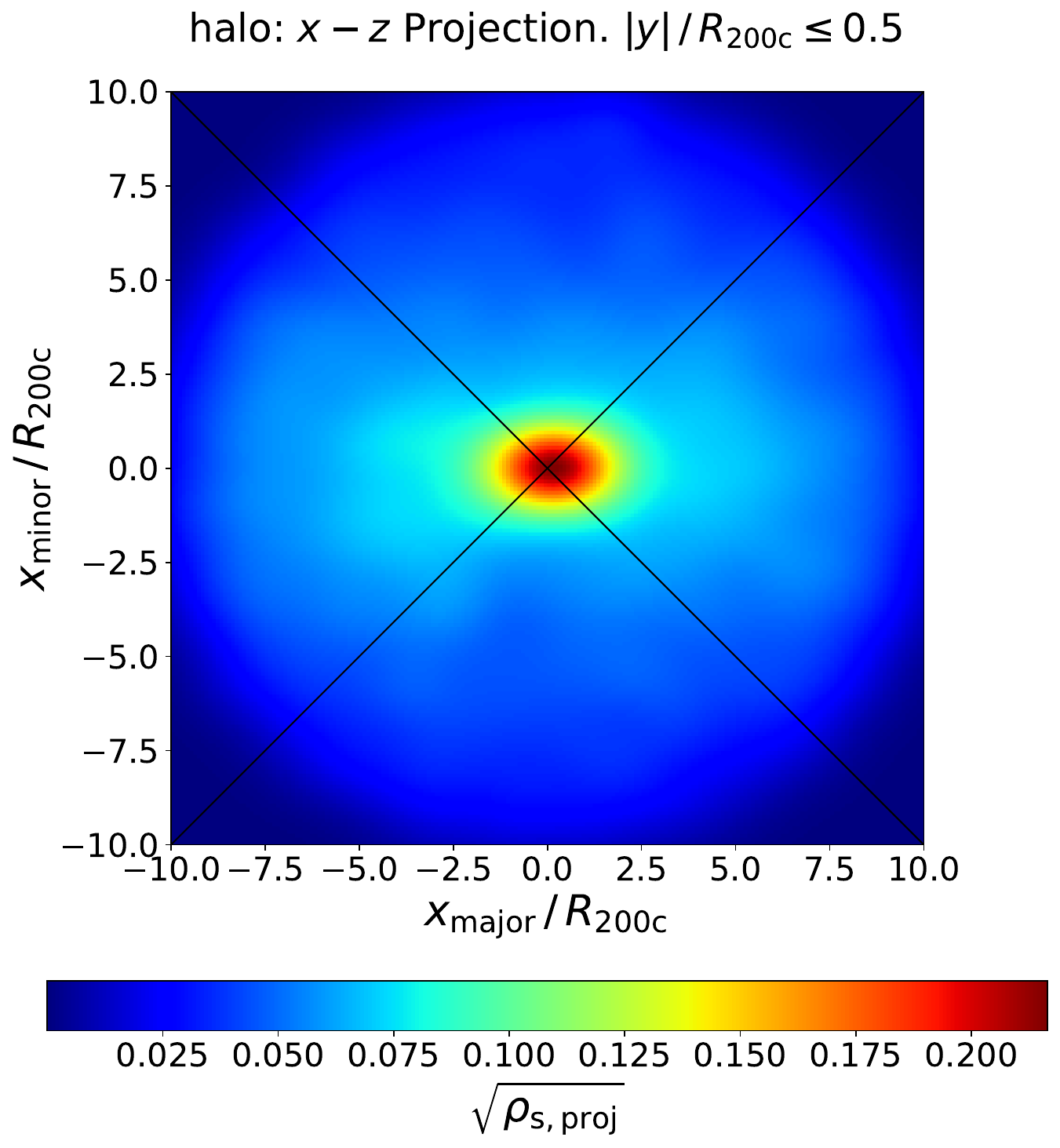}
    \end{minipage}
    \hspace{0pt}
    \begin{minipage}[b]{0.30\textwidth}
        \centering
        \includegraphics[width=\textwidth]{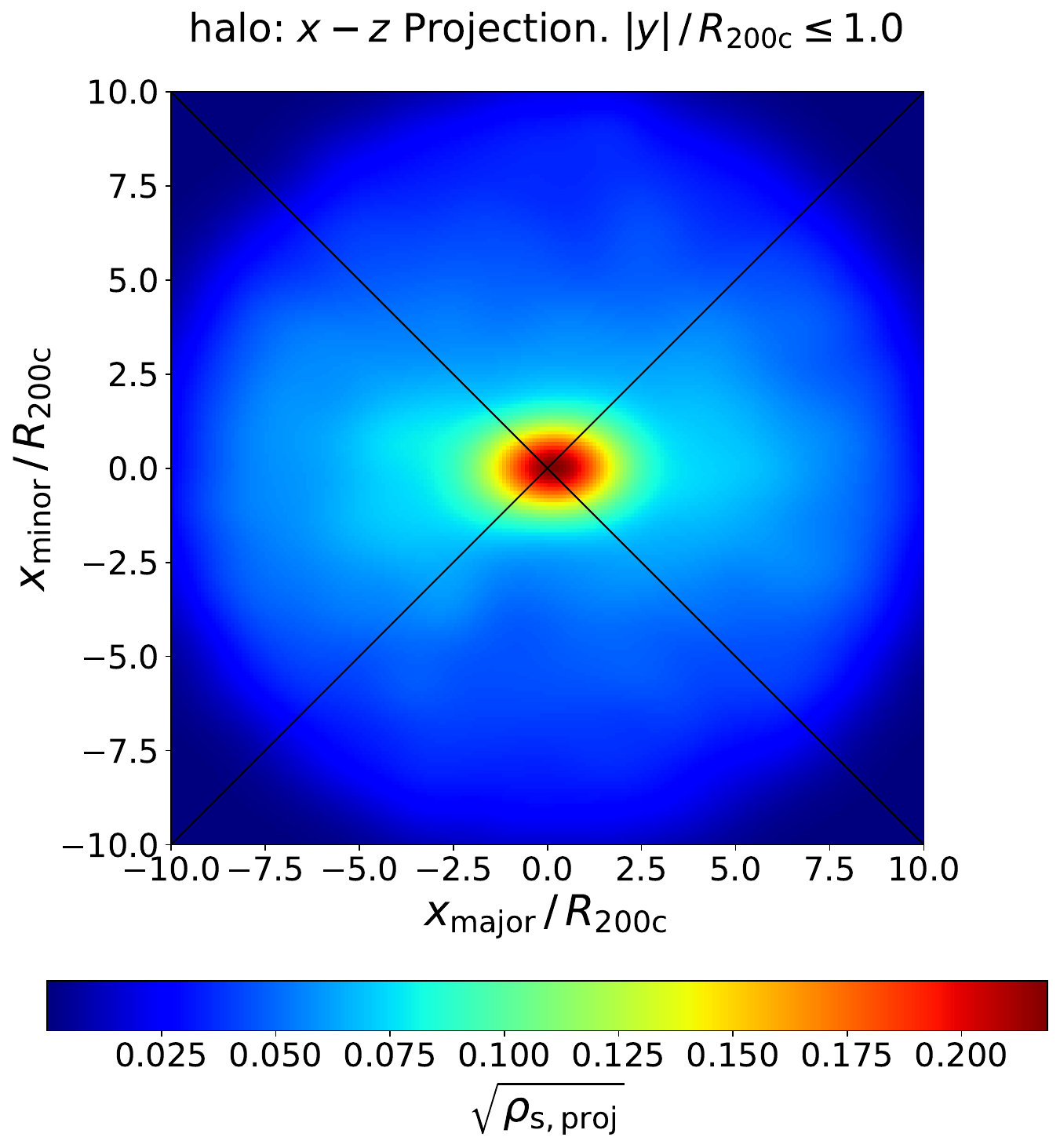}
    \end{minipage}
    \\[6pt]  
    \begin{minipage}[b]{0.30\textwidth}
        \centering
        \includegraphics[width=\textwidth]{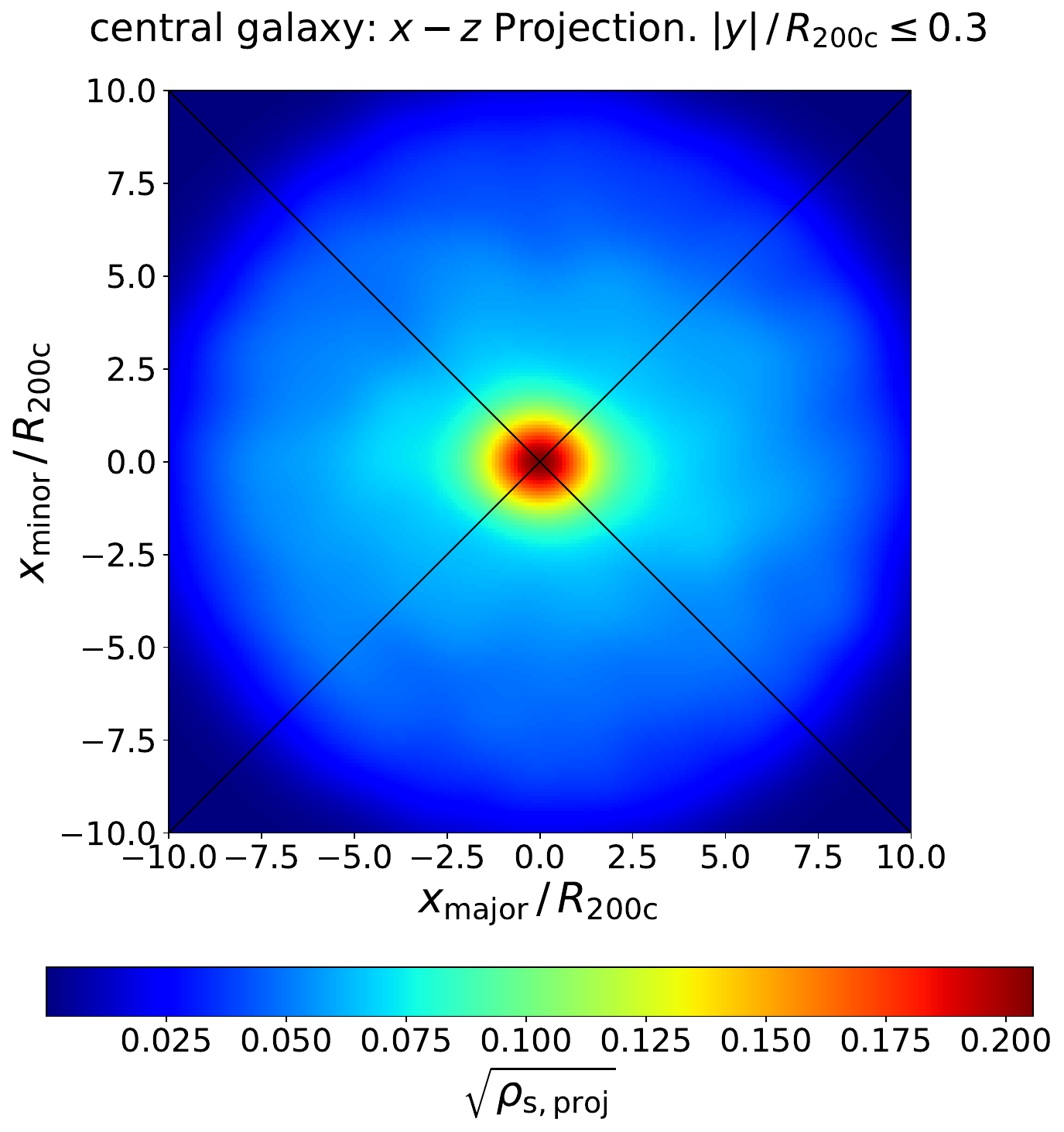}
    \end{minipage}
    \hspace{0pt}
    \begin{minipage}[b]{0.30\textwidth}
        \centering
        \includegraphics[width=\textwidth]{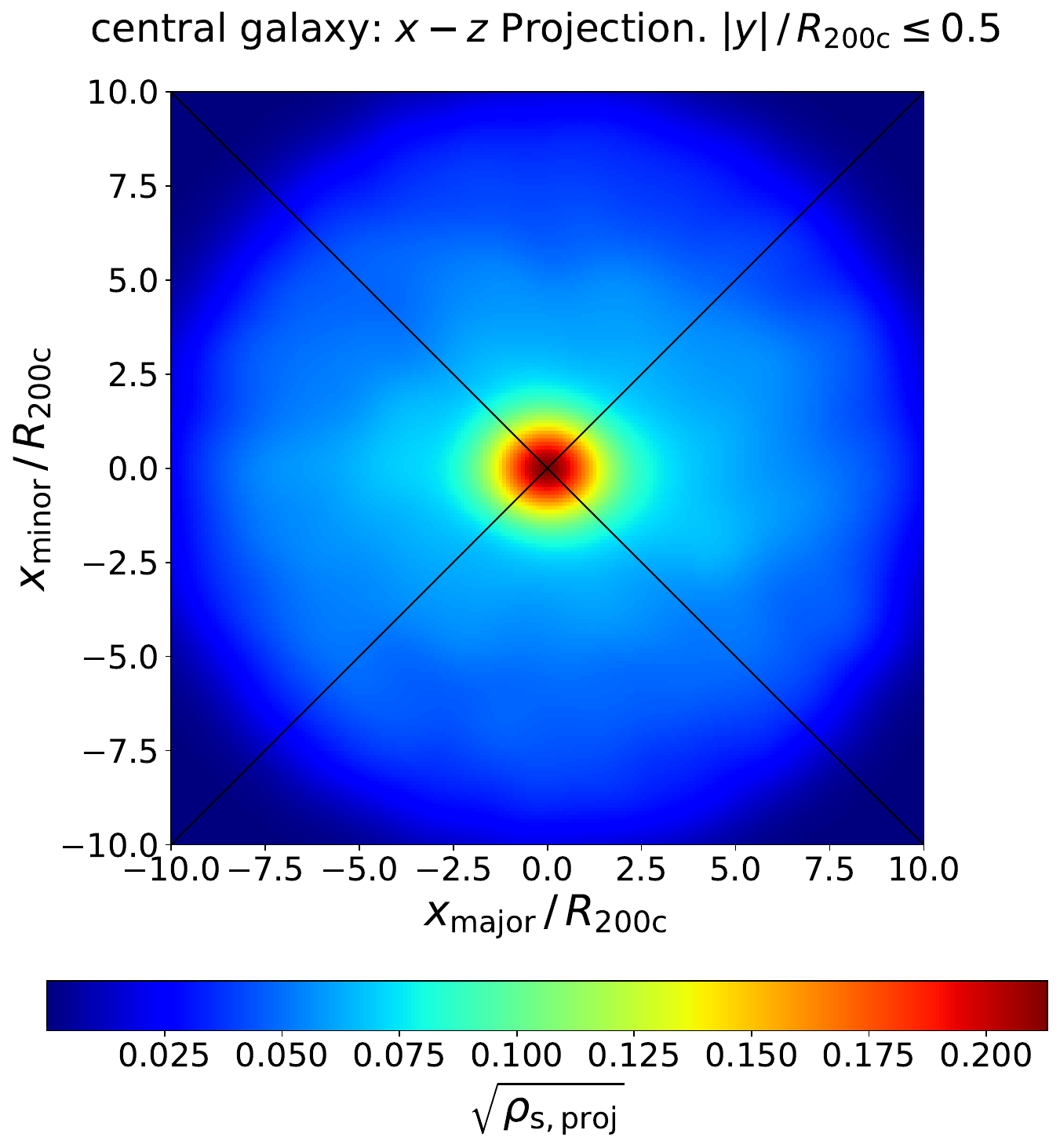}
    \end{minipage}
    \hspace{0pt}
    \begin{minipage}[b]{0.30\textwidth}
        \centering
        \includegraphics[width=\textwidth]{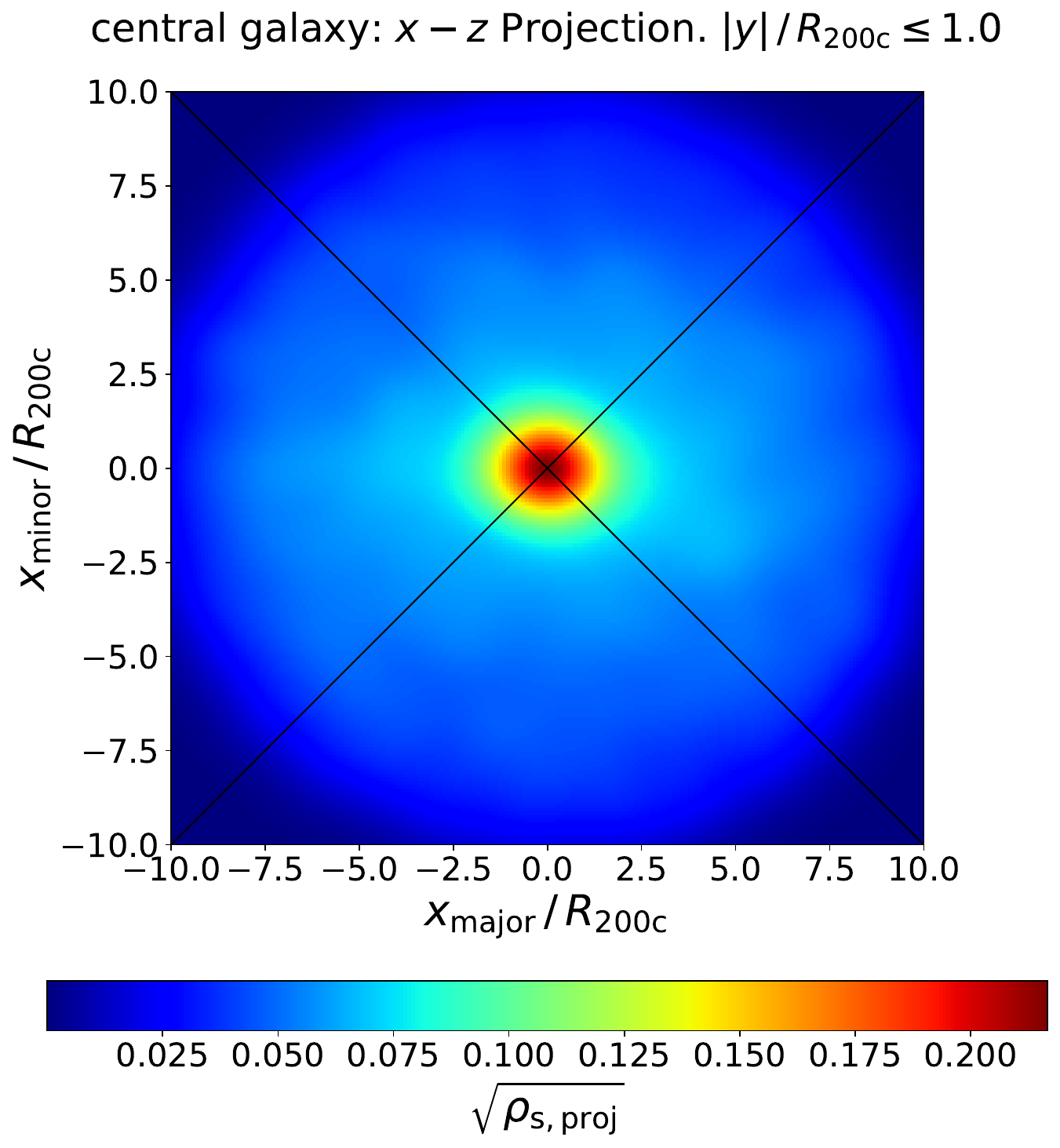}
    \end{minipage}
    \caption{Projected number density distribution of satellite galaxies stacked along the intermediate axis of host ellipsoids for systems with $M_\mathrm{200c} \geq 10^{11}M_{\odot}$.
    \textit{Top row}: distributions relative to halo ellipsoids.
    \textit{Bottom row}: distributions relative to central galaxy ellipsoids.
    Panels from left to right correspond to axis-aligned slices of thickness $0.6R_\mathrm{200c}$, $R_\mathrm{200c}$, and $2R_\mathrm{200c}$, respectively.
    The black diagonal lines denote cones with an apex angle of $90^{\circ}$. Halo-related anisotropy is significantly stronger than that associated with central galaxies.}
\label{fig:gal_proj_151}
\end{figure}

To better visualize the anisotropic distribution of satellite galaxies within host halos, we use galaxy group/cluster samples from the SIMBA simulation at $z=0$ with $M_{\rm 200c} \ge 10^{11} M_\odot$. For these samples, we slice satellite galaxies centered on the central galaxy, along the intermediate axis of the host halo and along the intermediate axis of the central galaxy separately (i.e., we perform slicing along the intermediate axis of the host halo and the intermediate axis of the central galaxy respectively). We then stack these galaxies to obtain the projected distribution map of satellite galaxy number density, as shown in Figure \ref{fig:gal_proj_151}. We discuss the slicing results with thicknesses of $0.6R_{\rm 200c}$, $R_{\rm 200c}$, and $2R_{\rm 200c}$ respectively. These three representative thicknesses correspond to shallow, moderate, and deep projection layers within the halo, and we employ this set of slices to validate the stability of the satellite anisotropy signal and rule out a strong dependence of our results on slice thickness selection.

As shown in the top row of Figure \ref{fig:gal_proj_151}, within the major-minor axis coordinate system of the host halo, the satellite galaxy distribution is significantly biased toward the major axis direction. This is a direct manifestation of the anisotropic satellite galaxy distribution mentioned in this work. Meanwhile, we can also observe that, given the significant strength of the anisotropic signal, the selection of a cone angle of $45^{\circ}$ is not a mandatory requirement; however, it plays an important role in acquiring more valid samples for robust analysis.

In contrast, as shown in the bottom row of Figure \ref{fig:gal_proj_151}, within the major-minor axis coordinate system of the central galaxy, the anisotropy of the satellite galaxy distribution is significantly weakened. This is fully consistent with our conclusion in Section \ref{sec:conclusions}, namely that the anisotropic distribution of satellite galaxies is dominated by the triaxial geometry of the host halo at halo scales, rather than by the central galaxy within it.

\section{Alternative Definitions of Aligned and Misaligned Samples} 
\label{sec:AS_MS_classification}
This section is dedicated to investigating the robustness of our results with respect to the adopted criterion for classifying AS and MS based on the angle between the major axis of the central galaxy and the major axis of its host halo. For comparison with Section \ref{sec:SDA_BCG_halo}, we adopt the same halo selection, using halos with $M_\mathrm{200c}\ge10^{13} M_\mathrm{\odot}$ and their corresponding central galaxies from the SIMBA simulation at $z=0$. In that section, we defined AS as samples where the central galaxy-halo major-axis angle is less than $40^{\circ}$, and MS as samples where this angle exceeds $50^{\circ}$. However, it is essential to verify whether the choice of these angle cuts affects the key conclusions regarding the anisotropic signals of satellite galaxies presented above.

To address this, we perform supplementary analyses using two alternative sets of alignment angle thresholds. The first alternative classification adopts stricter criteria: AS is defined as samples with a central galaxy-halo major-axis angle less than $30^{\circ}$, and MS as samples with an angle greater than $60^{\circ}$. The second alternative uses a symmetric threshold: AS corresponds to angles less than $45^{\circ}$, and MS corresponds to angles greater than $45^{\circ}$. For each alternative classification scheme, we replicate the same analyses as those in Section \ref{sec:SDA_BCG_halo}, focusing on the major-to-minor axis satellite galaxy count ratios as a function of radius.

\begin{figure}[htbp]
    \centering
    \begin{minipage}[b]{0.51\textwidth}
        \centering
        \includegraphics[width=\textwidth]{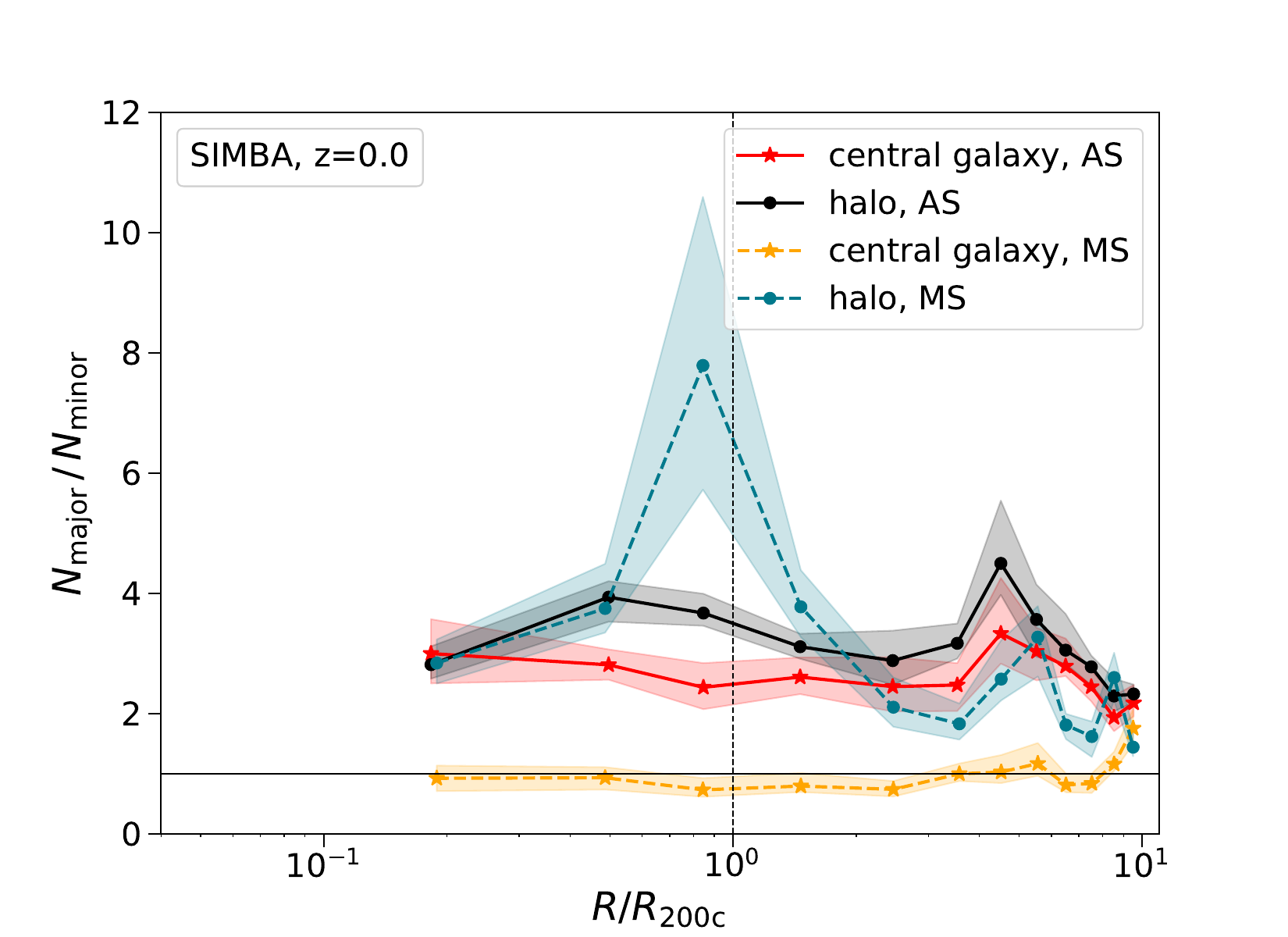}
        \small 
    \end{minipage}
    \hspace{-20pt}
    \begin{minipage}[b]{0.51\textwidth}
        \centering
        \includegraphics[width=\textwidth]{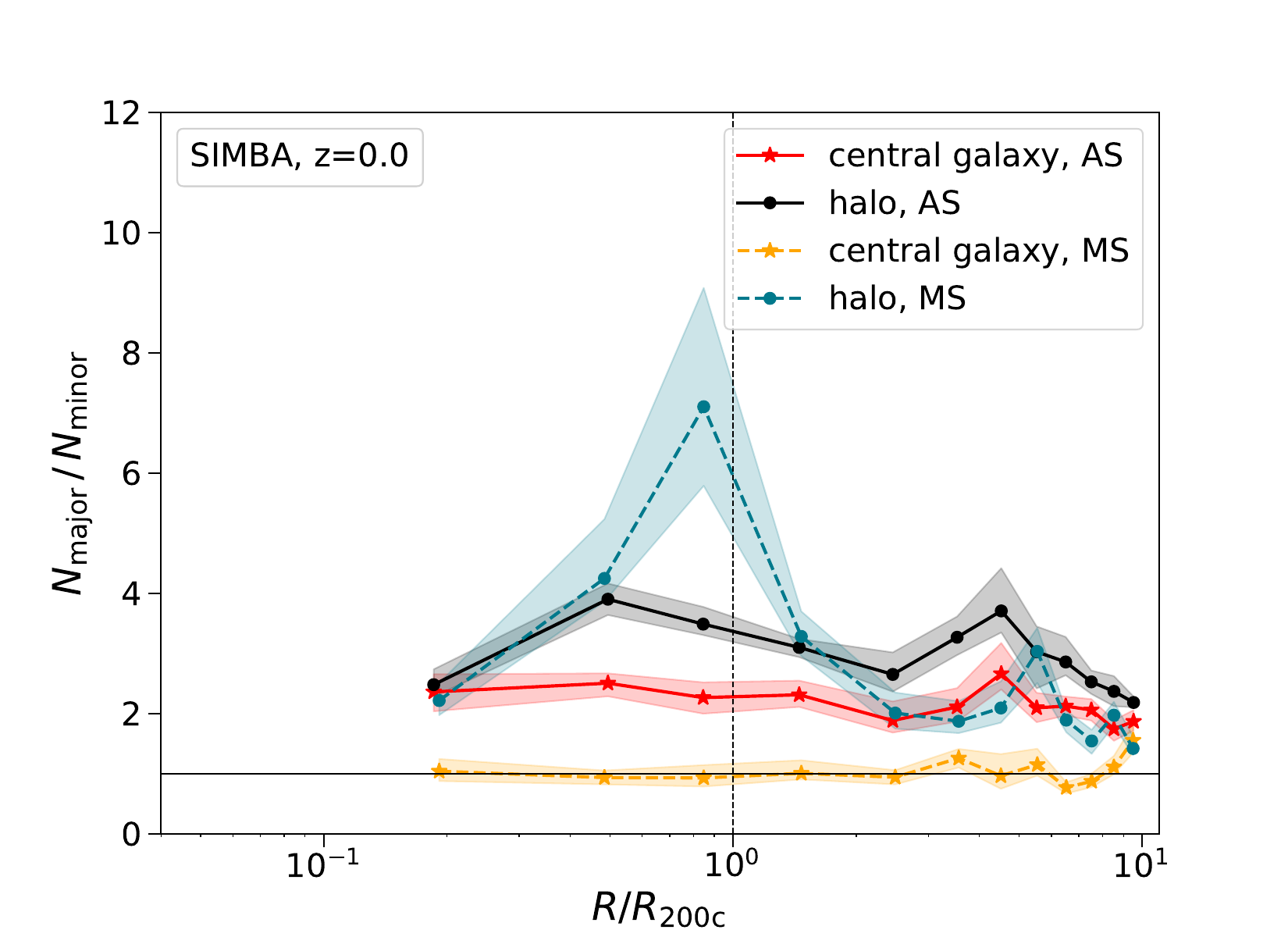}
        \small
    \end{minipage}
    \caption{Major-to-minor axis satellite galaxy count ratios for halos and central galaxies in the AS and MS as a function of radius, based on the SIMBA simulation at $z=0.0$. \textit{Left column}: AS defined as central galaxy-halo major-axis angle $<30^\circ$ and MS as angle $>60^\circ$; \textit{Right column}: AS defined as angle $<45^\circ$ and MS as angle $>45^\circ$. All axis definitions, curve color coding, and visual conventions are identical to Figure \ref{fig:Nx_Nz_R_halo_BCG_z}. Alternative definitions of AS and MS do not significantly affect the qualitative conclusions on the anisotropy strength across different samples.} \label{fig:Alternative_AS_MS}
\end{figure}

As shown in Figure \ref{fig:Alternative_AS_MS}, all alternative central galaxy-halo alignment angle thresholds do not lead to significant quantitative changes in the anisotropic signals of satellite galaxies reported in Figure \ref{fig:Nx_Nz_R_halo_BCG_z}. Specifically, the overall trends of the major-to-minor axis satellite count ratios across different radial bins remain consistent across all tested angle classification criteria. This demonstrates that our main conclusions regarding the anisotropy of satellite galaxy distributions are robust to the specific choice of central galaxy-halo alignment angle thresholds, confirming the reliability of the classification scheme adopted in the main text.

\section{Orientation Change Rates of Major and Minor Axes}
\label{sec:omega_major_minor}
As shown in Section \ref{sec:SDA_BCG_halo}, a considerable fraction of central galaxies have major axes misaligned with those of their host halos, and these samples are classified as MS in our study. Consistent with the previous analysis, we focus on halos with $M_\mathrm{200c}\ge10^{13} M_\mathrm{\odot}$ and their corresponding central galaxies from the SIMBA simulation at $z=0$. To preliminarily illustrate the physical origin of this misalignment, we calculated the orientation change rates of the major and minor axes for host halos and their central galaxies across three sample subsets: the total sample, the AS, and the MS. Here, the orientation change rate $\omega$ is defined as the angular shift of a triaxial principal axis divided by the corresponding time interval. This geometric quantity accommodates two physical origins: coherent bulk rotation, or intrinsic triaxial deformation that reorients axes without global rotation.

For each target galaxy group/cluster at $z=0$, we traced its counterparts in five snapshots at higher redshifts: $z=\{0.017, 0.034, 0.051, 0.068, 0.085\}$. For each pair of adjacent snapshots, we computed the angles between the axes of the halo and its central galaxy, and further derived the orientation change rates based on these angles. The final orientation change rate for each system was obtained by averaging the orientation change rates calculated from the five pairs of adjacent snapshots.

\begin{figure}[htbp]
    \centering
    \begin{minipage}[b]{0.9\textwidth}  
        \centering
        \includegraphics[width=\textwidth]{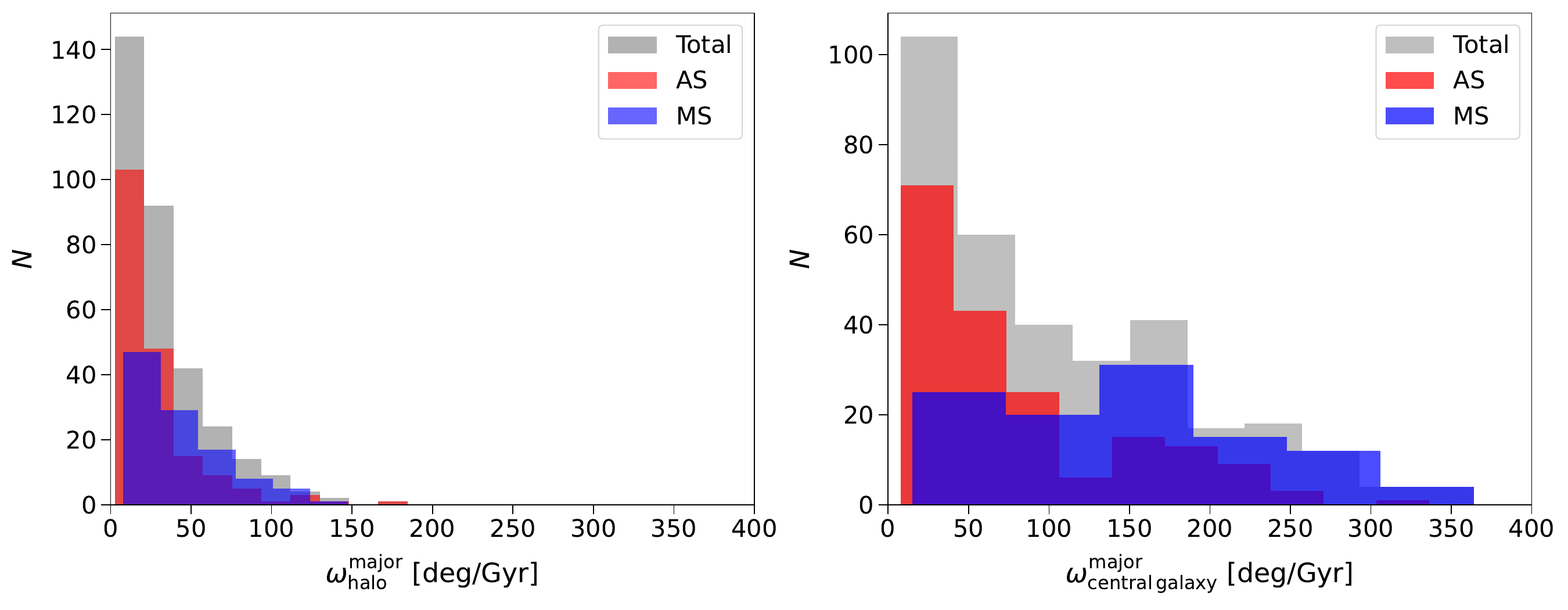}  
        \small 
    \end{minipage}
    \vspace{-1mm}  
    
    \begin{minipage}[b]{0.9\textwidth}
        \centering
        \includegraphics[width=\textwidth]{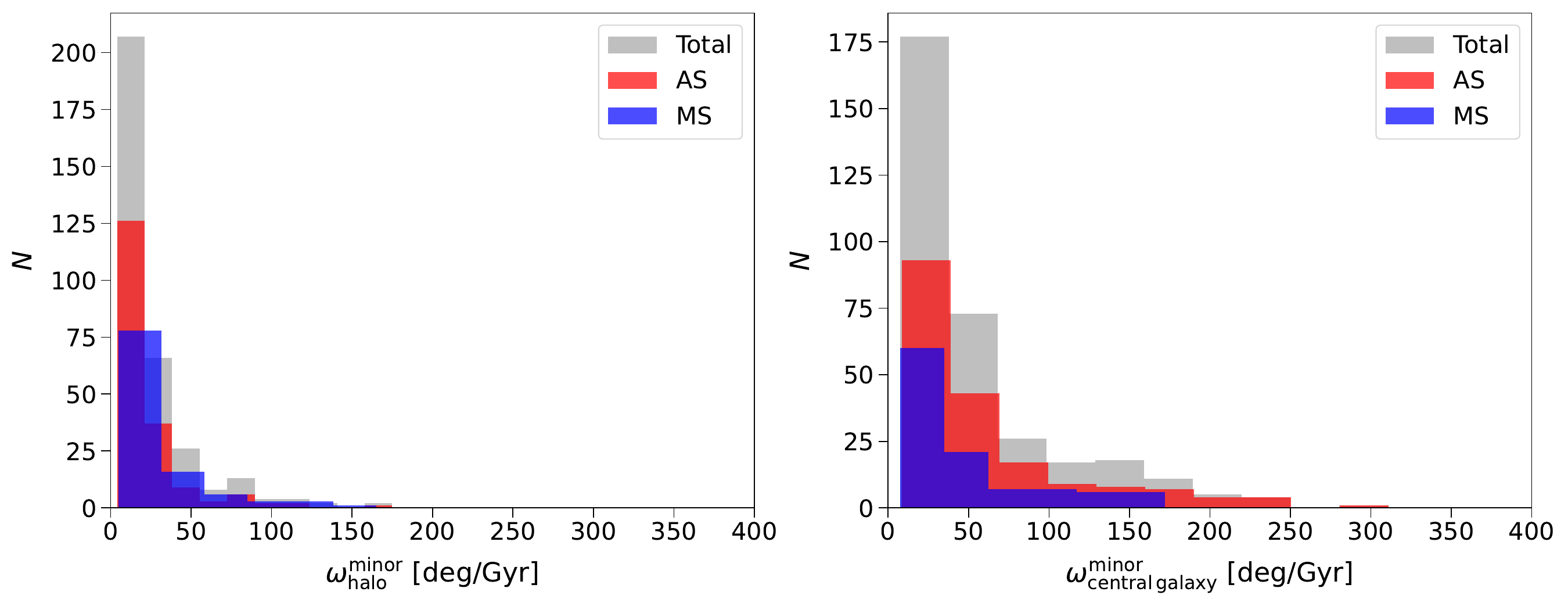}
        \small
    \end{minipage}
    
    \caption{Orientation change rates of the major and minor axes for host halos and their central galaxies. \textit{Top row}: orientation change rate of the major axis for the halo and its central galaxy. \textit{Bottom row}: orientation change rate of the minor axis for the halo and its central galaxy. Halo major and minor axis orientation change rates are significantly lower than those of central galaxies.}
    \label{fig:omega}
\end{figure}

The orientation change rates of the major and minor axes between halos and their central galaxies for the total sample, the AS, and the MS are presented in Figure \ref{fig:omega}. Statistically, the orientation change rates of the major axes of halos and central galaxies in the MS are larger than those in the AS. This larger difference in orientation change rates makes it more difficult for the major axes of halos and central galaxies in the MS to maintain aligned orientations, unless they evolve synchronously with identical orientation change rates. However, across all three subsets, the orientation change rates of both the major and minor axes of halos are significantly smaller than those of their corresponding central galaxies, which greatly suppresses the likelihood of synchronous evolution. In contrast, the lower orientation change rates of the AS contribute to more stable alignment between the orientations of halos and their central galaxies. On the other hand, we also find that for both halos and central galaxies, the orientation change rate of the major axis is faster than that of the minor axis. This result may share the same origin as the stability of the minor axis orientation of disk galaxies, i.e., the important role of angular momentum in regulating galaxy morphology \citep{2007MNRAS.378.1531K, 2025A&A...703A..67M}.

\begin{acknowledgments}
Z. Z. and Y. C. would like to thank Dr. Liang Gao, Dr. Peng Wang, Dr. Shihong Liao, and Dr. Lan Wang for their helpful discussions. This work has been supported by the National Key Research and Development Program of China (No.\ 2022YFA1602903), the National Natural Science Foundation of China (Nos.\ 12588202 and 12473002),  and the China Manned Space Program with grant no.\ CMS-CSST-2025-A03. 
 W.C. gratefully thanks Comunidad de Madrid for the Atracci\'{o}n de Talento fellowship no. 2020-T1/TIC19882 and Agencia Estatal de Investigaci\'{o}n (AEI) for the Consolidaci\'{o}n Investigadora Grant CNS2024-154838. He further acknowledges the Project PID2024-156100NB-C21, financed by MICIU/AEI/10.13039/501100011033/FEDER, and the science research grants from the China Manned Space Project.
\end{acknowledgments}

\vspace{5mm}

\software{\texttt{CAESAR}\footnote{https://github.com/dnarayanan/caesar}, \texttt{sklearn} \citep{scikit-learn}, \texttt{DisPerSE} \citep{2011MNRAS.414..350S, 2011MNRAS.414..384S}}

\bibliography{new}{}
\bibliographystyle{aasjournal}

\end{document}